\documentclass{article}

\usepackage{arxiv}

\usepackage[utf8]{inputenc} 
\usepackage[T1]{fontenc}    
\usepackage{hyperref}       
\usepackage{url}            
\usepackage{booktabs}       
\usepackage{amsfonts}       
\usepackage{nicefrac}       
\usepackage{microtype}      
\usepackage{lipsum}
\usepackage[inline]{enumitem}
\usepackage{diagbox}
\usepackage{threeparttable}
\usepackage{graphicx}
\usepackage{multirow}
\usepackage{amssymb}
\usepackage{amsmath}
\usepackage{verbatim}
\usepackage{comment}
\usepackage{xcolor}
\usepackage{makecell}

\title{Survey for Trust-aware Recommender Systems: A Deep Learning Perspective}

\author{
  Manqing~Dong \\ 
  University of New South Wales\\
  Sydney, Australia\\
   \And
 Feng~Yuan \\
  University of New South Wales\\
  Sydney, Australia\\
  \And
  Lina~Yao \\
  University of New South Wales \\
  Sydney, Australia\\
  \And
  Xianzhi~Wang \\
  University of Technology Sydney\\
  Sydney, Australia\\
  \And 
  Xiwei~Xu \\
  Data61, CSIRO\\
  Sydney, Australia\\
  \And 
  Liming~Zhu \\
  Data61, CSIRO\\
  Sydney, Australia\\
}

\begin{document}
\maketitle

\begin{abstract}
A significant remaining challenge for existing recommender systems is that users may not trust the recommender systems for either lack of explanation or inaccurate recommendation results. Thus, it becomes critical to embrace a trustworthy recommender system. This survey provides a systemic summary of three categories of trust-aware recommender systems: social-aware recommender systems that leverage users' social trust relationships; robust recommender systems that filter untruthful noises (e.g., spammers and fake information) or enhance attack resistance; explainable recommender systems that provide explanations of recommended items. We focus on the work based on deep learning techniques, an emerging area in the recommendation research. 
\end{abstract}


\section{Introduction}


Users face with a significant challenge of information overload as the Web information continuously grows. Recommender systems provide information, products, or services to meet users' personalized tastes and preferences to alleviate the above issue. Given those advantages, recommender systems have been popular in various domains and widely deployed in e-commerce applications \cite{zhang2019deep}.
For example, when a user looks for a pair of sleep-buds at Amazon\footnote{\url{https://www.amazon.com/}}, the website may recommend a case for the sleep-buds as ``Frequently bought together" and a list of products that other customers frequently ``buy after viewing this item."
According to a report\footnote{\url{http://rejoiner.com/resources/amazon-recommendations-secret-selling-online/}}, 35\% of Amazon.com's profit owes to its recommendation algorithms. 

Despite the success of recommender systems, a significant remaining challenge is that users may not trust the recommender systems for either lack of explanation or inaccurate recommendation results. For example, a user may not trust a stranger's taste even though they have similar history records; moreover, the system may recommend an item that is intentionally highly rated by malicious users. All these make a trustworthy recommender system urgent and important.
In this work, we define and focus on three aspects of trust in recommender systems:

\textbf{Social-awareness.}
With the development of social media, social-aware (sometimes called trust-aware) recommendation has attracted more attention.
Recent studies suggest users' ratings area has a positive correlation with the average of their social neighbors for both trust-alike relationships or trust relationships~\cite{guo2015trustsvd}. On the one hand, based on the phenomenon that users' tastes are often influenced by their friends \cite{pan2017trust}, leveraging the trust relationship has great potential to provide a trustworthy recommender system to the user, and also improve the recommendation quality by predicting the user's taste. On the other hand, adding such information alleviates the cold start problem of traditional recommender systems. 
Generally, social-aware recommender systems include memory-based methods and model-based methods.
Memory-based methods usually generate predictions for a user by leveraging the ratings of his/her direct or indirect trusted friends \cite{yang2017social}. Thus, the performance of such methods largely depends on trust propagation models. In addition, these methods are usually time-consuming and therefore are not suitable for handling large scale applications because they need to calculate similarities over an entire rating matrix and the whole trust network \cite{yang2017social}. 
Model-based methods (e.g., matrix factorization and deep learning) are widely used \cite{ma2017explicit}. Compared with memory-based methods, which uses trust ties to infer users' neighbors and then promote the accuracy of similarity calculation among users, most model-based methods simultaneously map users and items into low-dimensional feature spaces and then train a prediction model by optimizing some objective functions overrating and trust data \cite{yang2017social}.

\textbf{Robustness.}
While more people are relying on online product reviews to make purchase decisions \cite{fei2013exploiting}, reviews and social relations are increasingly subject to attacks from spammers and bots.
Moreover, social relations have different interpretations in different contexts. For example, two people may reach a consensus on movies but have diverse opinions on purchasing clothes \cite{wu2018socialgcn}.
There are two directions of research in improving the robustness of recommender systems.
The first direction is towards filtering out noisy or malicious feedback from the data before executing the recommendation algorithm. Research in this direction aims to find statistical patterns of users/ratings to identify malicious content \cite{zou2013belief}, use supervised classification methods based on feature-engineering \cite{yang2016re}, or use unsupervised clustering approach to eliminate suspicious users \cite{bhaumik2011clustering}.
The second direction aims to develop noise attack-resistant algorithms \cite{si2018shilling}. Related research mainly constructs noise-sensitive algorithms or introduce auxiliary information (e.g., human-made noise) into the recommendation~\cite{dong2017hybrid}.

\textbf{Explainability.}
Users tend to trust a recommendation when provided the appropriate information to understand the recommendation process and results.
Explainable recommender systems \cite{zhang2018explainable} not only provide users with personalized recommendations but also generate descriptions on why the items are recommended. Therefore, explainability improves both the trustworthiness and transparency of recommendation results. In general, explainable recommender systems can be classified according to two orthogonal criteria: information source and methodology. Existing work utilizes a variety of contents, such as features of items/users \cite{zhao2016exploring},  textual reviews \cite{zhang2014explicit}, product images \cite{lin2019explainable}, and social connection \cite{wang2014also}. And various approaches have been used to generate explanations, e.g., matrix factorization \cite{chen2016learning}, graph-based models \cite{heckel2017scalable}, topic models \cite{ren2017social}, deep learning \cite{seo2017interpretable}, and association rule mining \cite{davidson2010youtube}. In addition, much work exists that combining different information sources and methods. In this survey, we limit our scope to only deep learning models and give a thorough analysis of the recent work.

Deep learning techniques have exploded during the recent years. The 2019 Turing Award also recognizes the significant contribution of deep learning to various machine learning tasks. A recent survey on deep learning-based recommender systems \cite{zhang2019deep} points out the neural architecture's ability of being end-to-end differentiable and providing suitable inductive biases. 
One significant contribution of deep learning approaches is about the representation learning. For example, many studies use deep learning to learn compact representations from auxiliary data such as content, tag, images, or social graph relationships, and then use the compact representations for prediction or combined with traditional matrix factorization methods \cite{pan2017trust,monti2017geometric}.

Until now, considerable work has been conducted on applying deep learning into the trust-aware recommendation, including social-aware recommender systems \cite{yang2014survey}, robust recommender systems \cite{gunes2014shilling}, and explainable recommendation \cite{zhang2018explainable}. However, to the best of our knowledge, there is a lack of systemic survey on current deep learning-based trust-aware recommendation methods. 
This survey aims to review the trust issue in recommender systems from a deep learning perspective to fill the gap. We outline three aspects of trust, i.e. social-awareness, robustness, and explainability. For each aspect, we present the literature review and summarize the related deep learning-based techniques. 

The rest of the survey is organized as following.
From section \ref{sec:social-aware} to \ref{sec:explainable}, we introduce three categories of trust-aware recommender systems, respectively. In each section, we first provide an overview of the related methods and then introduce the technical details of deep learning-based approaches by the type of algorithms.
Then, in section \ref{sec:potential trends}, we summarize several challenges of current trust-aware recommendation techniques and provide insights into this field. 
The following sections assume readers have the basic understanding of deep learning techniques and concepts of recommendation techniques.


\section{Preliminaries}
\subsection{A Brief Overview about Recommender Systems}
\subsubsection{Tasks}
Traditional tasks about the recommendation problem include rating prediction and top-k recommendation. 
For the \textbf{rating prediction} (or link prediction) problem, the "rating" could be a binary value, e.g., 1 for click or purchase and 0 for no actions, or a numerical value between a interval (e.g. 1 to 5). For both conditions, predicting the ratings is to quantify the user's preferences as ratings. 
For example, for the matrix factorization based methods \cite{koren2009matrix}, the rating is normally predicted by $\hat{r}=u^{\top}v$, where $u$ and $v$ are learned latent factors for users and items. And the ways of learning the latent representation are various. In \cite{mnih2008probabilistic}, the authors suppose the latent representations are following some Gaussian distributions, and they learn such distributions by the maximum a posteriori estimation. While in \cite{he2017neural}, the latent representations are learned with neural networks. 
For the \textbf{top-k recommendation} problem, the goal of recommender systems are to predict a item list for each user to satisfy his/her taste. The goal of learning-to-rank is to define a personalized ranking function, and then learn loss functions to improve the ranking performance in the top-k recommendation problem. For example, in \cite{hu2008collaborative}, a ranking-oriented approach has been proposed to measure confidence for each user-item pair and improve the matrix factorization method for top-k recommendation. A Bayesian Personalized Ranking (BPR) algorithm is proposed to direct learn the ranking relation based on implicit feedback for top-K recommendation \cite{rendle2009bpr}.

\subsubsection{Methods}
From the perspective of the used techniques, recommender system can be roughly categorized into content-based, collaborative filtering-based and hybrid methods.
\textbf{Content-based methods} use user's or item's profiles to provide a recommendation list. 
For example, with utilizing the user profiles, the system can simply display recently visited items to facilitate the user returning to these items, or filter out from a recommendation system an item that the user has already purchased or read \cite{pazzani2007content}. In \cite{chen2014context}, they authors incorporate not only topic modeling to mine item content but also social matrix factorization to handle ratings and social relationships for recommendation.
\textbf{Collaborative filtering methods} make recommendations by learning from user-item historical interactions, either explicit (e.g. user's previous ratings) or implicit feedback (e.g. browsing history), which employ the preferences of a set of users so similar to the target user in the recommendation process. 
To this aim, several research efforts have been made to propose similarity measures so as to identify these users\cite{parvin2019tcfaco}, e.g. leveraging user-user relations or inter-item similarities. 
CF methods discover hidden preferences of users from past activities of users, i.e., the user-item rating matrix, to make recommendations. However, CF approaches, including matrix factorization methods \cite{mnih2008probabilistic,koren2009matrix}, suffer from data sparsity and cold start issues\cite{lam2008addressing}. For example, matrix factorization techniques cannot effectively learn the latent feature vectors for users with only a few ratings or newly added items. Nonetheless, matrix factorization techniques offer a flexible framework to incorporate additional sources of information to alleviate data sparsity and cold start issues\cite{yu2018joint}.

\subsubsection{Challenges}
Several challenges about recommender system are the \textbf{data sparsity problem}, the \textbf{cold-start problem}, the \textbf{scaling issues}, and the \textbf{trust problem} discussed in this survey. 
For example, in real-world applications like e-commerce websites, there are a vast number of users and items, but the user-item rating matrix which stores the numerical ratings on items by users is usually very sparse, which is called data sparsity problem. The cold-start problem may occur for some new users, thus the system cannot provide satisfactory recommendations due to the insufficient information. Different from data sparsity problem, the recommender could also face scaling issues when use some memory-based techniques. For example, it is computationally expensive to generate recommendation using user-based collaborative filtering methods. This is caused by the computation of similarity between all the users or items. It is obvious that the larger the numbers of users and items are, the more time-consuming it will be when computing similarities \cite{parvin2019tcfaco}. 

\subsection{Deep Learning for Recommendation}
Deep learning (DL)-based techniques, which have shown the effectiveness in representation learning, have been adopted in many application fields, such as computer vision\cite{voulodimos2018deep} and natural language processing\cite{pouyanfar2018survey}. 
Such attractive features also impact on the research for recommender systems, where a large volume of work has dedicated to deep learning based recommender systems in recent years. Please check \cite{zhang2019deep} for details about deep learning based recommender systems.
To roughly classify the DL-based recommender systems, the DL could be used for: representation learning, predictive learning and generative learning.

\textbf{Representation learning.}
Many studies use deep learning to learn compact representations from auxiliary data such as content, tag, images, or social graph relationships, and then use the compact representations for prediction or combined with traditional matrix \cite{pan2017trust}.
For example, the autoencoder can be used for learning the compact representation for items and users with the bottleneck layer \cite{sedhain2015autorec,strub2016hybrid}; convolutional neural network (CNN) can be used for extracting information from images or social graphs \cite{wu2019hierarchical,monti2017geometric}; recurrent neural network (RNN) can learn the textual information or capture the dynamic user interest\cite{song2019session,bansal2016ask}; etc.

. 

\textbf{Predictive learning.}
A predictive model is either a classification model for predicting the user's preference or a regression model that generates the potential ratings. 
For example, the prediction for ratings could be a simple multilayer perceptron with learning the representations from the users and items\cite{guo2017deepfm}. The encoder part of the autoencoder, which is trying to reconstruct inputs, can be also regarded as a predictor for ratings\cite{sedhain2015autorec}. Attention mechanisms, which learn the weights for inputs, can be used for enhancing the performance by emphasis more on the helpful parts of inputs\cite{sun2018attentive,ying2018sequential}.

\textbf{Generative learning.}
The generative learning for recommender systems is majorly using generative models, e.g. variational autoencoders (VAE) and generative adversarial neural network (GAN), for producing the potential ratings or scores. 
For example, the generative adversarial neural network, which include a generator and a discriminator, can produce samples with similar distribution toward real instances by training two operator simultaneously in a minimax game framework\cite{goodfellow2014generative}. This property can be used for generating the ratings or dealing with missing values when meet recommendation problems\cite{he2018adversarial,wang2017irgan}. 
For example, Wang et al. \cite{wang2018neural} proposed using GAN to generate negative samples for the memory network based streaming recommender system. In \cite{wang2019minimax}, the authors proposed UGAN to capture the pattern of the original data input and generate similar user profiles, which provides a promising way to mitigate the adverse impact of missing data.

\section{Social-aware Recommender Systems}
\label{sec:social-aware}
\subsection{Overview}
Social relations have proven helpful in boosting recommendation performance and thus has attracted much attention these years \cite{fan2019graph}.
Social recommendation techniques make use of the user-user trust social links to complement the sparse rating data and thus improve the user preference prediction by considering not only the user rating behavior but also the preference of the user trusted neighbors. 
Since users usually are interactive with people around them, social relations can greatly help users filter information and alleviate the cold start problem. 

\subsubsection{Problem Definition}
Suppose $\{u|u\in U\}$ is a list of $M$ users, $\{i|i\in I\}$ is a list of $N$ items, and $R$ is the user-item rating matrix, where $r_{ui}\in R$ is the rating by user $u$ for item $i$. 
We denote by $t_{uv}$ ($t \in T$) the trust relationship between user $u \in U$ and user $v \in V$, where $T$ is the user-user trust relationship matrix. 
Besides, we denote item description and user feedback by $D_{u}$ and $D_{i}$. Each element in $D_{u}$ and $D_{i}$ is a feature vector denoted by $d_{*}$. 
Rating-based recommendation tasks aim to predict a rating $\hat{r}_{ui}$ of user $u$ on an unknown item $i$, 
\begin{equation}
\label{eq:r-pred}
    \hat{r} = f(R, T, D_{u}, D_{i})
\end{equation}
where $f$ links the given information to ratings in a fixed range, e.g., [0,5].

In contrast, rank-based tasks aim to provide a user $u$ top-$K$ items with the highest score $s_{ui}$, which capture users' preference. 
\begin{equation}
    \hat{s} = f(R, T, D_{u}, D_{i})
\end{equation}

\subsubsection{Traditional Methods}
Traditional methods for social-aware recommendation include memory-based methods and model-based methods.
The \textbf{memory-based} methods deduce ratings of a targeted user via trust propagation based on ratings of its friends \cite{wu2019dual}.
For example, Jamali and Ester \cite{jamali2009using} combine TrustWalker \cite{jamali2009trustwalker} with neighborhood collaborative filtering. They first run random walks on the trust network and then perform a probabilistic item selection strategy to generate recommendations.
Similarly, Zhang et al. \cite{zhang2017collaborative} extract reliable social information from user feedback and use top-k identified friends to infer the user preferences.
Matrix factorization is probably the most widely used technique for \textbf{model-based} social-aware recommendation. 
Wen et al. \cite{wen2018network} learn vector representations of social relations via node2vec\cite{grover2016node2vec} and then combine them with rating history to conduct matrix factorization. 
Zhao et al. \cite{zhao2017social} present a trust-based Bayesian personalized Ranking approach to incorporate trust friendship. They assume the friends preference will affect the users decisions, i.e. the user will give higher ranks to items that preferred by their friends.
Guo et al. \cite{guo2015trustsvd} use a SVD++\cite{koren2008factorization} based methods with considering the user preference and friend's influence.
Ahn et al. \cite{ahn2018binary} provide the theoretical support for considering social relationships in recommender systems.

Deep learning-based social-aware recommendation methods diverge in three types: 
\textbf{regularization methods} minimize the distances of latent features between trusted users and maximize the latent features' distances between distrusted users to reflect social proximity; 
\textbf{ensemble methods} generate a new rating from the ratings of both a user and its social network; 
and \textbf{co-factorization methods} assume users should share the same user preference in rating and social space \cite{zhang2018social}.

We summarize various techniques for social-aware recommender systems, including autoencoder, recurrent neural network (RNN),  graph neural network (GNN), generative models (GM), and hybrid methods, in Table~\ref{tab:social_aware}.

\begin{table}[ht]
\caption{ A Summary for Social-aware Recommender Systems}
\centering
    \begin{tabular}{|c|p{70pt}|p{90pt}|p{80pt}|p{80pt}|}
    \hline
    Method & Regularization & Ensemble & Co-factorization & Others \\ \hline
    Autoencoder & \cite{nisha2019social} & \cite{pan2017trust},\cite{wang2019trust},\cite{wu2018collaborative},\cite{deng2016deep} & \cite{nisha2019social} & \cite{rafailidis2017recommendation}\\ \hline
    RNN & - & \cite{sun2018attentive,song2019session} & - & -\\ \hline
    GNN  & - & \cite{fan2019graph,song2019session,wu2019graph,wu2019dual} & \cite{wu2018socialgcn} & \cite{ying2018graph}\\ \hline
    GM  & - & \cite{fan2019deep},\cite{xiao2019variational} & - & -\\ \hline
    Hybrid Methods & - & \cite{wu2019hierarchical},\cite{bao2018contextual},\cite{liu2018social2},\cite{gao2017unified},\cite{geng2015learning} & - & \cite{monti2017geometric,zhao2017social,liu2015learning} \\ \hline
    Others & - & \cite{chen2019social},\cite{rafailidis2019neural},\cite{xiao2017neural},\cite{ren2017social} & - & \cite{fan2018deep,wang2017item,wen2018network,liu2018social} \\ \hline
    \end{tabular}
\label{tab:social_aware}
\end{table}

\subsection{Autoencoder-based Methods}
Autoencoder is a type of artificial neural network for learning compressed representations (encodings) for a set of high-dimensional data \cite{deng2016deep}.
Autoencoders can help recommendation with either learning latent factors of users and items (encoder) or reconstructing user's preferences (decoder). 
For the former one, the learned latent representations are normally further cooperated with other methods for prediction. 
For example, Deng et al. \cite{deng2016deep} use the learned latent factors as the initialization of matrix factorization. 
For the latter one, the decoder of the autoencoder is inferring the potential ratings from the user rating records through a narrow network, where the bottleneck of the network is representing the latent representation of the user rating records \cite{pan2017trust}.
For the social recommendation problem, where we have both user-item rating matrix and user-user trust networks, we also divide the autoencoder based methods into the two aforementioned categories: use \textbf{reconstructed input for recommendation} and use \textbf{latent representation for recommendation}.

\subsubsection{Using Reconstructed Input for Recommendation}
This type of method learns from dense representation and reconstructs input as predictions for recommendation. A key problem for this type of method in social-aware recommendation problems is the way of learning and aggregating representations from the social and rating information. 

\begin{figure}[ht]
\centering
    \begin{minipage}[t]{5cm}
    \includegraphics[width=4cm]{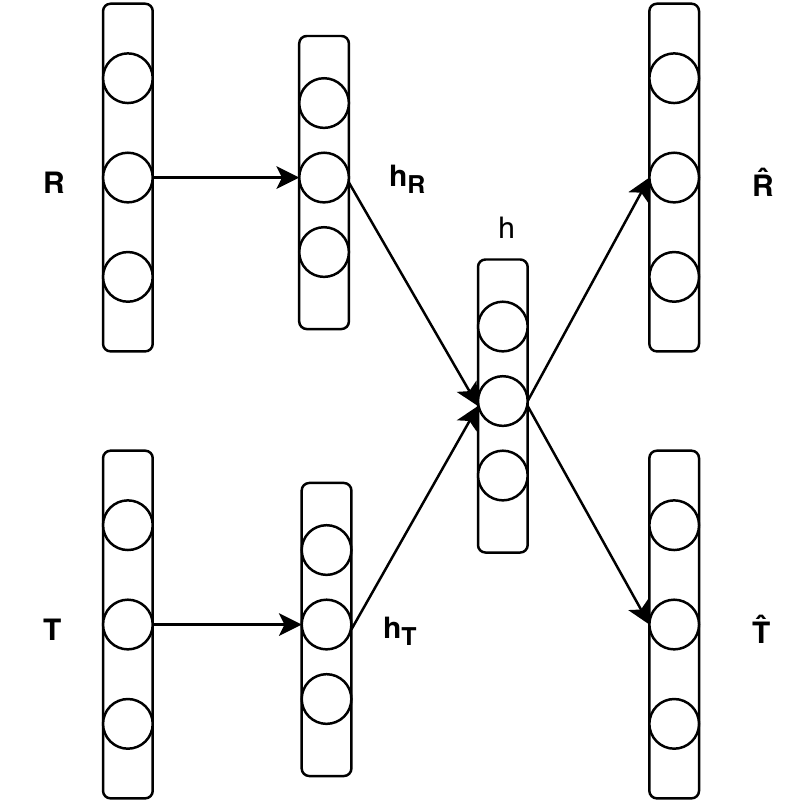}
    \centering{(a)}
    \end{minipage}
    \begin{minipage}[t]{5cm}
    \includegraphics[width=4cm]{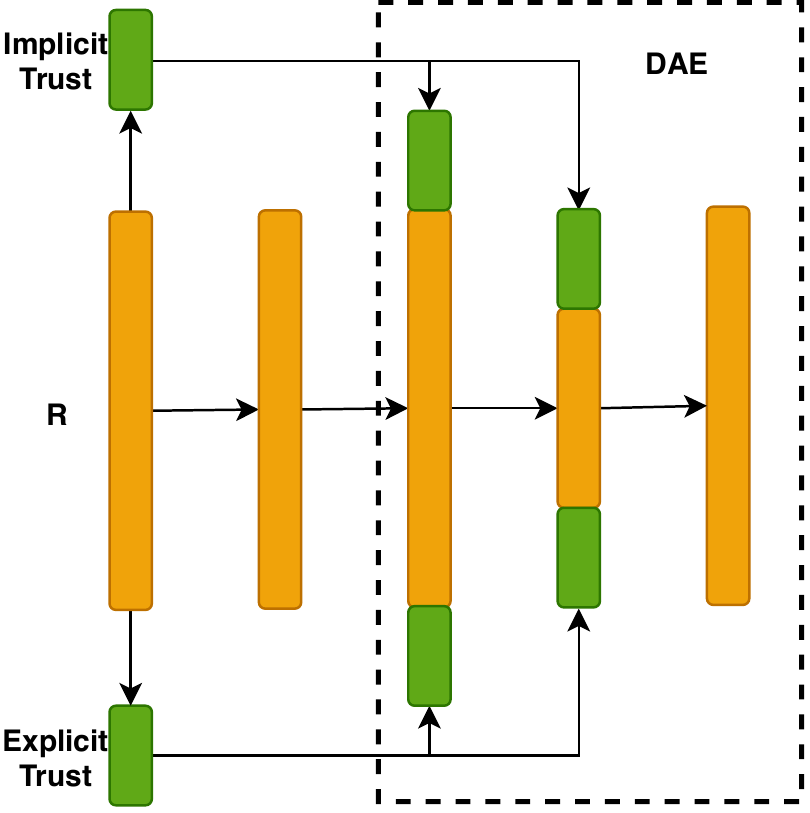}
    \centering{(b)}
    \end{minipage}
    \begin{minipage}[t]{5cm}
    \includegraphics[width=4cm]{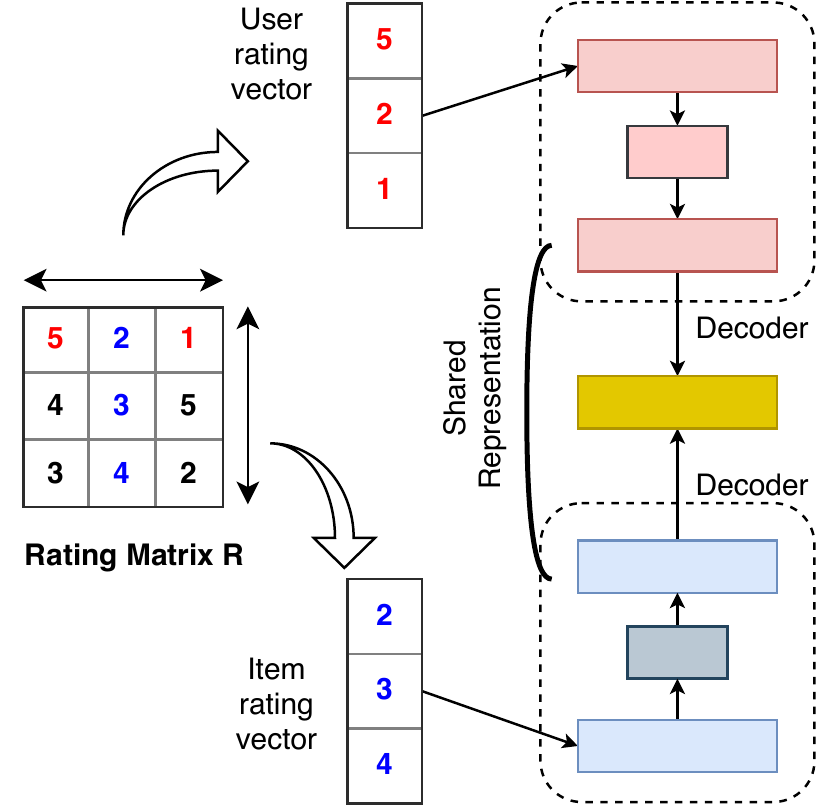}
    \centering{(c)}
    \end{minipage}
    \begin{minipage}[t]{5cm}
    \includegraphics[width=\textwidth]{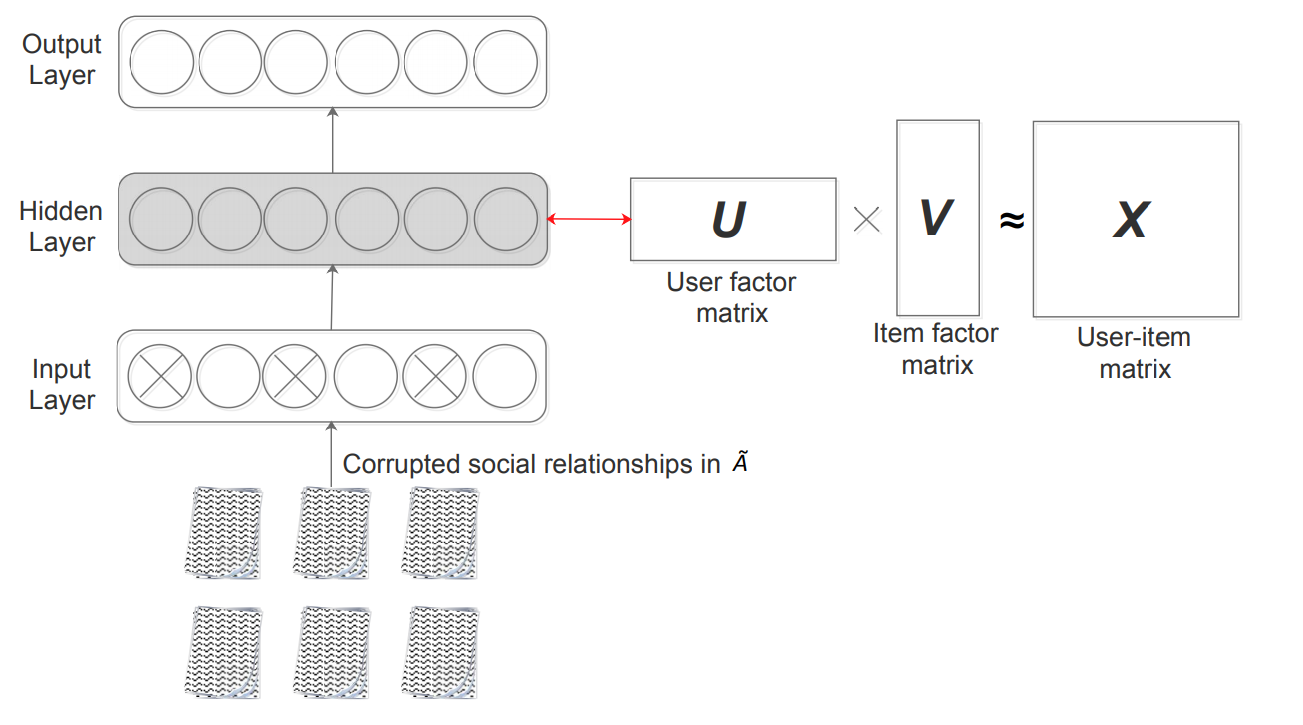}
    \centering{(d)}
    \end{minipage}
    \begin{minipage}[t]{5.5cm}
    \includegraphics[width=\textwidth]{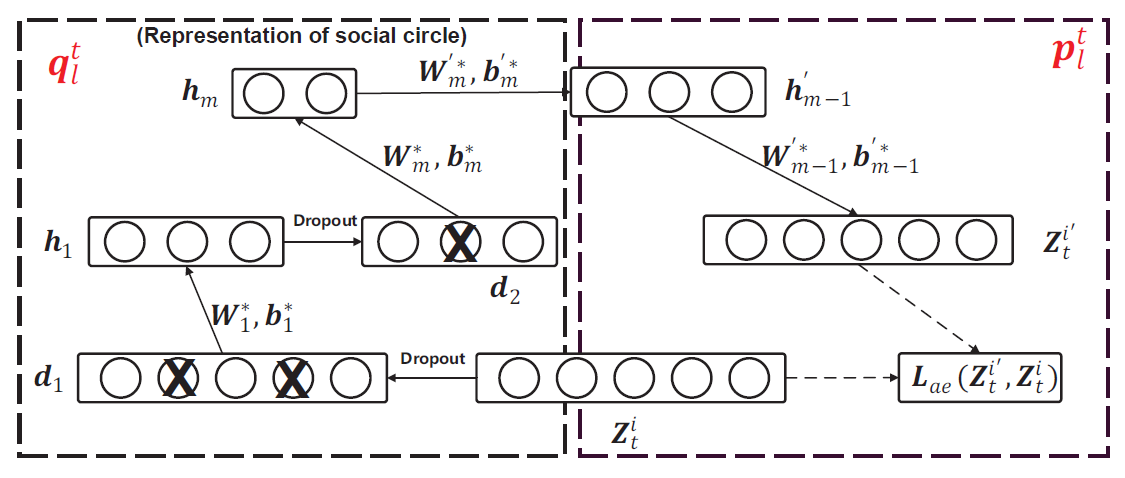}
    \centering{(e)}
    \end{minipage}
    \begin{minipage}[t]{4.5cm}
    \includegraphics[width=\textwidth]{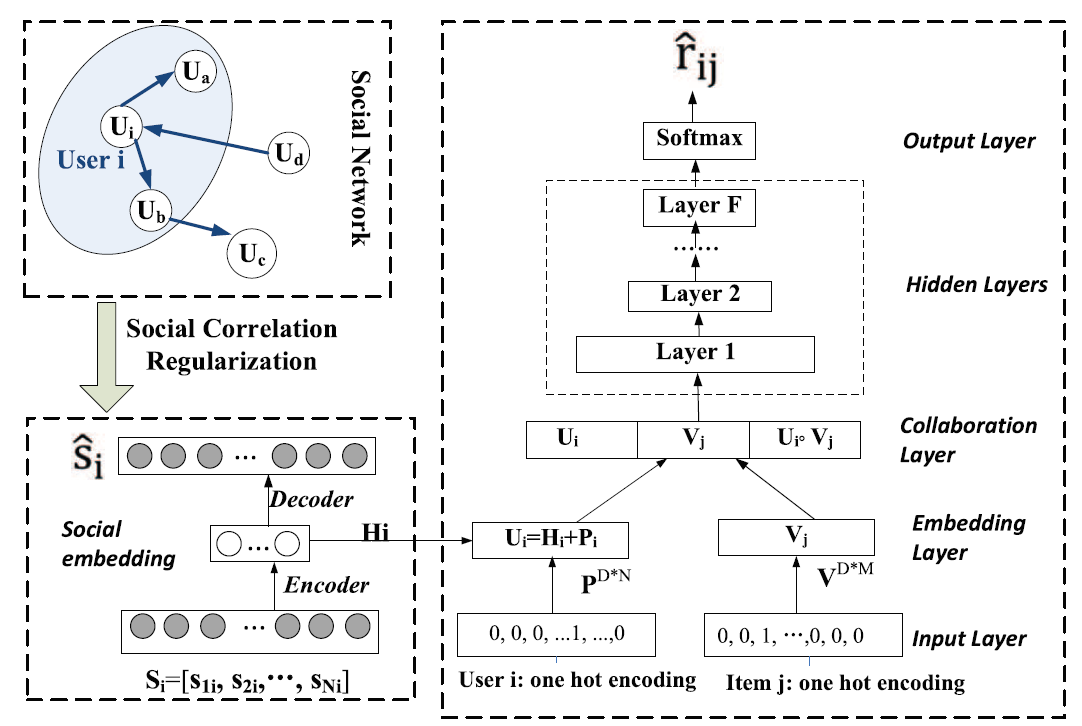}
    \centering{(f)}
    \end{minipage}
    \caption{Autoencoder for social-aware recommendation. The figures are from (a) Pan et al. \cite{pan2017trust}, (b) Wang et al. \cite{wang2019trust}, (c), Nisha et al. \cite{nisha2019social}, (d) Rafailidis et al. \cite{rafailidis2017recommendation}, (e) Liu et al. \cite{liu2018social2}, and (f) Wu et al. \cite{wu2018collaborative}.}
\label{fig:autoencoder_social}
\end{figure}

One way is to learn an ensemble representation of the two types of information for prediction.
For example, Pan et al. \cite{pan2017trust}  balance the contributions of the two representations learnt from social relationships and rating history via a weighted layer (Figure~\ref{fig:autoencoder_social} (a)).
They then use a correlative regularization to exchange information \cite{pan2017trust} and the unified latent representation to predict user ratings and trust relationships. Wang et al. \cite{wang2019trust}, instead, directly concatenate the two latent representations for recommendation (Figure~\ref{fig:autoencoder_social} (b)).

Another idea is to assume that a user's social representation share the same representation with the user's rating representation. For example, Nisha et al. \cite{nisha2019social} generate a list of trusted users and minimize the distance between their social representation to learn users's social representations. They first use an autoencoder to encode users' rating patterns and item's history rating patterns and then decode the learned representations for recommendation (marked as yellow in Figure~\ref{fig:autoencoder_social} (c)).
They also use a regularization item to control the distance between the user's social representation and the user's rating representation when training the autoencoder.

\subsubsection{Using Latent Representation for Recommendation}
Another idea for utilizing autoencoder for social-aware recommendation is to combine the learned latent representation with other methods. 
For example, Rafailidis et al. \cite{rafailidis2017recommendation} learn user latent representations from the social relationships via deep autoencoders and then use the latent factors in matrix factorization (Figure~\ref{fig:autoencoder_social} (d)). Liu et al. \cite{liu2018social2} use stacked denoising autoencoder (SDAE) \cite{vincent2010stacked} to learn the social information. The input of SDAE is from K friends of a given user, and each friend is represented by a vector. By aggregating the information from all friends, a condense vector is used for representing the user, see Figure~\ref{fig:autoencoder_social} (e). Such representation vector is further combined with other methods for recommendation. Similarly, \textit{Wu et al.}\cite{wu2018collaborative} use autoencoder to extract the compact representation of the social network and predict ratings by aggregating user information and item information via several fully connected neural network layers (Figure~\ref{fig:autoencoder_social} (f)). 

\subsection{RNN-based Methods}
Recurrent Neural Network (RNN) has shown its power in dealing with sequential data, e.g., textual \cite{lai2015recurrent} and time-series data \cite{pinheiro2014recurrent}. 
RNN-based methods generally target in dynamic user behaviors \cite{wu2016personal}, preferences\cite{donkers2017sequential,wu2017recurrent} or the side information \cite{wu2016joint}. Such methods majorly uses RNN to capture the sequential information and then learn a temporal or concrete representation for further uses. For example, Song et al.\cite{song2019session} capture the users' current preferences with RNN based methods and concatenate such information with user's history clicks as the representation for users. 

\begin{figure}
    \centering
    \begin{minipage}[t]{9cm}
    \includegraphics[width=\textwidth]{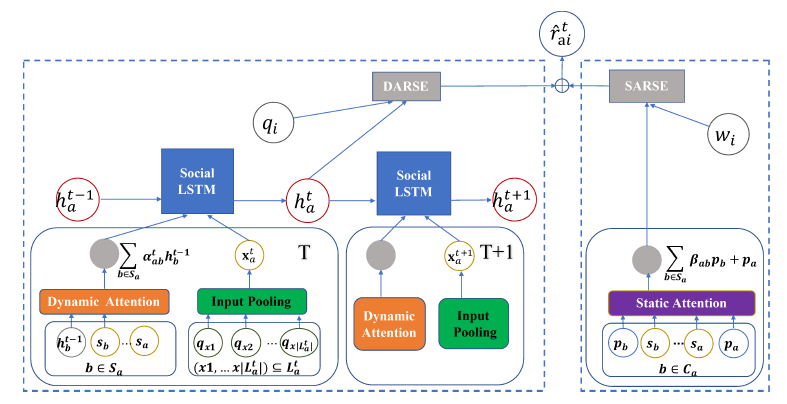}
    \end{minipage}
    \caption{Recurrent neural network for social-aware recommendation. (Sun et al. \cite{sun2018attentive}) }
\label{fig:social_rnn}
\end{figure}

For social-aware recommendation problem, RNN can extract user's temporal preferences and temporal biases of friends. For example, Amy liked painting last year and so as her friend Sarah. She was influenced by Sarah's preferences that she bought a painter with the same brand as Sarah's. This year, Amy starts to learn guitar, then she may infer other friends' preferences who are good at playing the guitar. 
Sun et al. \cite{sun2018attentive} propose a recurrent network based model with attention for temporal recommendation (see Figure~\ref{fig:social_rnn}). The method includes a static part, which captures the insistent user preference, and a dynamic part that captures the dynamic user preference. 
For the static part, the static social attention module is applied for selecting the static social relationships for each user, and then aggregates these social relationships together for enriching the user's representation vector.
For the dynamic part, a LSTM module is implemented for capturing the complex temporal latent representation of users, i.e. consider the social influences into the temporal preference modeling. Each part will predict a user preference score, and the final rating prediction is the sum of the scores from the two parts.

\subsection{GNN-based Methods}
\begin{figure}
\centering
    \begin{minipage}[t]{7cm}
    \includegraphics[width=\textwidth]{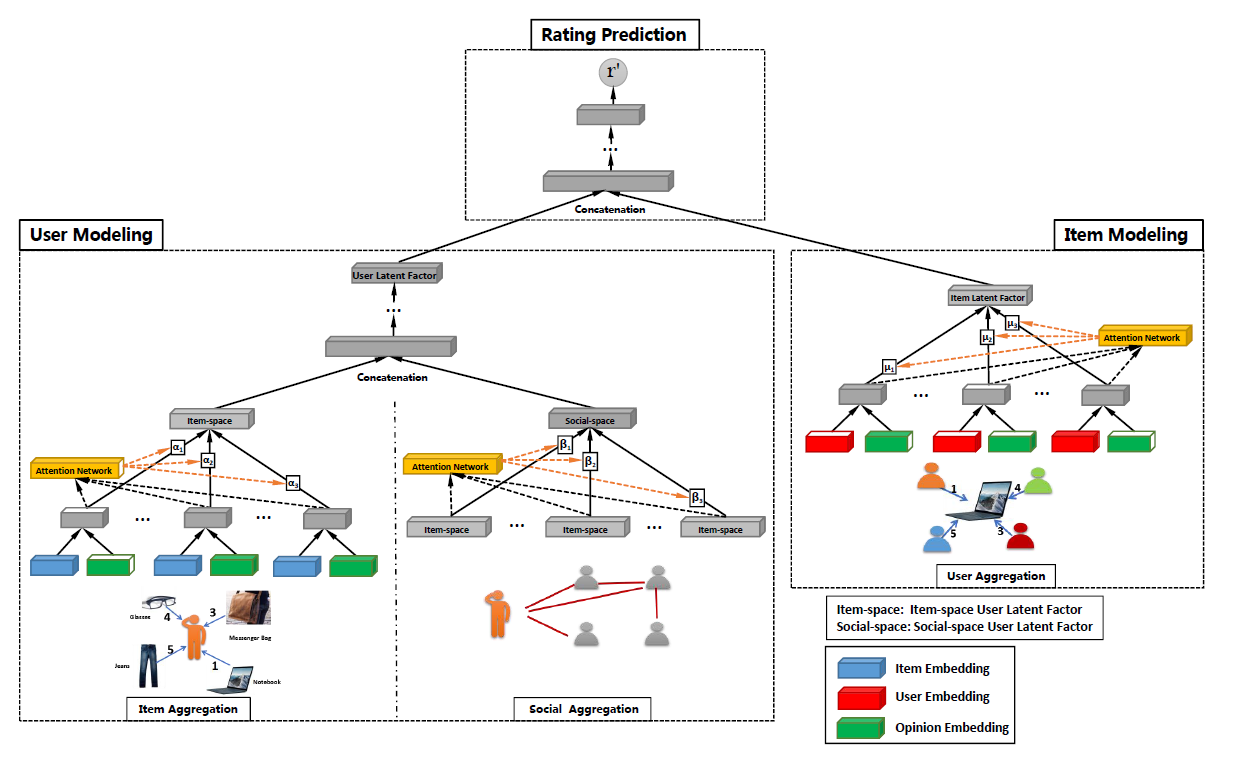}
    \centering{(a)}
    \end{minipage}
    \begin{minipage}[t]{8.5cm}
    \includegraphics[width=\textwidth]{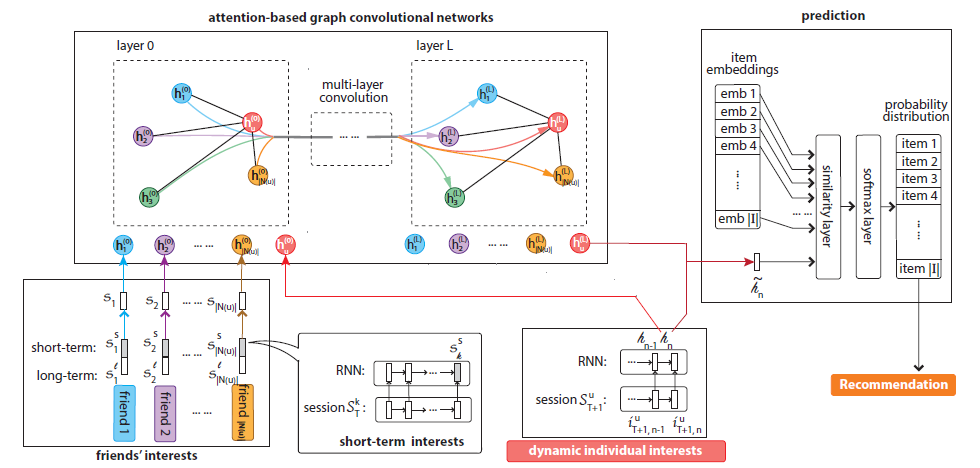}
    \centering{(b)}
    \end{minipage}
    \caption{Graph Neural Network for social-aware recommendation from (a) Fan et al.\cite{fan2019graph} and (b) Song et al. \cite{song2019session}.}
    \label{fig:paper_gnn}
\end{figure}

\textbf{Graph neural network (GNN)} has shown the effectiveness in learning on graphical data by the power of integrating node information and topological structures.
As such, for social-aware recommendation problems, GNN has a great potential for mining the social graph structures and user-item graph; where a key is utilizing the GNN to learn the latent factors of users and items \cite{fan2019graph}.
For example, Fan et al. \cite{fan2019graph} consider to learn the user latent factor and item latent factor via GNNs first, and then concatenate the two latent factors for the final rating prediction (Figure~\ref{fig:paper_gnn} (a)). For the user modeling, the user latent factor is the concatenation of item aggregation and social aggregation. The process of item aggregation is aggregating the user rating histories toward different items (the representation for each item is the combination of item-vector and the item rating) with attention algorithms; and the process of social aggregation is aggregating the user's friends rating histories (for each friend the representation is the reconstruction of user-item vector) with attention algorithms. For the item modeling, the item latent factor is the aggregation of other users' historical ratings toward the target item. 


Instead of static modeling of social relationships, the inference by friends may also change along with time. In this regard, Song et al. \cite{song2019session} consider a dynamic situation that users' interests are dynamically influenced by the social relationships -- the preferences of the friends may change differently among different periods. They propose a session-based social recommendation algorithm, which models dynamic interests and dynamic social influences. The whole structure is shown in Figure~\ref{fig:paper_gnn} (b). The model captures the user's current preferences by a RNN module, which models the user's historical actions, such as clicks. For modeling the friends' interests, the authors considered both short-term and long-term preferences, where short-term preference is modeled by RNN with capturing the current session's preference and long-term preferences capture the average interests. Then, each friend is represented by a concatenation of the short-term and long-term preferences. To learn social-aware user representation, the authors use an attention algorithm (which is learned by the similarity between the target user and the friends) to leverage the importance of the social relationships and then aggregate them with different weights. Then, they combine the social-aware user representation with the dynamic user interest. The probability distribution of recommending items is the softmax of the similarities between item embeddings and user hybrid representation.  

\begin{figure}
    \centering
    \begin{minipage}[t]{6cm}
    \includegraphics[width=\textwidth]{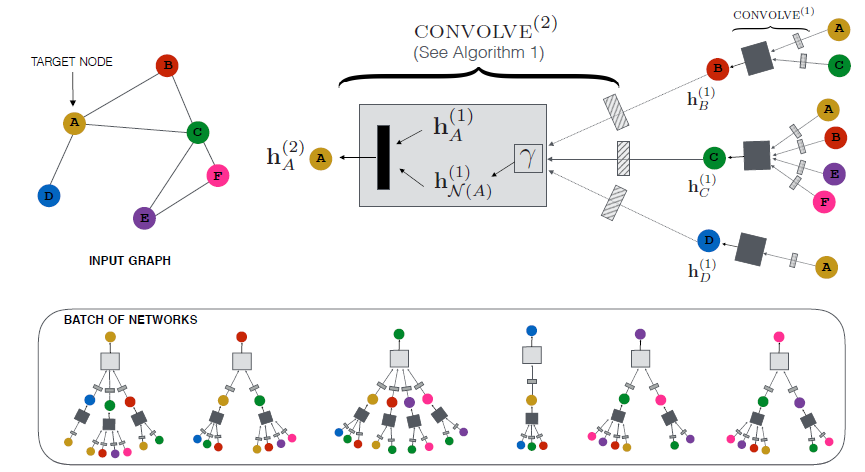}
    \centering{(a)}
    \end{minipage}
    \begin{minipage}[t]{9cm}
    \includegraphics[width=\textwidth]{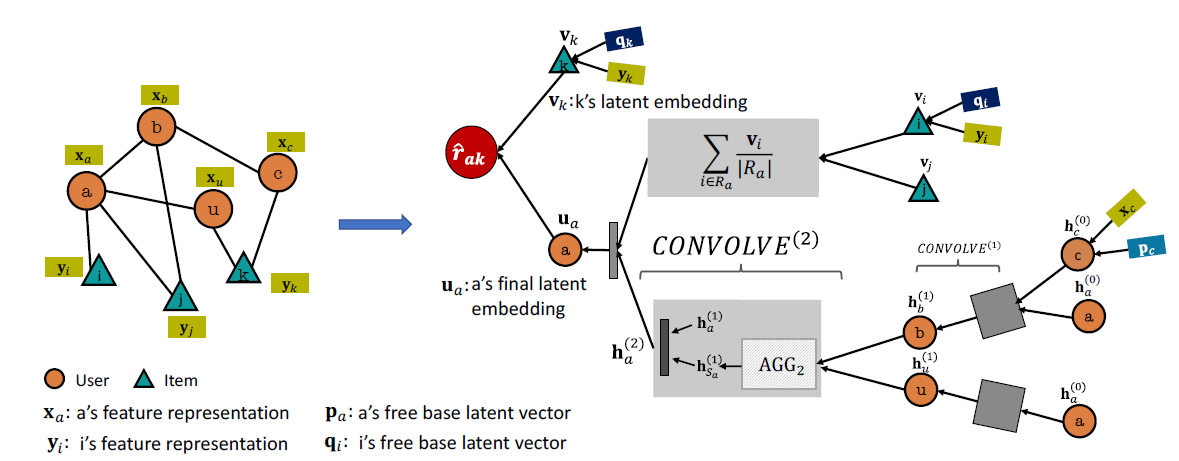}
    \centering{(b)}
    \end{minipage}
    \begin{minipage}[t]{14cm}
    \includegraphics[width=\textwidth]{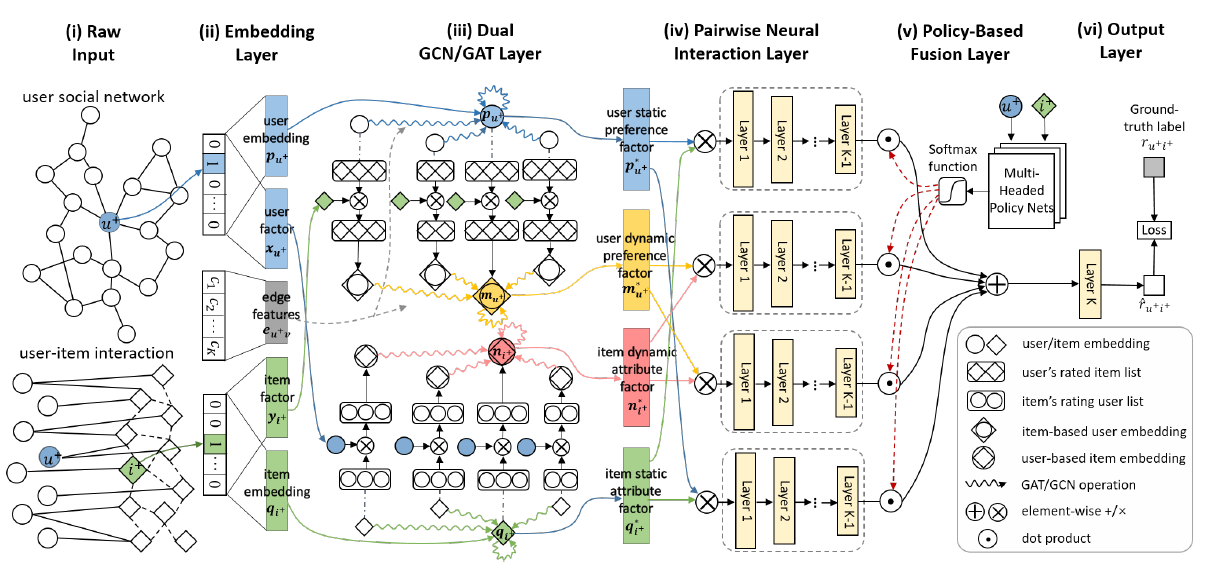}
    \centering{(c)}
    \end{minipage}
    \caption{Graph Convolutional Network for social-aware recommendation from (a) Ying et al. \cite{ying2018graph}, (b) Wu et al. \cite{wu2018socialgcn}, and (c) Wu et al. \cite{wu2019dual}. }
    \label{fig:social_GCN}
\end{figure}

As one type of graph neural network, \textbf{Graph Convolutional Network (GCN)} has been widely applied in recent social-aware recommendation studies due to the effectiveness in mining social relationships. Different to graph neural network, GCN generates the node embedding in a view of message passing or information diffusion\cite{wu2018socialgcn}, which can encode the graph structure information as low-dimensional representations. Specifically, the embedding for each node is the aggregation of the information from the neighborhoods; the embedding of the neighbors is further learned from the neighbors of the neighbors, and so on. 
For example, in \cite{ying2018graph}, the authors represent the node in graph with a two-layer graph convolutional neural network (see Figure~\ref{fig:social_GCN} (a)). In details, the embedding for a node (i.e., an item) is the aggregation of feature information (e.g., visual, textual features) from the node's local graph neighborhood. Each aggregation module learns how to aggregate information from a small graph neighborhood, and by stacking multiple such modules. Then, these embeddings are then used for recommender system candidate generation via nearest neighbor lookup or as features in machine learning systems for ranking the candidates. 
In \cite{wu2018socialgcn}, the authors assume that the user preference is modeled as the inner product of user and items embeddings. They first initialize the user embedding and item embedding, where each is the combination of descriptive features (such as user profiles and item descriptions) and free basic latent vector. Then, they model the diffusion of the user preferences as in layer-wise diffusion manner. Figure~\ref{fig:social_GCN} (b) shows the details.
Similarly, Wu et al.\cite{wu2019dual} assume that the representation of a user can be learned from a user-specific latent vector and the user's history rating records, which is called item-based user embedding; and the representation of an item can be learned from an item-specific latent vector and the users who rated it, which is called user-based item embedding (check Figure~\ref{fig:social_GCN} (c)). For learning the hybrid representation from embeddings with graph structures, the authors tried to use Graph Attention Network (GAT) based approach. Traditional GCN treat each neighbor equally and aggregate the embedding information without balancing them. Thus, the authors implement GAT, which leverages the attention mechanism to balance the significance of the neighbors, enhance the model to focus on important input, and improve the robustness of the model with filtering noise. In details, for an item (a user), the authors first learn its static attribute factor (user static preference factor) from the item embedding (user embedding and user relationships) and then learn a user-context-aware item factor (item-context-aware user factor) as a dynamic attribute factor (dynamic preference factor). And then they fuse the four factors (the static factor and dynamic factor for user and item) into a synthetic representation by a policy-based fusion layer, and the prediction is made based on it. 

\subsection{Generative Models}
Generative models classically describe models of joint distribution with data and labels. It will be taken to produce new samples with mechanisms sampling from the real data. The two most common types of generative models are generative adversarial nets (GANs)\cite{goodfellow2014generative} and Variational Autoencoders (VAEs)\cite{kingma2013auto}.  

\subsubsection{GAN-based Methods}
\begin{figure}
    \centering
    \begin{minipage}[t]{13cm}
    \includegraphics[width=\textwidth]{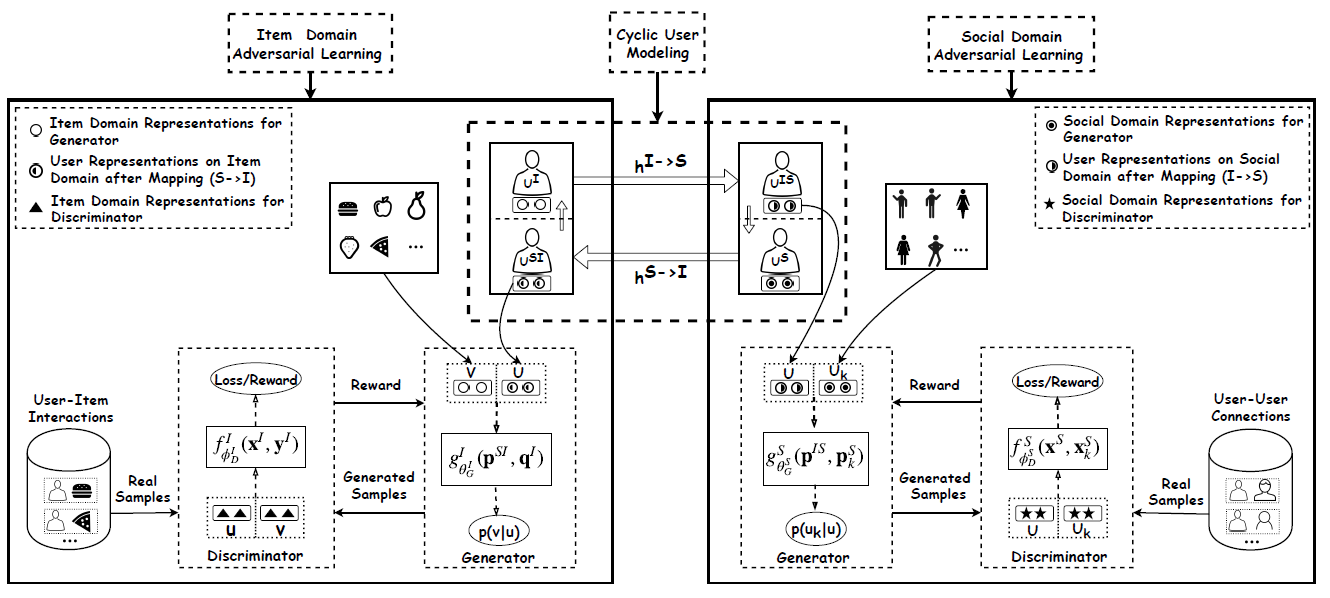}
    \centering{(a)}
    \end{minipage}
    \begin{minipage}[t]{6cm}
    \includegraphics[width=\textwidth]{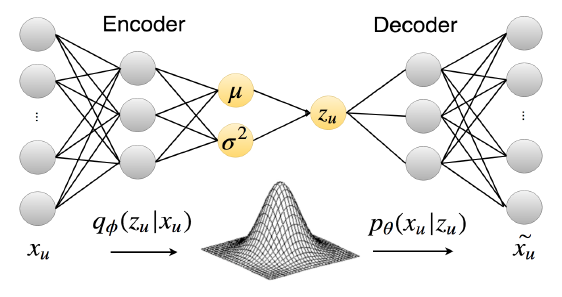}
    \centering{(b)}
    \end{minipage}
    \begin{minipage}[t]{8cm}
    \includegraphics[width=\textwidth]{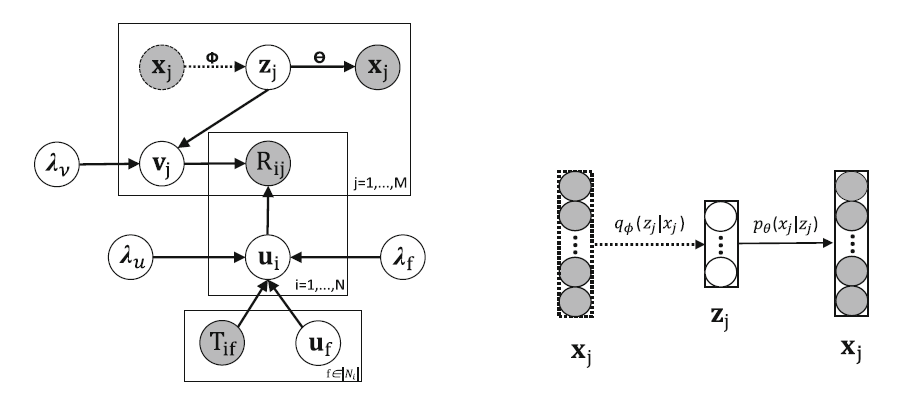}
    \centering{(c)}
    \end{minipage}
    \caption{Generative models for social-aware recommendation from (a) Fan et al.\cite{fan2019deep}, (b) Karamanolakis et al. \cite{karamanolakis2018item}, and (c) Xiao et al. \cite{xiao2019variational}}
    \label{fig:generative_social}
\end{figure}

Generative Adversarial Network (GAN), which includes a generator and a discriminator to conduct adversarial learning, has shown the effectiveness across various domains due to the ability of learning data probabilistic distribution and generating new samples. 
The generator attacks the discriminator by generating new samples with similar distribution as real samples; and the discriminator is distinguishing the source of the samples, i.e. whether the sample is coming from real cases or generated cases. Then, a min-max game is played between the two processors, which can promote both of the two models. This is known as adversarial training. When it comes to the recommendation problem, generative models are normally applied for i) predicting missing values and ii) enhancing the representation of items and users. For example, Wang et al. \cite{wang2019minimax} 
use a generative model to generate simulated user preference distribution of real data. The performance of the generative model is improved by maximizing the classification loss, and the training will be stopped until producing promising generative predicted ratings. For the social-aware recommendation problem, the critical problems are the way of learning information from trust relationships and the way of combining such trust information with rating history. Fan et al. \cite{fan2019deep} design two adversarial learning modules for enhancing the user representations in the user-item rating part and the social part. In details, for each part, a discriminator is designed for distinguishing the real instances and the generated samples, and a generator is designed for modeling the actual conditional distribution for a given user. For adaptively enhancing the representations in two parts, they utilize a bidirectional mapping between the two parts, where in each iteration, the user's social representation will be updated by a nonlinear mapping operation from rating pattern representation, and then updated by the domain adversarial trainer; likewise, the user's rating pattern representation will be updated with the trained social representation, and then be following with the domain-specific training, see Figure~\ref{fig:generative_social} (a) for details.

\subsubsection{VAE-based Methods}
Similar to autoencoder-based methods, variational autoencodesr can be used for predicting missing values or learning comprehensive representations. For the former task, the VAE predicts item ratings/scores by inferring the latent factors \cite{karamanolakis2018item}. It consists of three parts: a bag-of-items vector $i_{u}$ by user $u$ is provided as input to the decoder; a latent user vector $z_{u}$ is sampled from a Gaussian distribution with parameters specified by the encoder; and a new bag-of-items vector $\tilde{i}_{u}$ see \ref{fig:generative_social} (b). For the latter, the VAE learns representation for users or items. For example, Xiao et al.\cite{xiao2019variational} consider three types of information for recommendation: user-trust relationship, user-item rating history, and item content information. For representing the users, each user is represented by the aggregation of trusted users. For representing the items, the content information for items is considered, where a variational autoencoder model is used to learn the latent patterns for content information. Such latent patterns are used to represent the items. Then, they consider using traditional matrix factorization method for predicting the user-item ratings. The graph structure can be noted in figure~\ref{fig:generative_social} (c). 

\subsection{Hybrid Methods}
For bridging the advantages of the above models for enhancing the recommendation performance, several hybrid models are proposed. The hybrid models may adopt hybrid algorithms or multiple types of input for recommendation. 

\begin{figure}
    \centering
    \begin{minipage}[t]{5cm}
    \includegraphics[width=\textwidth]{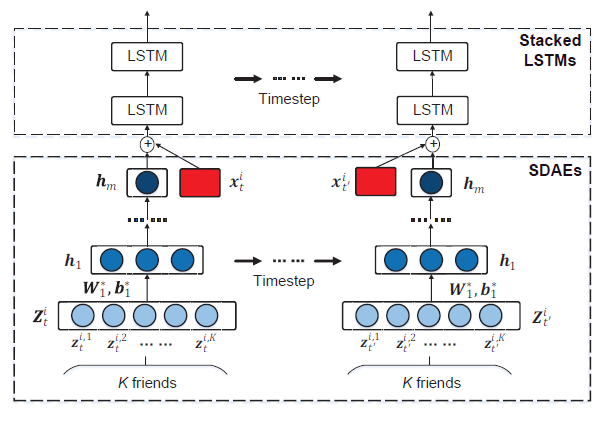}
    \centering{(a)}
    \end{minipage}
     \begin{minipage}[t]{10cm}
    \includegraphics[width=\textwidth]{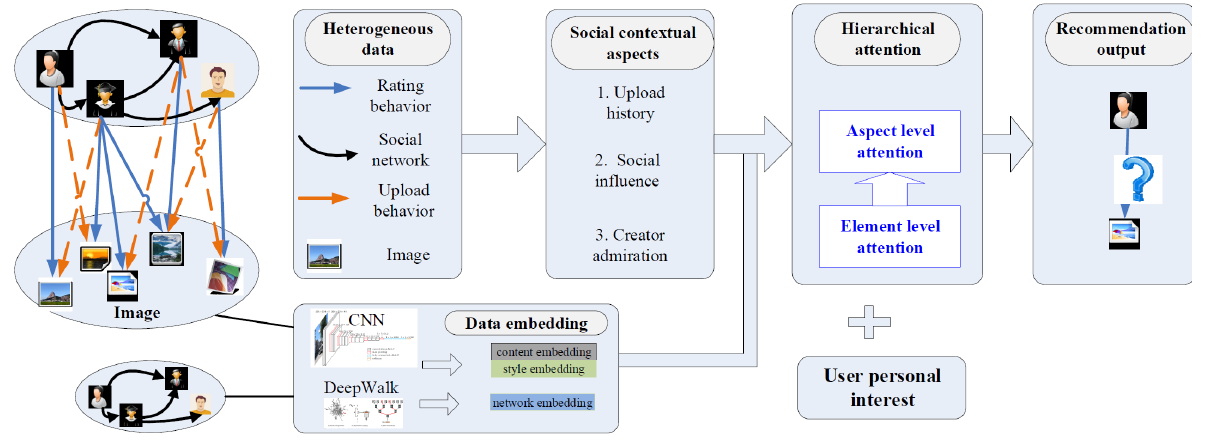}
    \centering{(b)}
    \end{minipage}
    \begin{minipage}[t]{4cm}
    \includegraphics[width=\textwidth]{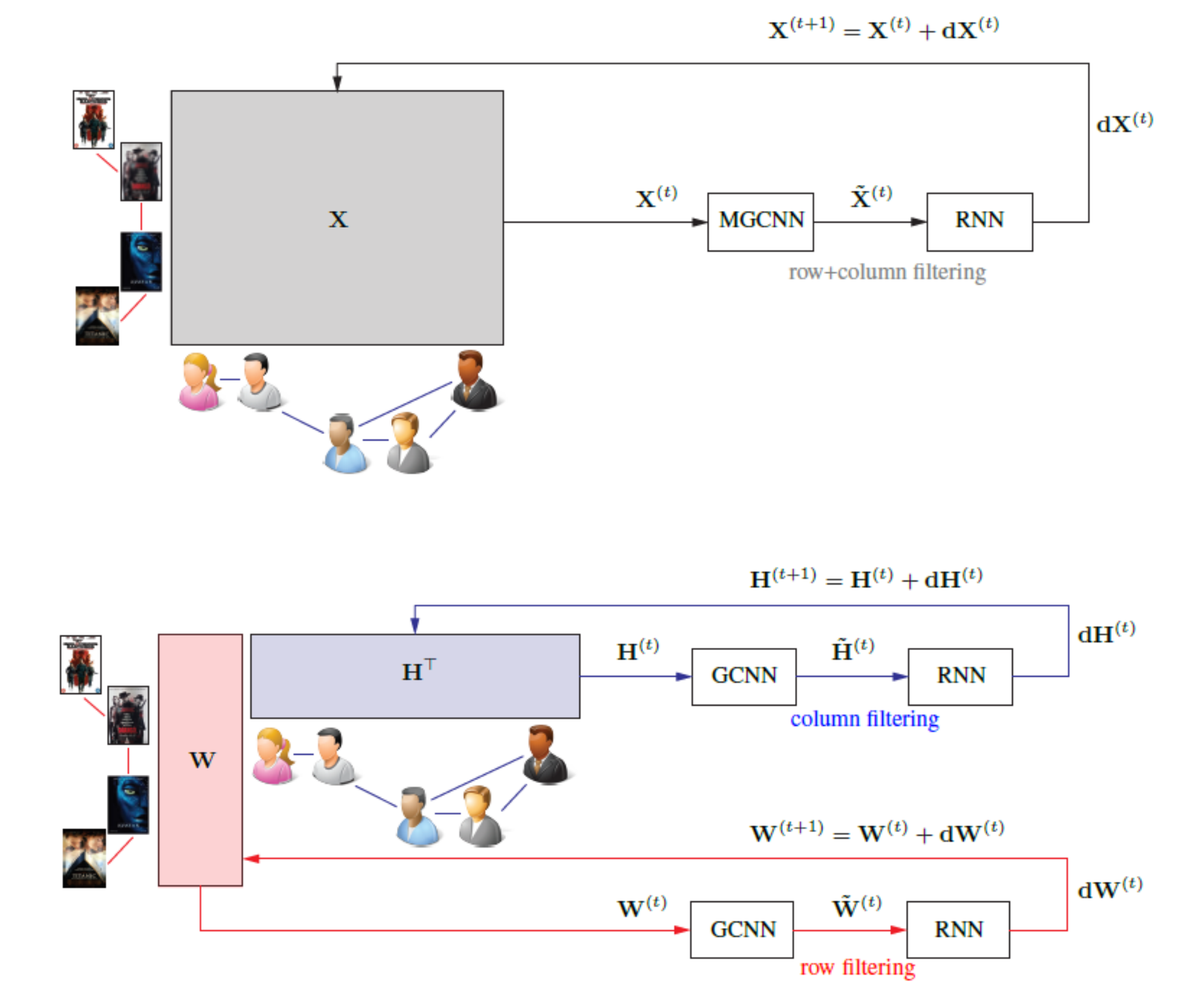}
    \centering{(c)}
    \end{minipage}
    \begin{minipage}[t]{11cm}
    \includegraphics[width=\textwidth]{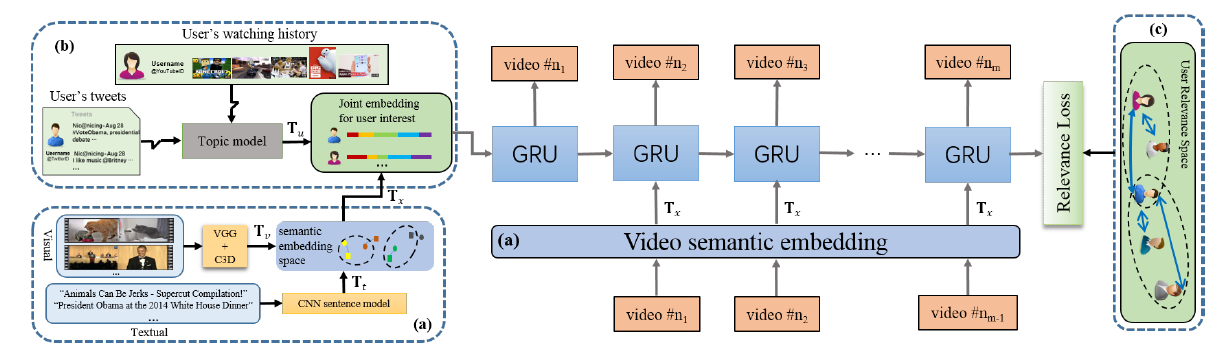}
    \centering{(d)}
    \end{minipage}
    \caption{Hybrid methods for social-aware recommendation from (a) Liu et al. \cite{liu2018social2}, (b) Wu et al. \cite{wu2019hierarchical}, (c) Monti et al. \cite{monti2017geometric}, and (d) Gao et al. \cite{gao2017unified}.}
    \label{fig:hybrid_social}
\end{figure}

According to the research targets and characteristics of features, multiple algorithms may apply. For example, RNN based methods are widely applied for dealing with session-based recommendation problem, and autoencoder is good at extracting a condense representation of input. Then for solving a session-based trust-aware recommendation problem, such two methods can be applied together. Liu et al. \cite{liu2018social2} consider that user's preferences are changing over time. They use a stacked denoising autoencoder to learn the user representation, which is the aggregation of the friends' representations, at each time step. Then, such representation is used as the input of the LSTM module for predicting user's current preference. A stacked LSTM for predicting the user's whole time preferences. See Figure~\ref{fig:hybrid_social} (a). 

For recommendation problems such as image recommendation and movie recommendation, the auxiliary information will be considered.
Wu et al. \cite{wu2019hierarchical} target an image recommendation problem. For the image information, they use CNNs for learning the embedded representation. Besides for the images, they consider three auxiliary information: user upload history, social influence, and creator admiration. They use an attention model to leverage such three aspects. The user's preference is represented by the aggregation of all the information. And the rating prediction is based on the product of item embedding and user's preference vector, see Figure ~ \ref{fig:hybrid_social} (b). 
Zhao et al. \cite{zhao2017social} design a heterogeneous social-aware movie recommender system by exploiting multi-modal movie contents (i.e. images and corresponding descriptions), users' social relations and their relative preference feedback. The goal of the model is providing top-K movies for recommendation, where the rank for movies is a ranking score of the given user and item, which is calculated on the representations for the user and item. For learning a sharing item representation with both movie images and descriptions, they use a multi-modal learning approach: a deep convolutional neural network for the images and a deep recurrent neural network for the descriptions. The representations for users, which is the aggregation of friend relationships, are learned via DeepWalk\cite{perozzi2014deepwalk}. 
Monti et al. \cite{monti2017geometric} propose two different ways of predicting the user-item rating matrix. The first one is Recurrent Multi-Graph CNN (RMGCNN) architecture, see the top figure of Figure\ref{fig:hybrid_social} (c), which operates on the user-item matrix and operates simultaneously on the rows and columns. Both of the users and items are learned via Multi-Graph CNNs: the users are modeled by their relationships, and the items are modeled with the images. Then, the whole rating matrix $X$ is learned by RNN model step by step, until provides stable predictions. For the second method, which is called Separable Recurrent MGCNN (sRMGCNN), operates separately on the rows and columns of the matrix, see the bottom figure of Figure\ref{fig:hybrid_social} (c). 
Gao et al. \cite{gao2017unified} consider a video recommendation problem. They propose a dynamic RNN to capture the user dynamic preference by considering the video information, user interest, and user social relationships. The video semantic embedding includes visual features and textual features, which are learned by pre-trained deep models. User interest modeling is based on the user view history, which is learned by topic modeling. As for the user social relationship mining, see Figure~\ref{fig:hybrid_social} (d).

\subsection{Others}
\subsubsection{Attention-based Methods}
Informally, a neural attention mechanism enables the neural network focus on a subset of its input (or features), i.e. assigns different weights to the input. For example, for the machine translation problem, it allows the machine translator to look over all the information the original sentence holds, then generate the proper word according to current word it works on and the context\cite{luong2015effective}. Now the attention mechanism is popular in many other areas, such as object recognition and image caption\cite{xu2015show}. There is limited work only emphasises on discussing the effectiveness of applying an attention mechanism into recommender systems; but on the contrast, as an enhancing module, the attention can work well when incorporated with other models for recommendation\cite{zhang2019deep}. 

For the social-aware recommender system, the attention mechanism could be used for balancing the friends influences. For example, \textit{Chen et al.} \cite{chen2019social} consider the problem that the influence of user's friends should be different and dynamic. For different items, the user may infer different friends preferences. Thus, they propose a hierarchical attention module for the recommendation. First, each friend's representation is built on the user embedding and friend embedding, and is learned by attention mechanism. Then, for different friends, the authors also applied an attention mechanism for balancing friends' influences to get a final user representation. This user representation is then multiplied with item representation as a score for ranking.
\textit{Rafailidis and Weiss} \cite{rafailidis2019neural} propose a similar structure that considers a subset of friends and uses an attention mechanism for social collaborative filtering.

\subsubsection{Combining with Matrix Factorization}
Traditional matrix factorization methods is predicting the ratings/scores by multiplying the latent representations of users $h_{u}$ and items $h_{i}$, i.e. $\hat{r}_{ui}=h_{u}^{\top}\cdot h_{i}$. The difference between deep matrix factorization methods and traditional ones is that the latent representation learning is implemented with deep learning techniques. For tackling social-aware recommendation problem, a common way is combining the social influences into the user representation. 
Fan et al. \cite{fan2018deep} learn the social embeddings via node2vec \cite{grover2016node2vec}, and then use multilayer perceptron for learning the embeddings. Each user is represented by the social embeddings and an initialized latent factors. Such user representation is further be used in probabilistic matrix factorization \cite{mnih2008probabilistic}.
Bao et al. \cite{bao2018contextual} use an attentive way for learning social influences. They first use autoencoder to learn compact representation of neighbors, where each neighbor is represented by $h_{v}$. Then the social influences for user $u$ is modeled by $h_{T}=\Sigma_{v\in V}\alpha_{v}\cdot h_{v}$; $\alpha$ is the attention value. Then the user latent representation is given by $\hat{r}_{ui}=(\beta h_{u}+ (1-\beta)h_{T})^{\top}\cdot h_{i}$, where $\beta$ is a self-defined hyper-parameter. 
In \cite{wang2017item}, the authors argue that the matrix factorization method could be represented with a shallow neural network model. They consider a cross-domain recommendation problem\cite{elkahky2015multi}, where they are trying to recommend a top-K items list from information-domain to users in social-domain. Specifically, they use a deep collaborative filtering model to predict the user preference $\hat{s}_{ui}$, which is calculated by the latent user representation $h_{u}$ and latent item representation $h_{i}$. 
Both of the latent representations are learned from the initial embedding vector and attribute vectors (i.e. the hashtags). Then, the prediction of user-item interaction is further enhanced by integrating the social relationships: the intuition that users with strong connections are more likely to share similar tastes on items. The processing is minimizing the user latent representation gap between strongly connected users.

\subsubsection{Others}
Xiao et al. \cite{xiao2017neural} propose a SVD++\cite{koren2008factorization} based model for recommendation. The inputs for the network are user's representation $h_{u}$, item's representation $h_{i}$ and user's social representation $h_{t}$. They incorporate the social relationships by considering both its latent representation and the interaction with items.
The following layer is represented as the concatenation of the above information, i.e. $[h_{u}, h_{i}, h_{t}, f(h_{u},h_{i}), f(h_{i},h_{t})]$, where $f$ is representing several neural networks. The prediction is made after a few fully connected layers.

\subsection{Summary}

To summarize, deep learning based social-aware recommendation algorithms have shown their effectiveness in different tasks. Different to traditional methods, deep learning based techniques need less manual extracted features and have the advances in grasping complex latent feature interactions. It is a trend to incorporate traditional recommendation methods with deep learning methods, e.g., graph neural network, which can leverage both advantages. 
Although effective, current deep learning based social-aware recommendation algorithms have the following challenges. 
\begin{itemize}
    \item The quality and the quantity of the social links. For example, in most of the recommender systems, it is hard to get explicit and reliable links since a few amount of users indicate their social relationships. 
    \item Most of the current works model the trust relationships with shallow model and ignore the high-order interactions among each users' friends. It's possible for a user to take all the opinions of his friends into account and then come out his own thinking rather than linearly combine all of them.
    \item The assumption that user shares similar tastes with friends may mislead the recommendation. For example, a user can connect with people who have different shopping preferences.
    \item Most of existing approaches ignore that users have different knowledge in different domains. 
\end{itemize}

Table~\ref{tab:social_dataset} shows some widely used datasets for the social-aware recommender systems\cite{guo2015librec}, which are taken from popular social networking websites\footnote{The links for each dataset can be found in the Librec website \url{https://librec.net/datasets.html}}. 

\begin{table}[h]
\centering
\caption{Commonly used datasets for social-aware recommendation}
\begin{tabular}{c|ccccc}
    \hline
        Dataset  & Ciao\tnote{1} & Epinions\tnote{2} & FilmTrust\tnote{3} & Flixster\tnote{4} & Douban\tnote{5}  \\ \hline
        \# of Users   & 7,317 & 18,088   & 1508 &    147,612 & 129,490 \\ \hline
        \# of Items   & 104,975 & 261,649 & 2071 &   48,794 & 58,541\\ \hline
        \# of Ratings & 283,319 & 764,352 & 35,497&  8,196,077 & 16,830,839 \\ \hline
        Density (ratings) & 0.0368\% & 0.0161\% & 1.1366\%& 0.1138\% & 0.2220\%\\ \hline
        \# of Social relations & 111,781 & 355,813 & 1,853 & 7,058,819 & 1,711,780\\ \hline
        Density (social relations) & 0.2087\% & 0.1087\% & 0.0815\% & 0.0324\% & 0.0102\%\\ \hline
    \end{tabular}
\label{tab:social_dataset}
\end{table}

\section{Robustness of Recommender Systems}
\label{sec:robust recommendation}
\subsection{Overview}
The recommender systems promote the efficiency and benefits for both customers and merchants. Although effective, the recommendation schemes are vulnerable to shilling attacks or noise. For example, merchants may hire a group of spammers to insert their profiles and fake ratings into the systems, which will affect the performance of the recommendation\cite{gunes2014shilling} and also the customer's trust on recommender systems. 
Detecting such attacks and designing a robust recommender system are very important. Generally, researches in this field include shilling attack detection techniques and robust recommender systems. 

\subsubsection{Attack Types}
The shilling attacks can be classified by the type of attackers, the intent for the attack, the knowledge of the attack, etc. 
For example, according to the intention, the attacks can be categorized into push attack, nuke attack, and random attack\cite{si2018shilling}. The first intends to increase the popularity of the items while the second intends to decrease the popularity.
According to the knowledge-cost, the attacks diverge into high-knowledge attack, i.e., the attackers get some knowledge about other normal users, and low-knowledge attack. 

The attack profile consists of the history rated items and generally includes four parts: the target items, selected items, filler items and unrated items.
The target items $I_{T}$, which will be either "push" or "nuke" ratings, are rated with a rating function $\sigma_{T}$. The selected items $I_{S}$ are rated by the attacker with particular intentions, e.g. the group attacks. The filler items $I_{F}$ include randomly chosen items that to make the profile look normal and harder to detect. Also, We denote the unrated items by $I_{U}$\cite{si2018shilling}. Different parts may have different generative functions for getting the ratings. 
\begin{figure}[ht]
    \centering
    \includegraphics[width=12cm]{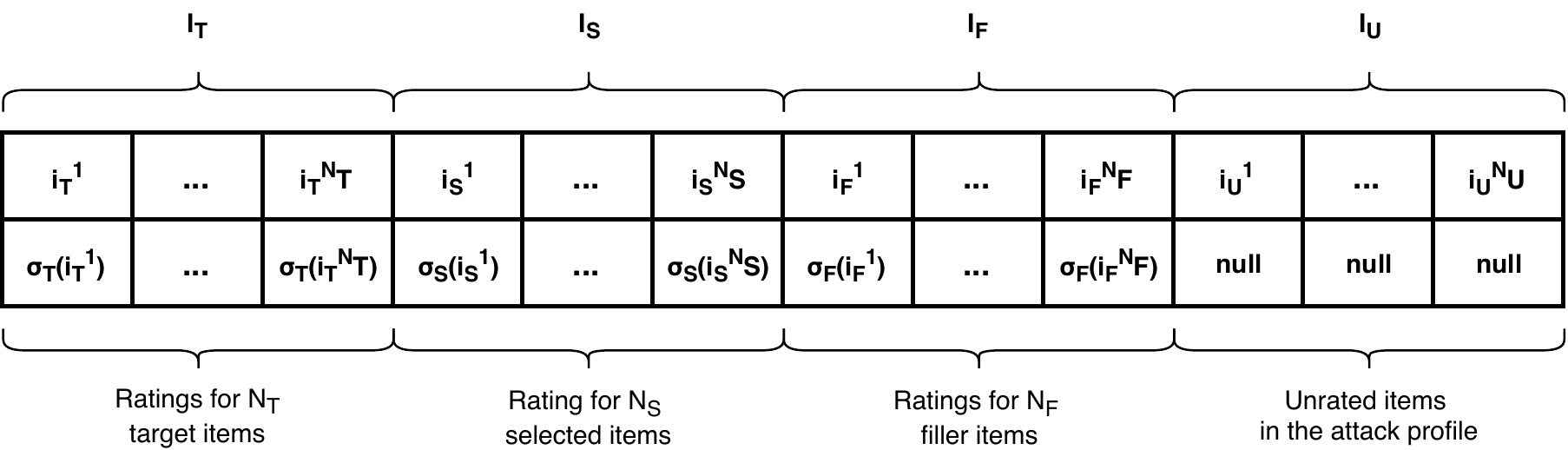}
    \caption{The general components of an attacker profile }
    \label{fig:attack_types}
\end{figure}

\subsubsection{Research Fields}
The two directions to reduce the effects of shilling attacks on recommender systems include shilling attack detection techniques and robust algorithms. 
The former is first detecting the attacks, filtering the attack profiles, and then constructing the recommender systems. 
The latter refers to construct attack-resistant recommender systems, i.e., robust recommendation methods\cite{si2018shilling}.

\textbf{Shilling Attack Detection Algorithms} are mainly discussing the way of detecting malicious user files. According to the research targets, the shilling attack detection algorithms can be used for detecting point (personal) attacks or collaborative (group) attacks. The point attacks may represent an irregularity or deviation that happens randomly and may have no particular interpretation. Also, according to the nature of input data, we may utilize sequential (e.g., textual information and time series) or non-sequential data ((e.g., images, user profiles). 
Techniques used for shilling attack detection can be roughly classified into statistical methods, supervised classification methods, semi-supervised methods, and unsupervised clustering methods. The \textit{statistical methods} are focusing on detecting the outlier items. For example, statistical testing is used for identifying the differences between the sample distribution.
Zou et al. \cite{zou2013belief} introduce a probabilistic inference network and  the Belief Propagation (BP) algorithm \cite{lecun2015deep} to perform inference efficiently.
For the \textit{supervised classification methods}, most of the work conducts feature engineering first and then design the algorithms. Features such as rating deviation, the similarity with top neighbors are considered. For example, Yang et al. \cite{yang2016re} propose three new features, i.e., the filler size with maximum, minimum and average ratings, on filler or selected items to identify the attack profiles. The features are analyzed with statistical tests and classified by a variant of AdaBoost method. 
\textit{Unsupervised clustering approach} is normally clustering the users into groups and then eliminate the suspicious users. 
For example, Bhaumik et al. \cite{bhaumik2011clustering} apply k-means clustering on user profiles and identify the small clusters as attacker groups.

\textbf{Robustness of Recommender Systems} are focusing on developing attack-resistant systems, which is trying to reduce the influence of shilling attacks. Current robust recommendation algorithms mainly lie on two aspects: constructing robust algorithms or taking trust relationships into consideration. We have discussed the latter one in the previous section. Thus, here, we majorly introduce the former one: robust algorithms for recommendation. Some work adopts matrix factorization for the methods. For example, 
Alonso et al. \cite{alonso2019robust} use matrix factorization based method to identify the shilling attacks. According to the observation that fake ratings are occurred during a short interval of time, they assume that malicious profiles will affect the reliability of the model in an anomalous way. They use two matrix factorization models to obtain the real prediction errors and the estimated prediction errors; the second error is used for evaluating the prediction reliability. 
Zhang et al. \cite{zhang2017robust} incorporate the R1-norm into the loss function to improve the robustness. They claim that the squared error function is sensitive to large residuals. 
Yu et al. \cite{yu2017novel} design a robust matrix factorization model with kernel mapping and kernel distance.
He et al. \cite{he2018adversarial} consider to improve the robustness of the recommender system by adding an adversarial module to the training. 

\subsection{Deep Learning-based Shilling Attack Detection Algorithms}
For shilling attack detection problem, a key point is to evaluate the suspiciousness of users, posts and reviews.  
One advantage of using deep learning for detecting the shilling attacks is its capability of capturing complex structures in the data. Also, there is a need for large-scale detection techniques as the volume of data increases in the real world cases. Different from traditional methods, deep learning-based approaches need less manually defined features and thus solve the problem in an end-to-end trainable way\cite{chalapathy2019deep}.

Some use convolutional neural networks for mining the local features.
Convolutional neural networks work as feature extractor; it can learn from local representations and map these representations into higher or lower dimensional representations for further uses\cite{ren2019learning}.
Many studies mine textual information for detecting the suspicious ratings/reviews\cite{zhao2017towards}. For example, Zhang et al. \cite{zhang2018dri} propose a deep model to identify the review spam. They assume the fraudulent users lack real experience, while the normal users are with real experience; then, the textual information should indicate different patterns between the fraudulent users and normal users.
Li et al. \cite{li2015learning} use word vector to represent the textual features, and use CNN to learn the semantic representation. 
Some work also considers the user's behaviors \cite{wang2017detecting}. Wang et al. \cite{wang2017handling} consider the cold-start problem for new-coming users. They use CNN to learn the embedding from both textual and behavioral information. 

Some use RNN-based methods for targeting the sequence input. 
Recurrent Neural Network (RNN), which has the function of memory, has shown its efficiency for processing sequential information. The following work, such as Gated Recurrent Unit (GRU)\cite{chung2014empirical} and Long Short-Term Memory (LSTM)\cite{hochreiter1997long} are further designed for solving the gradient vanishing problems. 
As for detecting malicious ratings or reviews, the RNN based methods are used for learning sequential patterns such as texts and time series. For example, Ren et al. \cite{ren2016deceptive} use CNN to learn from words and use a bidirectional-GRU for learning the sentences. The learned representation is further be used for prediction. Similarly, Wang et al. \cite{wang2018detecting} use LSTM to learn from texts.

Some consider hybrid methods or multiple inputs. 
For example, Wang et al. \cite{wang2016learning} propose a hybrid method that learns from both the review content and product information. They use a tensor factorization algorithm to learn the latent representations from reviews and products. Then, the learned representations are further combined with deep learning-based classifiers. Dong et al. \cite{dong2018opinion} use autoencoder to extract latent representations from textual information and user behavioral patterns. 
Aghakhani et al. \cite{aghakhani2018detecting} improve model performance by adding adversarial noises. 

\subsection{Deep Learning for Robustness of Recommender Systems}
\subsubsection{Introducing noises to recommender systems}
An intuition way for enhancing the robustness of recommender systems is adding noises in the recommendation training process. By doing so, the model is forced to learn robust parameters to improve the denoising capability.  

\textbf{Randomly generalized noises.}
Some may add man-made noise into the input. For example, the model may be added with extra bias terms; the inputs are corrupted before feeding in the models. By doing so, the model is forced to learn the most informative and robust parameters to improve the robustness of the recommender systems. One example is the denoising auto-encoder (DAE) \cite{dong2017hybrid}, which corrupts the inputs with adding noises. 
Wu et al. \cite{wu2016collaborative} propose the collaborative denoising auto-encoder (CDAE) that has similar ideas of DAE. They first corrupt the inputs, i.e. the ratings, with Gaussian noises, feed the inputs into neural nets and get a dense representation of such corrupted inputs. The decoder of the model tries to recover the original values of the dense representation and thus predicts the ratings for recommendation. 
Strub et al. \cite{strub2016hybrid} also corrupt inputs by stacked denoising autoencoders\cite{vincent2010stacked}. Besides, they incorporate the side information, such as the user profile (age and gender) and the movie category, to enhance the robustness of the model. 
Wang et al.\cite{wang2016collaborative} propose collaborative recurrent autoencoder that integrates RNNs and denoising autoencoder for recommendation. 
They design a robust recurrent network to process the item textual information and overcome the shortage of using man-crafted features. In details, the recurrent network is designed in the autoencoder way, i.e. the layers in the encoder and decoder are recurrent networks. By doing so, the recurrent autoencoder can learn both the sequential information and the dense representation of inputs. 
The learned dense representation is regarded as the item representation and further combined with user representation for rating prediction. 
Furthermore, in case of over-fitting, they design a denoising and a beta-pooling approach.

\textbf{Adversarial noises.}
In recent work, some attempt to add adversarial noise to the model to improve the robustness. 
For example, He et al. \cite{he2018adversarial} consider to solve the problem by adversarial training based on Bayesian Personalized Ranking.
In details, they corrupt the model parameters with adversarial noises; the adversarial personalized ranking is made by minimizing the training loss and maximizing the adversarial loss, i.e. to identify the worst case of the corruption. The model is optimized via stochastic gradient descent.
Similarly, Yuan et al.\cite{yuan2019adversarial} propose an adversarial training framework for recommendation. The model is designed based on the collaborative denoising autoencoder. Different to traditional CDAE method that corrupts the inputs, they insert a noise mixing layer into the autoencoders. The addressed adversarial training strategy includes: a training step to obtain optimal parameters; and a re-training step to minimize the training loss while maximize the adversarial noise loss. 
Wang et al. \cite{wang2018neural} consider a session-based recommendation problem and design a memory network for storing the long term and short term user preferences. They use generative adversarial nets to generate negative samples to improve the model parameter inference.
Wang et al. \cite{wang2019minimax} 
propose a generative adversarial model for recommendation. Similar to generative adversarial network, the model includes two modules: the generative model simulates the real user profiles by capturing the patterns from the raw datasets; while the discriminative model tries to identify such generated samples from the real ones. In such case, the generator do similar work as the malicious users, and the discriminator identifies the malicious content; thus, they promote the both performance and improve the robustness of the recommender system.


\subsubsection{Other Methods}
The attention mechanism is able to filter the uninformative information from the input and thus prevent the side effects of noise. 
Many works consider to mine informative patterns in user history records. 
Jhamb et al. \cite{jhamb2018attentive} examine from user preferences that they propose Attentive Contextual Denoising Autoencoder and use attention mechanism for encoding contextual attributes of user preferences.
Zhou et al. \cite{zhou2018atrank} consider from the heterogeneous user behaviors; they use self-attention algorithm for predicting the user preferences by aggregating the different contributions of the user behaviors.
Loyala et al. \cite{loyola2017modeling} study the user transitions in different sessions by RNN-based method and use attention module for learning the more expressive portions of the sequences.
Ying et al. \cite{ying2018sequential} also use attention mechanism for a dynamic situation of user preferences. Two attention layers are utilized to learn user's long-term preferences and learn from both long-term and short-term preferences, separately. Liu et al. \cite{liu2018stamp} provide similar ideas for incorporating long-term and short-term preferences. 
Some works consider from other patterns. 
Seo et al. \cite{seo2017representation} mine the potential benefits from textual features. They build vector representations of user and
item using attention-based CNNs, where the attention mechanism is used for extracting keywords before the CNN modules; such vector representations will be further used to predict the ratings. 
Tay et al. \cite{tay2018latent} focus on the user-item relationships and use an attention module that can visualize the model and enhance the model performance with capturing the significant patterns.
Chen et al. \cite{chen2017attentive} propose two attention modules that one for selecting informative components of multimedia items and one for scoring the item preferences.

Besides, incorporating auxiliary knowledge from other domains, such as social relationships (we have discussed in section\ref{sec:social-aware}), can also help improve the robustness of the recommendation.

\subsection{Summary}
Shilling attacks detection has been the traditional research field to tackle the robustness of recommender systems, by filtering or removing malicious profiles. Some other work designs robust machine learning methods to neutralize the malicious profiles impact, e.g. adding man-made noise in the system to improve the robustness. 
Both ways show the improvement in the recommendation performance. However, there are still several challenges in this area. 
\begin{itemize}
    \item It is hard to unify such two methods in a trainable end-to-end model for leveraging both capabilities. And most of the deep learning methods are sensitive to the data resource and the cross-domain conditions. 
    \item Most of the current work does not consider the dynamic conditions, i.e., anomalous behavior may change over time.
    \item The anomalies are rare entities in real life. Thus, it is challenging to obtain labels. 
\end{itemize}

\section{Explainable Recommender System}
\label{sec:explainable}
\subsection{Overview}
As another approach to enable trust-aware recommendation, explainable recommender systems address the problem from a different perspective\cite{zhang2018explainable}. Unlike other personalized recommendation algorithms, explainable recommender systems offer reasons on why the systems provide users with such recommendations and also give guidance to system designers to improve the recommendation results. It not only improves the effectiveness and user satisfaction of recommendation systems but also enables the systems to generate trustworthy recommendations. Recently, a host of explainable recommendation approaches have been proposed, including but not limited to matrix factorization, deep learning, association rule mining, topic modeling, knowledge-graph models. Despite such variety, these methods can be divided into two groups in general. \textit{Post-hoc} \cite{tintarev2011designing, sharma2013social} methods do not modify the recommendation algorithm itself, but attempt to explain the results, such as ``this item is the most popular" and ``people like the same item as you do also bought." These methods often can not explain the recommendation mechanism, and the diversity of explanations is limited. On the contrary, \textit{Embedded methods} \cite{chen2018neural,chen2018visually,chen2017attentive} design explanation-oriented recommendation models so that the recommendation process itself can automatically generate explanations which are normally selected from the side information, e.g., texts or images. In this section, we narrow down our focus on only deep learning models, which belongs to the family of embedded explanation methods. Deep learning has recently become very successful in recommendation tasks\cite{zhang2019deep}. Similar to other embedded methods, we find that the majority of recent work based on deep learning leverages text or image information, e.g., user reviews, product photos, and movie posters for explanation generation. Most of them are proposed to explain a specific recommendation model, but, recently, some work addresses the recommendation explainability from a model-agnostic perspective \cite{wang2018reinforcement,mcinerney2018explore}. For instance, Wang et.al.\cite{wang2018reinforcement} employs reinforcement learning to explain any recommendation model. Therefore, in this section, we classify the previous work into five categories: \begin{enumerate*}\item traditional explanation based on collaborative filtering; \item explanation using a piece of textual sentence; \item explanation via visual contents; \item explanation via temporal dynamics; \item other deep learning-based explainable recommendation models. \end{enumerate*}


\begin{table}[ht]
    \label{tab:explainable_recsys_summary}
    \caption{Summary for Deep Learning-Based Explainable Recommender Systems}
    \centering
        \begin{tabular}{|c|>{\centering}m{20pt}|>{\centering}m{40pt}|>{\centering}m{40pt}|>{\centering}m{40pt}|>{\centering}m{40pt}|>{\centering}m{40pt}|>{\centering\arraybackslash}m{35pt}|}
        \hline
            \backslashbox{Method}{Data Type} & RBM & Attentive CNN & Attentive RNN & Generative RNN & GAN & Memory Networks & Others \\ \hline
            Ratings Only & \cite{abdollahi2016explainable} &&&&&& \\ \hline
            Textual Reviews && \cite{seo2017interpretable}, \cite{chen2018neural}, \cite{wu2019context}, \cite{chen2019dynamic} & \cite{cong2019hierarchical} & \cite{costa2018automatic}, \cite{li2017neural}, \cite{lu2018like}, \cite{ouyang2018improving}, \cite{zhao2018you}, \cite{suzuki2018toward} &&& \cite{wang2018reinforcement},\cite{lin2018neural} \\ \hline
            Images && \cite{chen2017attentive}, \cite{chen2018visually}, \cite{hou2019explainable} &&& \cite{kang2017visually}, \cite{kumar2019c} && \cite{bharadhwaj2018layer}, \cite{lin2019explainable}, \cite{wu2019hierarchical} \\ \hline
            Temporal Data && \cite{li2017neural}, \cite{tang2018personalized} &&&& \cite{chen2018sequential} & \\ \hline
        \end{tabular}
    \end{table}

\subsection{Explanation on Collaborative Filtering}
\begin{figure}
    \centering
    \includegraphics[width=0.9\linewidth]{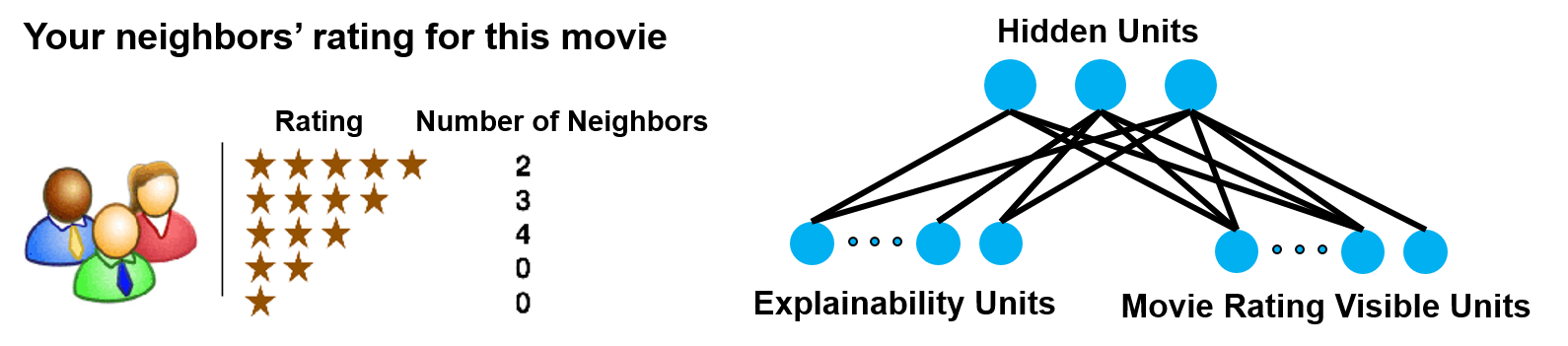}
    \caption{Explainable Restricted Boltzmann Machines: explanations with relevant users (left), conditional RBM for explainability (right).(Abdollahi and Nasraoui\cite{abdollahi2016explainable})}
    \label{fig:explainable_RBM}
\end{figure}

In the very early days of recommendation explanation research, collaborative filtering (CF) serves as the fundamental method for personalized recommendation. CF leverages the users' implicit or explicit feedback from which explanations can sometimes be generated in a very straightforward way. For example, in user-based CF, the system decides whether or not to provide a certain user with an item according to the ratings from his/her neighbors, which can be considered as a form of explanation. Similarly, in item-based CF, explanations are generated for a target user based on whether the rating given to an item is similar to the other already-given ratings. However, this is not the case for deep learning-based CF. Recently, with a number of deep learning-based recommendation models being proposed, the state-of-the-art performance in various recommendation tasks, such as rating prediction, top-N recommendation, and sequential recommendation, has been dominated by deep approaches, but most of them lack the ability to explain their recommendation results because what the deep neural networks have learned is normally hard to interpret. Abdollahi and Nasraoui \cite{abdollahi2016explainable} focuses on the interpretability of Restricted Boltzmann Machines (RBM) based CF recommendations without relying on any auxiliary data, such as item content or user attributes. In this paper, similar to the idea of explainable user-based CF, for each target user, the authors introduce the concept of "explainability score" which is calculated from the rating distribution inside his neighbors that are determined using the cosine similarity. This score is ranges from zero to one. Only when a score is greater than zero, the item is explainable for a user. The higher the score, the more explainability is achieved. Then, the authors employ the conditional RBM model with an additional visible layer that has exactly the same number of hidden units as the number of items (see figure~\ref{fig:explainable_RBM}). The output value of each hidden unit in this layer is limited within $0$ and $1$, representing the above explainability score. In this way, the conditional RBM model tends to recommend items that are explainable. In essence, this approach provides explanations via user-based neighborhoods.

\subsection{Explanation on Textual Data}
In recommender systems, textual contents are a common and major source of auxiliary information such as user reviews and product descriptions. Numerous deep learning techniques have been adopted to exploit textual data, such as CNN \cite{seo2017interpretable,chen2018neural,chen2019dynamic}, RNN \cite{ouyang2018improving,zhao2018you,lin2019explainable}, and attention mechanism \cite{chen2017attentive,wu2019context,cong2019hierarchical}. Among the above methods, CNNs are often employed for deep feature extraction from text and are combined with an attention mechanism to generate explanations\cite{seo2017interpretable,chen2018neural}, while RNNs are often used for textual explanation generation\cite{suzuki2018toward,zhao2018you}. Here, we review these deep learning-based explainable recommendation models that exploit textual side information, and analyze their strengths and weaknesses.

\begin{figure}
    \centering
    \includegraphics[width=0.9\linewidth]{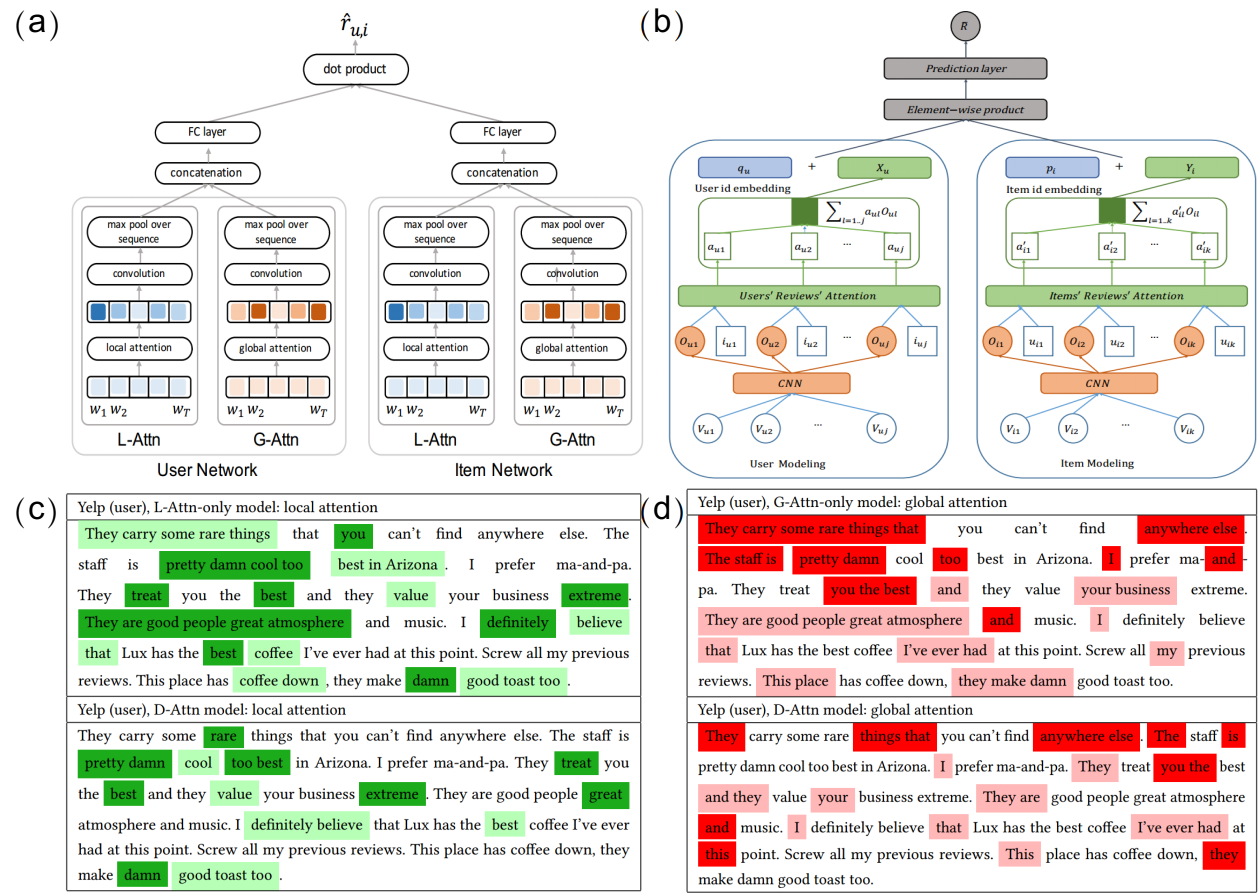}
    \caption{Examples of attentive CNN networks for explainable recommendation. (a)Dual-attention neural networks (Seo et al. \cite{seo2017interpretable}); (b)Neural attentional regression networks (Zheng et al. \cite{chen2018neural}); (c) and (d) Words in a user's review text annotated according to the learned attention scores(Seo et al. \cite{seo2017interpretable}).}
    \label{fig:explainable_text_cnn_att}
\end{figure}

The first group of methods combines CNN with attention mechanism to analyze the textual data. Seo et al. \cite{seo2017interpretable} aggregates all the review texts given by a user and an item respectively to form two sets of representations from which the abstract features of different users and items are learned via convolutional neural networks with dual attention mechanism (i.e., global and local attention), as shown in figure~\ref{fig:explainable_text_cnn_att}(a). The predicted ratings are then generated from the learned features similar to that of matrix factorization. In the meantime, the dual attention networks enable word focusing in the review texts. Different from the architecture in \cite{seo2017interpretable}, Chen et al. \cite{chen2018neural} proposed a neural attentional regression model with review-level explanations (NARRE) which employs the DeepCoNN network to process the reviews \cite{zheng2017joint}, shown in figure~\ref{fig:explainable_text_cnn_att}(b). Although NARRE uses only a single attention layer on the output of DeepCoNN network, it is not only capable of generating highly-accurate prediction ratings, but can also choose useful reviews that offer a form of review-level explanations to the target users. The above two methods use review texts as inputs only. In other words, they ignore the user-item interactions, and thus fail to completely model the users' rating behaviors. Wu et al. \cite{wu2019context} proposed a context-aware user-item representation learning model (CARL) to overcome such shortcomings. CARL fuses two different networks, one for review feature extraction and the other for user-item interaction feature extraction. To process reviews, CARL employs an attentive CNN neural network, and to model user-item rating interactions, it adopts a matrix factorization-like approach to learn user/item latent representations. The final predicted ratings are fused by a dynamic weighting scheme between the outputs of the two networks. Unlike the aforementioned static methods with attention applied only on textual data, Chen et al. \cite{chen2019dynamic} built a dynamic explainable recommender (DER) by combining a gated recurrent unit (GRU)-based network that models user dynamic ratings with a sentence-level CNN to profile an item by its reviews. DER applies attention on the mixture of a user's time-varying preference at a certain time and the sentence-level features of the item reviews, merging sentence embeddings under "user-aware" attention weights. Therefore, DER can provide explanations in a dynamic, personalized manner. Despite the model differences, in all the above approaches, the recommendation explanations are produced in the form of a group of words with high attention weights to help the user understand the recommendations, shown in figure~\ref{fig:explainable_text_cnn_att}(c)(d).

\begin{figure}
    \centering
    \includegraphics[width=0.9\linewidth]{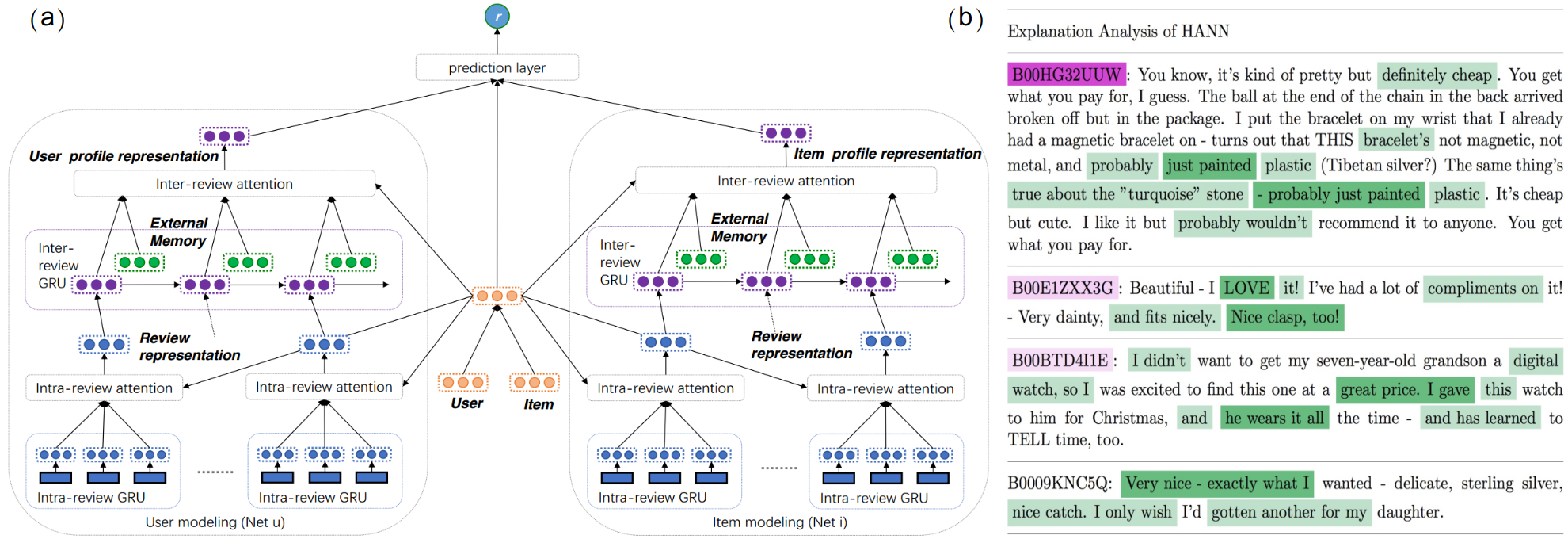}
    \caption{An Example of attentive RNN networks for explainable recommendation. (a)Hierarchical attention-based network (HANN); (b)Explanation analysis of HANN.(Cong et al. \cite{cong2019hierarchical})}
    \label{fig:explainable_text_rnn_att}
\end{figure}

The second group of methods exploits RNN \cite{mikolov2010recurrent}, a very effective family of deep neural networks for natural language processing. Some approaches introduce attention mechanism into RNNs for similar reasons as the above attentive CNN-based models to select highly-relevant words from the review texts as explanations \cite{cong2019hierarchical}. Most existing models take advantage of the generative ability of RNNs to produce user/item review explanations \cite{li2017neural,lu2018like,suzuki2018toward,costa2018automatic,ouyang2018improving,zhao2018you,lin2019explainable}. Cong et al. \cite{cong2019hierarchical} proposed a hierarchical attention-based network(HANN) that generates explanations by considering the contribution of reviews to the overall ratings at two levels, the word level and the review level, shown in figure~\ref{fig:explainable_text_rnn_att}. HANN is similar to Seo et al. \cite{seo2017interpretable} in that HANN replaces the CNN by GRU-RNN. HANN also splits the textual data into user reviews and item reviews that are further fed into two separate GRU-based deep neural networks, known as the user net and the item net. Dual attention is adopted, one at word level for intra-review attention and the other at review level for inter-review attention. Both nets are fused together by fully-connected layers to predict the ratings. In figure~\ref{fig:explainable_text_rnn_att}(b), the explanations are generated using the attention scores at both levels. The darker the pink color, the higher the attention score reaches at the review level. Word level attention scores are similarly denoted by the green color. In this way, HANN extracts useful words from the reviews to form the explanation and meanwhile, globally distinguishes the effectiveness of reviews on the final predicted rating scores.

\begin{figure}
    \centering
    \includegraphics[width=0.9\linewidth]{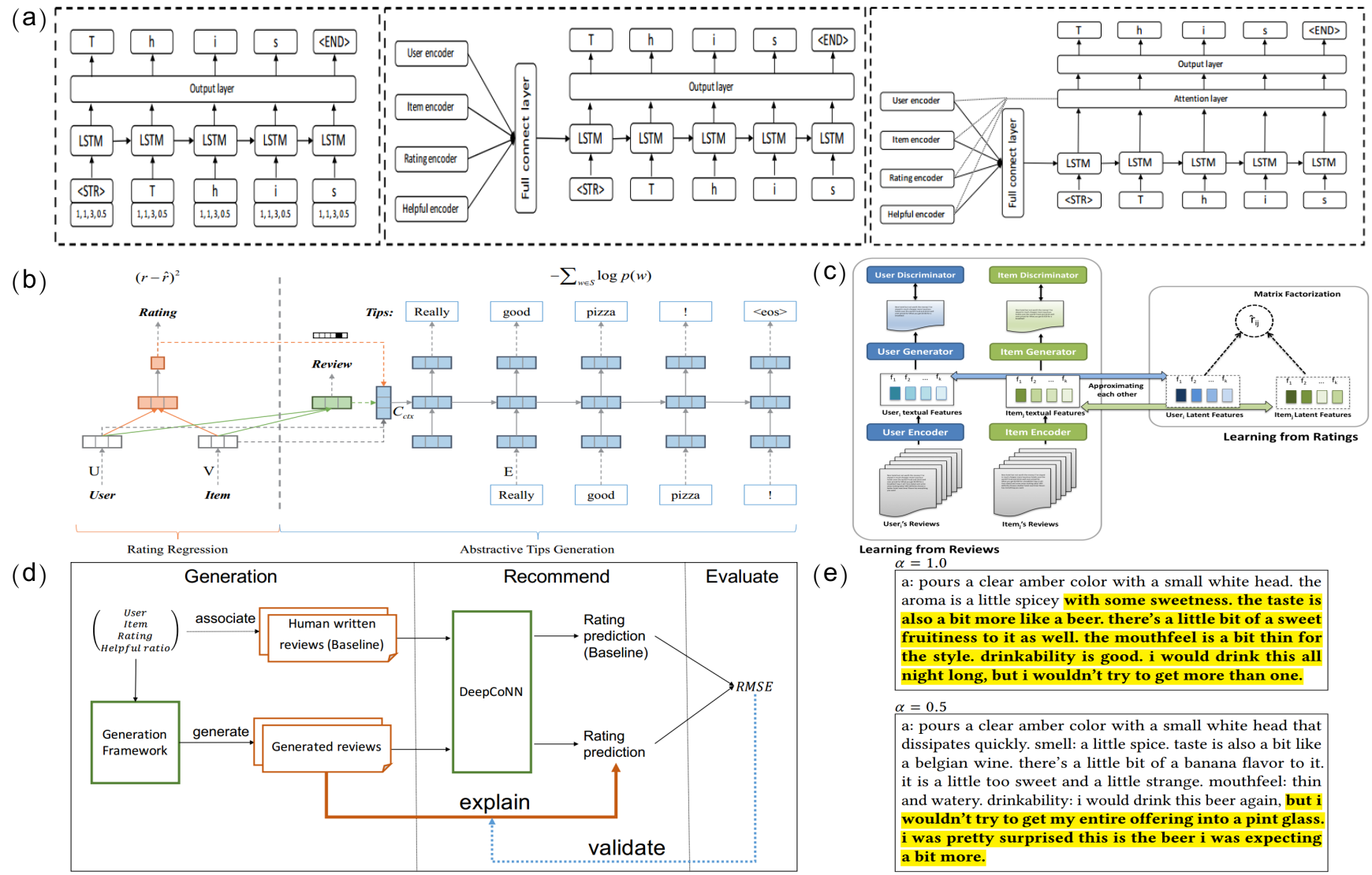}
    \caption{Examples of generative RNNs for explainable recommendation. (a) Generative concatenative networks, context-aware generation model and attention enhanced generating model (Ouyang et al. \cite{ouyang2018improving}); (b) Neural rating and tips generation (NRT) network (Li et al. \cite{li2017neural}); (c) Multi-task learning model for simultaneous rating prediction and review generation (Lu et al. \cite{lu2018like}); (d) Validation setup of recommendation performance of generated reviews (Ouyang et al. \cite{ouyang2018improving}); (e) Generated textual reviews for explanations. With different parameters, the model generates different explanations. (Costa et al. \cite{costa2018automatic})}
    \label{fig:explainable_text_rnn_gen}
\end{figure}

Due to the powerful text generation ability of RNNs, a host of RNN-based explainable recommender systems are proposed to leverage this feature. Costa et al. \cite{costa2018automatic} designed a character-level generative concatenative network based on LSTM cells \cite{hochreiter1997long}, where the ground-truth ratings serve as an auxiliary information and are concatenated into the input layer. Therefore, the model is able to generate reviews following the directions pointed by the rating scores (figure~\ref{fig:explainable_text_rnn_gen}(a)). By adjusting the hyper-parameters, the model is able to provide very natural explanations for human readers shown in figure~\ref{fig:explainable_text_rnn_gen}(e). Instead of generating reviews, Li et al. \cite{li2017neural} proposed a multi-task learning model, i.e. the neural rating and tips generation network (NRT). NRT takes ratings and reviews as context and produces abstract tips, as shown in figure~\ref{fig:explainable_text_rnn_gen}(b). User-item rating pairs are first used to learn the user/item latent factors via multi-layer perceptrons (MLP) that forms the rating regression network. These latent factors are then fed into a standard MLP-based review text generation network whose output layer, together with the predict ratings, serves as the context of a GRU-RNN based tip generation network. The multi-task objective is then composed of the rating regression loss, the review generation loss, and the tip generation loss. The generated tips are both concise and vivid enough to predict users' possible experience and feelings. Another multi-task recommendation model is proposed by Lu et al. \cite{lu2018like}. The authors utilize the adversarial sequence to sequence learning technique. The reviews are first encoded into latent feature vectors using a bidirectional GRU-RNN network and then decoded by a single GRU-RNN network, creating a review autoencoder structure. This autoencoder is adversarially trained with a CNN-based review discriminator to identify if a piece of given review is written by user $i$ on item $j$. Unlike NRT that considers ratings as the context of the explanation generation process, Lu et al. the latent textual features are fed into a matrix factorization rating prediction algorithm as the context. Both models are jointly trained using the alternating least squares (ALS) technique \cite{hu2008collaborative} to perform rating prediction and explanation generation. Furthermore, the RNN-generated reviews not only provide explanations but can also act as inputs for the recommender system. To test whether the generated reviews are more effective in recommendation than human-written reviews, using DeepCoNN \cite{zheng2017joint} as the recommender, Ouyang et al. \cite{ouyang2018improving} compare human-written reviews with synthetic reviews that are produced at both character and word levels by popular review generation models (figure~\ref{fig:explainable_text_rnn_gen}(d)). Results show that synthetic reviews can carry more consistent information appropriate to the demands of a recommender system than human-written reviews, justifying the feasibility and rationality of using generated reviews to explain recommendations. Instead of using reviews or tips as inputs, Zhao et al. \cite{zhao2018you} feed user/item side information (e.g., user/item tags, item title, user gender, etc.) into a recurrent attention generation network to produce reasons for the explainable recommendation in conversation applications. Similarly, Suzuki et al. \cite{suzuki2018toward} adopts a MLP network to encode multicriteria evaluation ratings (e.g., overall rating, location rating, service quality rating, price rating, etc.) into a latent vector which is then decoded by an attentive LSTM-RNN network into reviews so that personalized explanations are generated for the predicted ratings.

\subsection{Explanation on Visual Data}
Compared with textual data, visual contents often contain more information that can be exploited for recommendation explanations. Most previous image-based recommendation approaches transform images into latent representation vectors to be incorporated into recommendation algorithms \cite{he2016vbpr, liu2017deepstyle, wang2017your, yu2018aesthetic}. However, such approaches are hardly useful in ex planing why a particular item is recommended. Recently, some initial steps have been taken towards visual explainability of recommendations via exploiting the power of deep learning. Most existing recommendation models using visual data adopt CNN as the building block, given its popularity and success in processing visual data. Other deep learning techniques employed to enable explainable recommendations include the attention mechanism \cite{chen2017attentive, chen2018visually, hou2019explainable} and generative adversarial network (GAN) \cite{kumar2019c, kang2017visually}. Apart from generating explanations from the visual data itself, some approaches regard images as a source of auxiliary information to help explain the recommendations \cite{lin2019explainable, bharadhwaj2018layer, wu2019hierarchical}. Unlike the textual data based explainable recommendation models, all the above approaches explain the recommendation results in a straightforward manner.

\begin{figure}
    \centering
    \includegraphics[width=0.95\linewidth]{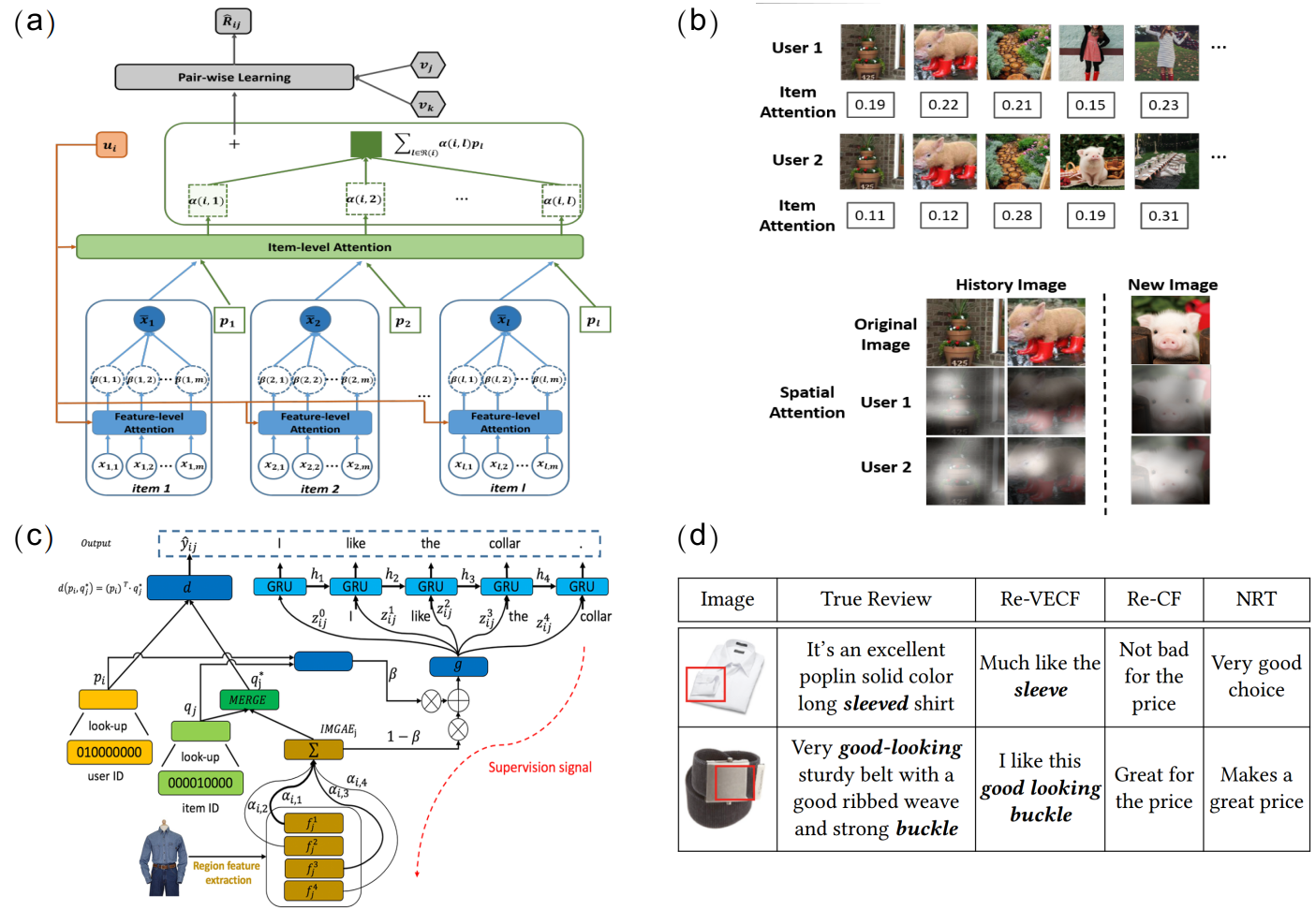}
    \caption{Examples of attentive CNN networks for explainable recommendation. (a) Attentive collaborative filtering (ACF) (Chen et al. \cite{chen2017attentive}); (b) Visualization results of the dual level attention given by ACF. At the item-level, the value under each item represents the attention weight of the item, while for the component-level, a heat map is used to represent the attention value, in which the darker the color is, the lower its represented attention value is (Chen et al. \cite{chen2017attentive}); (c) Review-enhanced visually explainable collaborative filtering (Re-VECF) \cite{chen2018visually}; (d) Generated reviews from Re-VECF compared with the true reviews \cite{chen2018visually}.}
    \label{fig:explainable_visual_cnn_att}
\end{figure}

\begin{figure}
    \centering
    \includegraphics[width=0.95\linewidth]{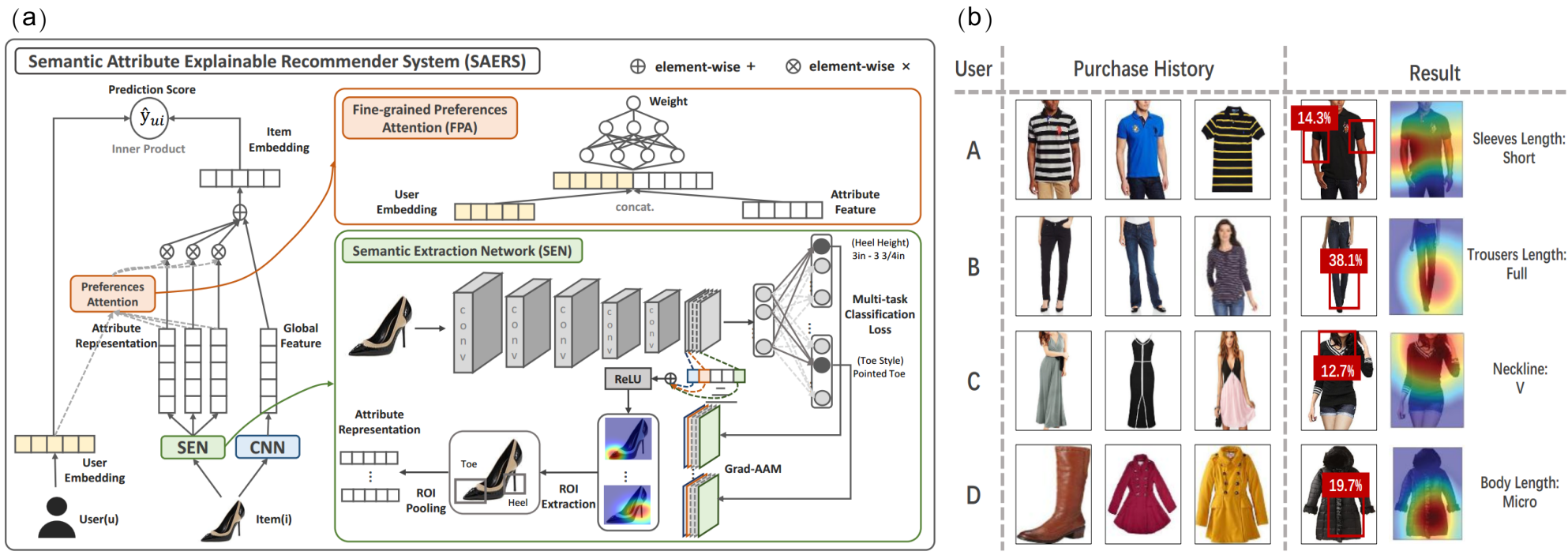}
    \caption{The semantic attribute explainable recommender system (SAERS). (a) The architecture of SAERS; (b) Examples of the visual explanations from SAERS. (Hou et al. \cite{hou2019explainable})}
    \label{fig:explainable_visual_cnn_att2}
\end{figure}

Similar to those models that leverage attention mechanism for textual data, the first type of visually-explainable recommender systems applies attentive deep neural networks to select a group of "physical regions" \cite{chen2017attentive, chen2018visually} or "semantic regions" \cite{hou2019explainable} from the images as the explanations. The earliest attempt we can find is from Chen et al. \cite{chen2017attentive}, in which work the attentive, collaborative filtering (ACF) model is proposed via hierarchical attention at both component level and item level, as shown in figure~\ref{fig:explainable_visual_cnn_att}(a). ACF combines the latent factor model with an attentive neural network that processes the features from items to provide top-N recommendations using implicit feedback. The item features are extracted by a CNN-based deep network, ResNet-152 \cite{he2016deep}, from images or video frames. After processed by the dual attention network, these features are merged with user latent factors via element-wise addition to reflect the users' detailed preferences. Bayesian Personalized Ranking (BPR) \cite{rendle2009bpr} is adopted as the last step to generate the final recommendations. Figure~\ref{fig:explainable_visual_cnn_att}(b) shows an example from the ACF's results. The explanations are given by the attention weights where the higher the weights, the more probable the user will like the entire images or the regions of images. Chen et al. \cite{chen2018visually} exploits both item images and user's textual reviews whose features are extracted via VGG-19 and GRU-RNN, respectively. The VGG-19 produced image features are split into a number of regions which are then passed through an attention layer for explanation and merged with the item latent factors to represent the items. The user and item latent factors are combined with the item representation vectors to serve as the inputs of the GRU-RNN network for review generation. This model is named review-enhanced visually explainable collaborative filtering (Re-VECF), shown in figure~\ref{fig:explainable_visual_cnn_att}(c). Unlike ACF that aims at user representations, Re-VECF focuses on item representations and uses a single attention layer with the adoption of element-wise multiplication to merge the image features and item latent factors. Figure~\ref{fig:explainable_visual_cnn_att}(d) illustrates some explainable recommendations, where the bolded italic words (e.g., sleeve) mean that the word generated by Re-VECF was also mentioned in the true review, and the word is aligned to the boxed area of the image learned by the attention mechanism. Different from the above two approaches that exert attention on "physical regions" of the images, Hou et al. \cite{hou2019explainable} proposed the Semantic Attribute Explainable Recommender System (SAERS) to understand users' semantic preferences via integration of the Fine-grained Preferences Attention (FPA) mechanism and the Semantic Extraction Network (SEN) for fashion recommendation. SAERS converts each attribute extracted from a particular region of the clothing images into one dimension in a semantic attribute visual space. As shown in figure~\ref{fig:explainable_visual_cnn_att2}(a), SEN consists of the CNN-based ResNet-50 \cite{he2016deep} network for semantic attribute classification and the Gradient-weighted Attribute Activation Maps (Grad-AAM) \cite{selvaraju2017grad} for location and extraction of attribute representations in a weakly-supervised manner. The authors then adopt FPA to align the user latent factors with the semantic attribute visual space. Each user latent factor is concatenated with one transferred semantic attribute representation vector, upon which the attention mechanism is applied to learn the user's preferences over different semantic attributes. Finally, BPR is used for recommendation, where the ratings are predicted via the inner product of user and item latent factors. Figure~\ref{fig:explainable_visual_cnn_att2}(b) shows several visual explanation examples from SAERS. The learned attention weights are demonstrated in the red boxes. The weights indicate how much the user prefers a particular attribute. For instance, when a dress is recommended to user C, the model explains to her that this dress has a V-shaped neckline, which is reasonable because according to user C's purchase history, she has bought three V neckline dresses before. Therefore, the recommendations are visually explained, improving the trust of the system.

The second type of visual explainable recommendation approaches exploits GAN. Similar to SAERS, Kang et al. \cite{kang2017visually} addresses the fashion recommendation by employing the Siamese CNNs \cite{hadsell2006dimensionality} to extract ``fashion-aware" image features to give the notations of "style." Although this is enough to provide explanations from a certain aspect, the authors further adopt conditional GANs \cite{mirza2014conditional} to generate images leveraging semantic inputs, where the product's top-level category is chosen as the condition. Thus, this approach can generate novel items that is likely to satisfy the users and are not in the training dataset so that the explanations for a series of recommended items can be summarized into such generated images. Kumar et al. \cite{kumar2019c} tackles the pairing problem in fashion recommendation via an enhanced conditional GAN model called $c^{+}$GAN. Given one piece of clothing image, the model recommends a set of items that best match the given clothing in a generative manner. $c^{+}$GAN modifies the generator with a classical mean squared error (MSE) loss and also a simplified perceptual loss using discrete cosine transform (DCT) coefficients of the generated as well as the target images. A simplified lensing technique \cite{sajjadi2018tempered} to the discriminator is applied to stabilize the generator training. Equipped with these techniques, $c^{+}$GAN is able to generate very meaningful fashion items as recommendation explanations.

\begin{figure}
    \centering
    \includegraphics[width=0.95\linewidth]{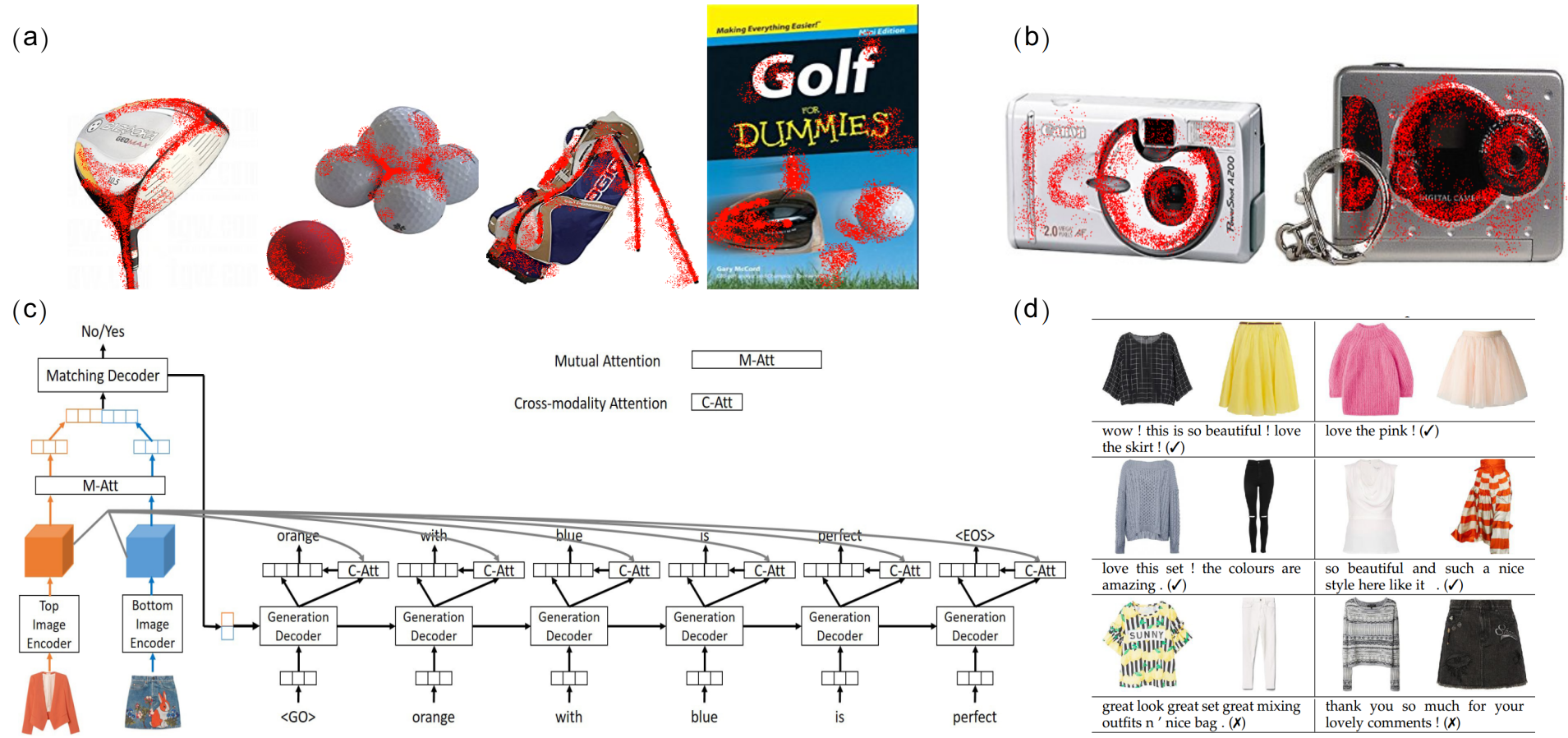}
    \caption{Examples of some other image-based recommendations. (a) and (b) Recommendation via layer-wise relevance propagation. The left-most image is the input query item and the rest are recommendations. On the original images, the generated heatmaps are shown (Bharadhwaj \cite{bharadhwaj2018layer}); (c) Overview of the neural outfit recommendation (NOR) model; (d) A few examples of recommendations and the corresponding generated comments for explanation from NOR (Lin et al. \cite{lin2019explainable}).}
    \label{fig:explainable_visual_others}
\end{figure}

Bharadhwaj \cite{bharadhwaj2018layer} adopts the content-based similarity approach to recommend images. The authors modify the VGG-16 network by layer-wise relevance propagation \cite{bach2015pixel}, which enables relevance conservation at each layer in a pixel-wise manner. Given the purchase history of the target user (query items), the model can generate a list of other items that most resemble the query items. The explanations are given by the highlighted pixels that are the most informative for inferring which items go along well with the query item, shown in figure~\ref{fig:explainable_visual_others} (a) and (b).

Other explainable recommender systems consider the visual features extracted via deep neural networks as the auxiliary information to generate non-visual explanations. Lin et al. \cite{lin2019explainable} applied multi-task learning to recommendation and proposed the neural outfit recommendation (NOR) model. NOR recommends outfits to users with abstractive comments generated as explanations. To achieve both tasks, as shown in figure~\ref{fig:explainable_visual_others} (c), NOR adopts two neural networks, i.e., the outfit matching network and the comment generation network. Equipped with the mutual attention mechanism, the outfit matching network utilizes CNNs for visual feature extraction. These abstract visual features are further transformed into rating scores to predict the most matched outfit. To generate textual explanations from the aforementioned visual features, the authors exploit the cross-modality attention over the above CNN network and a GRU-RNN network, leading to the comment generation network. Figure~\ref{fig:explainable_visual_others} (d) shows some examples from the recommendations produced by NOR. The recommendation explanations are given in the form of generated comments which are mostly focused on the general opinions on the matched pair of outfits. Wu et al. \cite{wu2019hierarchical} exploited the hierarchical attention mechanism in the field of image recommendation in social networks. The proposed model leverages heterogeneous data, e.g., users' rating behaviors, social network, upload behaviors, images. And from such complex relationships between users and images, the model represents these contextual factors as different sets of embeddings. A hierarchical attention network is then applied to attend differently to various embeddings. Images are represented using their content vectors, extracted by VGG-19, and their style vectors, generated by a CNN-based synthesis method \cite{gatys2015texture}. However, this model does not generate explanations straightforwardly, but the recommendations can be intuitively interpreted by the learned attention weights of different aspects.

\subsection{Explanation on Temporal Data}
\begin{figure}
    \centering
    \includegraphics[width=0.95\linewidth]{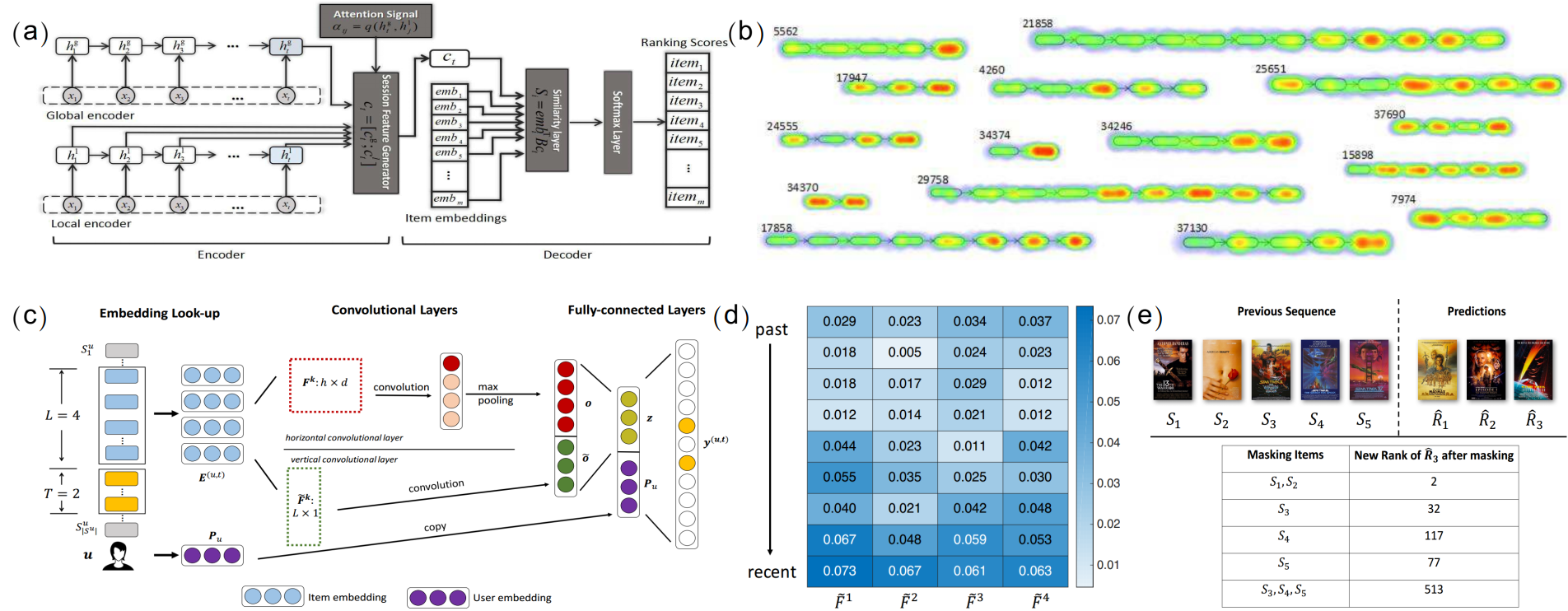}
    \caption{Examples of sequential explainable recommender systems. (a) The architecture of neural attentive recommendation machine (NARM) (Li et al. \cite{li2017neural}); (b) Heatmap visualization of items weights by NARM (Li et al. \cite{li2017neural}); (c) The network architecture of convolutional sequence embedding recommendation (Caser)(Tang et al. \cite{tang2018personalized}); (d) Illustration of the vertical convolutional filters in the Caser model (Tang et al. \cite{tang2018personalized}); (e) Illustration of the union-level pattern extraction ability of the horizontal convolutional filters in the Caser model (Tang et al. \cite{tang2018personalized}).}
    \label{fig:explainable_others_seq}
\end{figure}

\begin{figure}
    \centering
    \includegraphics[width=0.9\linewidth,height=11cm]{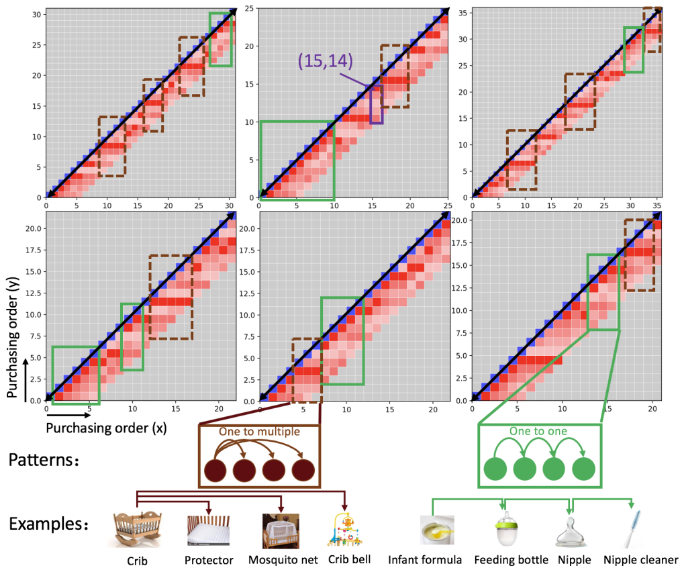}
    \caption{Illustration of sequential item-to-item transitions using MANN (Chen et al. \cite{chen2018sequential}).}
    \label{fig:explainable_others_seq_mem}
\end{figure}

Sequential recommendation takes advantage of the temporal characteristics that exist in user dynamic behaviors to improve recommendation effectiveness \cite{vinagre2015overview}. The temporal aspect provides another dimension to generate explanations for the recommendations. A user's history is no longer a collection of unordered items, but a sequence of time-aware items. The order itself can provide certain explanations. For instance, if item $i$ and $j$ are complementary, a user bought item $j$ at a certain time might be explained by the fact that this use had bought item $i$ sometime earlier. With such observation, Li et al. \cite{li2017neural} proposed the Neural Attentive Recommendation Machine (NARM) to learn the user's primary intention in the current session (figure~\ref{fig:explainable_others_seq}(a)). NARM employs GRU-RNN as the basic building block. NARM contains a global encoder that interprets the last hidden state in the RNN as the user's behavior feature, and a local encoder that interprets all the hidden states in the current session as the user's primary purpose feature.  Attention mechanism is applied to the local encoder to learn different weights for the hidden states so that the model can tell which past items contribute more to the future items. The two encoders are then combined as inputs of the decoder, which predicts the recommendation possibility of each candidate item. In figure~\ref{fig:explainable_others_seq}(b), we can clearly see the effect of the attention mechanism. In a particular session, the importance of items are reflected by the depth of the colors. This can give certain explanations on the next recommended items. To be specific, the users' decision on the next clicked items are more influenced by those near the end of the session than those at the start, which is consistent with people's purchasing or browsing behavior we have noticed in reality. Tang et al. \cite{tang2018personalized} proposed a Convolutional Sequence Embedding Recommendation Model (Caser) as another solution to the sequential pattern extraction problem. As shown in figure~\ref{fig:explainable_others_seq}(c), Caser embeds a set of recent items into a two dimensional matrix whose dimensions represent the time and latent feature respectively. Two convolutional filters, one vertical and the other horizontal, are then applied onto this matrix to learn sequential patterns that are expressed as local features. The two filters capture patterns at different levels. Horizontal filters aim for union-level patterns via unifying the data into multiple sizes, while vertical filters aim for point-level sequential patterns by calculating the weighted sums using the previous items' latent representations. For clarity, visualization of vertical filters is shown in figure~\ref{fig:explainable_visual_others}(d) that reflects the importance of different past items. Union-level sequential patterns can be effectively extracted by the horizontal filters (figure~\ref{fig:explainable_visual_others}(e)), where the recommended $\hat{R}_3$ (ground truth) is generated by the union of $S_3$, $S_4$ and $S_5$ due to the same genre they belong to. If any of $S_3$, $S_4$ and $S_5$ is masked in the horizontal filters, the ranking position of $\hat{R}_3$ is largely reduced. Chen et al. \cite{chen2018sequential} took advantage of the memory mechanism for long-term memory and integrated collaborative filtering into a memory-augmented neural network (MANN). MANN stores and updates users' historical records explicitly and is also able to extract the intuitive patterns of how users' future actions are influenced by their previous decisions and behaviors. Shown in figure~\ref{fig:explainable_others_seq_mem}, MANN can capture two different types of sequential patterns. "One-to-one" behavior pattern generates sequences where the next action is only influenced by the most recent action. "One-to-multiple" behavior pattern generates sequences where a set of continuous behaviors are influenced by the same previous action. Both patterns can be widely observed in practice. For example, when browsing web pages, one may keep following the related links on each page and form a "one-to-one" pattern, and when searching key words through a search engine, one may browse multiple pages related to the same key words, leading to a "one-to-multiple" pattern. The discovery of such patterns by MANN can act as the explanations for why a particular user will buy a certain item in the future.

\subsection{Other Approaches}

Wang et al. \cite{wang2018reinforcement} proposed a reinforcement learning framework for explainable recommendation, which is quite universal because instead of integrating a certain explanation mechanism into a recommendation model, it has no restrictions on the details of the model to be explained. The authors consider users, items, side information, and a recommendation model to be explained as the environment. Two couple agents are employed, one for explanation generation and the other for explanation discrimination. At each state, the generator agent gives a piece of explanation taking the user-item pairs as inputs, while the discriminator agent takes the generated explanation as input to predict rating scores. The reward of the agents is calculated by measuring how similar the agent-predicted scores are to the recommendation model-predicted scores and the presentation quality (e.g., readability and consistency) of the generated explanation. By taking the textual sentences as interpretable components, the authors then adopt a personalized attention-based neural network as an instantiation of the proposed framework and show that it can well explain the recommendations via sentence-level explanations. Lin et al. \cite{lin2018neural} integrates the rating score prediction task and the explainable word generation task into a unified neural network. In this model, neural collaborative filtering (NCF) \cite{he2017neural} is applied to the user-POI rating matrix to predict the rating scores. The reviews are transformed into syntax relations by utilizing the spaCy CNN dependency parsing model \cite{kiperwasser2016simple,goldberg2012dynamic}, which are further organized into pairs <opinion, aspect>. The learned user embeddings from NCF are then clustered based on cosine similarity. The textual explanations are then extracted from the pre-processed word pairs in the top-K users' reviews.

\subsection{Summary}

In this section, we reviewed deep learning-based explainable approaches for trust-aware recommendation. After introducing the general techniques for explainable recommendation, we focused on deep learning methods that leverage collaborative filtering, textual data, visual data, and temporal data.
Generally, deep learning aims to explain the recommendation results from the mechanism of how the recommendation process works. Most models leverage textual reviews, item images, and temporal information of the user-item interactions to deal with limited user-item ratings. Considerable work combine attention mechanism with deep neural networks (e.g., RNNs, CNNs and memory neural networks) and generate explanations from auxiliary data (i.e., reviews, images, sequential patterns). Another category of models adopts generative methods (e.g., RNN and GAN), to provide novel textual/image explanations. Other models include traditional content-based similarity methods, hybrid methods that exploit both texts and images, and reinforcement learning that controls the quality of explanations.
Overall, deep learning has demonstrated as a promising approach to explainability for trustworthy recommendation.

\section{Discussion}
\label{sec:potential trends}
Though effective, current researches face the challenges such as relying on sufficient labels, requiring manually tuning, and inflexibility for multi-tasks. We discuss potential solutions to some of the issues as follows. 

\subsection{Dynamic Trust in Recommender Systems}
In the real world cases, the trust information is changing over time. For example, trust friend relationships may change; we may have a different group of friends during different periods. Also, the preference for friendship may change over time\cite{chen2019dynamic}. Another example is for malicious users. We know that the malicious reviews or ratings may affect common users, but not all the reviews written by malicious reviewers are for sure the fake reviews\cite{si2018shilling}. Some fake reviewers may mystify the detection system by writing some common reviews but write fake reviews to the target products. Most work assumes that the reviews/ratings made by malicious users are fake samples; however, limited work considers solving the problem case by case because it is hard to get all the labels. Another challenge is that the dynamic system may be time-consuming for updating the whole system over time. 

\subsection{Embedding for Recommender System}
Embedding methods, which include node embedding, sequential embedding, and graph embedding, are widely applied into recommender systems. For example, several graph embedding methods learn social relationship as a type of graph information. Such embedding methods include node2vec\cite{grover2016node2vec}, Euclidean embedding\cite{li2017social}, UniWalk (explainable) \cite{park2017uniwalk}, deepWalk\cite{perozzi2014deepwalk}, and the recent model, graph neural networks\cite{defferrard2016convolutional}. 
Effective embedding has achieved significant improvement in recommender systems, including various application domains\cite{ouyang2019learning}.
The learned embedding vector is normally used for representing users/items or further combined with other representations. 
Some may construct the graph for items to model the user behaviors: each item is represented by a node in the graph, and the co-occurrence of items is denoted by edge. Then, the graph embedding methods are applied to learn the embedding\cite{ouyang2019learning}. 
Some may represent the user by their trust relationships, and then combine such representation with user rating behaviors for recommendation \cite{wu2018socialgcn,wu2019dual}.
Most related work has the limitation of ignoring the inner interactions between different types of information. Intuitively, a user's decision is affected by many factors. Friendship, product reviews (especially malicious reviews), and product description can all affect our decisions. Thus, the latent representation for the user should be better not simply the concatenation of representations in different domains, but a unified factor.  

\subsection{Deep Meta-learning for Recommender Systems}
For recommendation problem, there is no standard base model for dealing with different tasks. Like the aforementioned three trust-aware tasks, limited work combines different ideas, i.e., cover all the bases, for recommendation problems. Applying meta-learning or learning to learn may cover this limitation. Meta-learning, or learning to learn, is the science of observing the performance of different model configurations on various tasks, and then learning from the history observation, to provide guidance to the new tasks. This will not only improve the efficiency of the new task modeling but also enable the model with automatically learning capabilities. This is an inspiring area since most current works are based on hand-engineered way\cite{vanschoren2018meta}.
Most deep models only perform well on one task or a single dataset. This means we may cost a lot of manual efforts on designing the models instead of solving the problems. Thus, it is meaningful to design a module to let the machine learns itself with supporting the related information. 

\subsection{Blockchain for Decentralized Trust Management}
Current recommender systems are built upon the data from web users -- which contain both normal users and malicious users -- and thus become vulnerable to the real world frauds. For example, with the increasing number of fraudulent ratings or feedbacks, the truth for a recommender system will deviate from the genuine truth. A major reason for fraudulent behaviors is the easiness of getting publicly available information. With learning from normal users, fraudulent users can hide their intention and cheat the detection techniques, which will affect the robustness of recommender systems. 
One potential solution is leveraging the blockchain idea for trust management in recommender systems. 
Blockchain\cite{yao2019recommendations, crosby2016blockchain} is a shared ledger technology, and each participant shares a common view of truth. It uses a decentralized peer-to-peer network to manage the data, which can eliminate the potential risks of centrally stored data,
and all validated activities are permanently recorded. Each participant can get access to their own data, and even a system administrator cannot delete the records. This secures each transaction and thus eliminates the human error or fraud. 
Besides, most of the deep learning-based recommendation techniques are data-hungry and are centralized in computing. A recent idea is distributing the learning tasks of a deep learning method on blockchain, which can improve both efficiency and privacy \cite{weng2018deepchain}.

\color{black}

\section{Conclusion}
In this survey, we investigate three aspects of trust in recommender systems: social-awareness, robustness, and explainability, with a focus on deep learning-based on recommender systems. We describe how deep learning methods work for the trust-aware recommendation in representation learning, predictive learning, and generative learning. We notice that the growing research in deep learning have brought about a significant improvement in the performance of recommender systems in various tasks. Meanwhile, current research still faces severe challenges in adapting to labeled data, reducing tuning efforts, and enhance the flexibility in handling multiple tasks. We hope this survey could give readers a comprehensive understanding of state-of-the-art studies in deep learning-based recommendation and inspire more insights and contributions to this vibrant research domain.

\bibliographystyle{unsrt}
\bibliography{survey}

\begin{thebibliography}{100}

\bibitem{zhang2019deep}
Shuai Zhang, Lina Yao, Aixin Sun, and Yi~Tay.
\newblock Deep learning based recommender system: A survey and new
  perspectives.
\newblock {\em ACM Computing Surveys (CSUR)}, 52(1):5, 2019.

\bibitem{guo2015trustsvd}
Guibing Guo, Jie Zhang, and Neil Yorke-Smith.
\newblock Trustsvd: collaborative filtering with both the explicit and implicit
  influence of user trust and of item ratings.
\newblock In {\em Twenty-Ninth AAAI Conference on Artificial Intelligence}.
  AAAI, 2015.

\bibitem{pan2017trust}
Yiteng Pan, Fazhi He, and Haiping Yu.
\newblock Trust-aware collaborative denoising auto-encoder for top-n
  recommendation.
\newblock {\em arXiv preprint arXiv:1703.01760}, 2017.

\bibitem{yang2017social}
Bo~Yang, Yu~Lei, Jiming Liu, and Wenjie Li.
\newblock Social collaborative filtering by trust.
\newblock {\em IEEE transactions on pattern analysis and machine intelligence},
  39(8):1633--1647, 2017.

\bibitem{ma2017explicit}
Xiao Ma, Hongwei Lu, Zaobin Gan, and Jiangfeng Zeng.
\newblock An explicit trust and distrust clustering based collaborative
  filtering recommendation approach.
\newblock {\em Electronic Commerce Research and Applications}, 25:29--39, 2017.

\bibitem{fei2013exploiting}
Geli Fei, Arjun Mukherjee, Bing Liu, Meichun Hsu, Malu Castellanos, and
  Riddhiman Ghosh.
\newblock Exploiting burstiness in reviews for review spammer detection.
\newblock In {\em Seventh international AAAI conference on weblogs and social
  media}, 2013.

\bibitem{wu2018socialgcn}
Le~Wu, Peijie Sun, Richang Hong, Yanjie Fu, Xiting Wang, and Meng Wang.
\newblock Socialgcn: An efficient graph convolutional network based model for
  social recommendation.
\newblock {\em arXiv preprint arXiv:1811.02815}, 2018.

\bibitem{zou2013belief}
Jun Zou and Faramarz Fekri.
\newblock A belief propagation approach for detecting shilling attacks in
  collaborative filtering.
\newblock In {\em Proceedings of the 22nd ACM international conference on
  Information \& Knowledge Management}, pages 1837--1840. ACM, 2013.

\bibitem{yang2016re}
Zhihai Yang, Lin Xu, Zhongmin Cai, and Zongben Xu.
\newblock Re-scale adaboost for attack detection in collaborative filtering
  recommender systems.
\newblock {\em Knowledge-Based Systems}, 100:74--88, 2016.

\bibitem{bhaumik2011clustering}
Runa Bhaumik, Bamshad Mobasher, and Robin Burke.
\newblock A clustering approach to unsupervised attack detection in
  collaborative recommender systems.
\newblock In {\em Proceedings of the International Conference on Data Mining},
  page~1. Citeseer, 2011.

\bibitem{si2018shilling}
Mingdan Si and Qingshan Li.
\newblock Shilling attacks against collaborative recommender systems: a review.
\newblock {\em Artificial Intelligence Review}, pages 1--29, 2018.

\bibitem{dong2017hybrid}
Xin Dong, Lei Yu, Zhonghuo Wu, Yuxia Sun, Lingfeng Yuan, and Fangxi Zhang.
\newblock A hybrid collaborative filtering model with deep structure for
  recommender systems.
\newblock In {\em Thirty-First AAAI Conference on Artificial Intelligence},
  2017.

\bibitem{zhang2018explainable}
Yongfeng Zhang and Xu~Chen.
\newblock Explainable recommendation: A survey and new perspectives.
\newblock {\em arXiv preprint arXiv:1804.11192}, 2018.

\bibitem{zhao2016exploring}
Wayne~Xin Zhao, Sui Li, Yulan He, Liwei Wang, Ji-Rong Wen, and Xiaoming Li.
\newblock Exploring demographic information in social media for product
  recommendation.
\newblock {\em Knowledge and Information Systems}, 49(1):61--89, 2016.

\bibitem{zhang2014explicit}
Yongfeng Zhang, Guokun Lai, Min Zhang, Yi~Zhang, Yiqun Liu, and Shaoping Ma.
\newblock Explicit factor models for explainable recommendation based on
  phrase-level sentiment analysis.
\newblock In {\em Proceedings of the 37th international ACM SIGIR conference on
  Research \& development in information retrieval}, pages 83--92. ACM, 2014.

\bibitem{lin2019explainable}
Yujie Lin, Pengjie Ren, Zhumin Chen, Zhaochun Ren, Jun Ma, and Maarten
  De~Rijke.
\newblock Explainable outfit recommendation with joint outfit matching and
  comment generation.
\newblock {\em IEEE Transactions on Knowledge and Data Engineering}, 2019.

\bibitem{wang2014also}
Beidou Wang, Martin Ester, Jiajun Bu, and Deng Cai.
\newblock Who also likes it? generating the most persuasive social explanations
  in recommender systems.
\newblock In {\em Twenty-Eighth AAAI Conference on Artificial Intelligence},
  2014.

\bibitem{chen2016learning}
Xu~Chen, Zheng Qin, Yongfeng Zhang, and Tao Xu.
\newblock Learning to rank features for recommendation over multiple
  categories.
\newblock In {\em Proceedings of the 39th International ACM SIGIR conference on
  Research and Development in Information Retrieval}, pages 305--314. ACM,
  2016.

\bibitem{heckel2017scalable}
Reinhard Heckel, Michail Vlachos, Thomas Parnell, and Celestine D{\"u}nner.
\newblock Scalable and interpretable product recommendations via overlapping
  co-clustering.
\newblock In {\em 2017 IEEE 33rd International Conference on Data Engineering
  (ICDE)}, pages 1033--1044. IEEE, 2017.

\bibitem{ren2017social}
Zhaochun Ren, Shangsong Liang, Piji Li, Shuaiqiang Wang, and Maarten de~Rijke.
\newblock Social collaborative viewpoint regression with explainable
  recommendations.
\newblock In {\em Proceedings of the tenth ACM international conference on web
  search and data mining}, pages 485--494. ACM, 2017.

\bibitem{seo2017interpretable}
Sungyong Seo, Jing Huang, Hao Yang, and Yan Liu.
\newblock Interpretable convolutional neural networks with dual local and
  global attention for review rating prediction.
\newblock In {\em Proceedings of the Eleventh ACM Conference on Recommender
  Systems}, pages 297--305. ACM, 2017.

\bibitem{davidson2010youtube}
James Davidson, Benjamin Liebald, Junning Liu, Palash Nandy, Taylor Van~Vleet,
  Ullas Gargi, Sujoy Gupta, Yu~He, Mike Lambert, Blake Livingston, et~al.
\newblock The youtube video recommendation system.
\newblock In {\em Proceedings of the fourth ACM conference on Recommender
  systems}, pages 293--296. ACM, 2010.

\bibitem{monti2017geometric}
Federico Monti, Michael Bronstein, and Xavier Bresson.
\newblock Geometric matrix completion with recurrent multi-graph neural
  networks.
\newblock In {\em Advances in Neural Information Processing Systems}, pages
  3697--3707, 2017.

\bibitem{yang2014survey}
Xiwang Yang, Yang Guo, Yong Liu, and Harald Steck.
\newblock A survey of collaborative filtering based social recommender systems.
\newblock {\em Computer Communications}, 41:1--10, 2014.

\bibitem{gunes2014shilling}
Ihsan Gunes, Cihan Kaleli, Alper Bilge, and Huseyin Polat.
\newblock Shilling attacks against recommender systems: a comprehensive survey.
\newblock {\em Artificial Intelligence Review}, 42(4):767--799, 2014.

\bibitem{koren2009matrix}
Yehuda Koren, Robert Bell, and Chris Volinsky.
\newblock Matrix factorization techniques for recommender systems.
\newblock {\em Computer}, (8):30--37, 2009.

\bibitem{mnih2008probabilistic}
Andriy Mnih and Ruslan~R Salakhutdinov.
\newblock Probabilistic matrix factorization.
\newblock In {\em Advances in neural information processing systems}, pages
  1257--1264, 2008.

\bibitem{he2017neural}
Xiangnan He, Lizi Liao, Hanwang Zhang, Liqiang Nie, Xia Hu, and Tat-Seng Chua.
\newblock Neural collaborative filtering.
\newblock In {\em Proceedings of the 26th International Conference on World
  Wide Web}, pages 173--182. International World Wide Web Conferences Steering
  Committee, 2017.

\bibitem{hu2008collaborative}
Yifan Hu, Yehuda Koren, and Chris Volinsky.
\newblock Collaborative filtering for implicit feedback datasets.
\newblock In {\em 2008 Eighth IEEE International Conference on Data Mining},
  pages 263--272. Ieee, 2008.

\bibitem{rendle2009bpr}
Steffen Rendle, Christoph Freudenthaler, Zeno Gantner, and Lars Schmidt-Thieme.
\newblock Bpr: Bayesian personalized ranking from implicit feedback.
\newblock In {\em Proceedings of the twenty-fifth conference on uncertainty in
  artificial intelligence}, pages 452--461. AUAI Press, 2009.

\bibitem{pazzani2007content}
Michael~J Pazzani and Daniel Billsus.
\newblock Content-based recommendation systems.
\newblock In {\em The adaptive web}, pages 325--341. Springer, 2007.

\bibitem{chen2014context}
Chaochao Chen, Xiaolin Zheng, Yan Wang, Fuxing Hong, and Zhen Lin.
\newblock Context-aware collaborative topic regression with social matrix
  factorization for recommender systems.
\newblock In {\em Twenty-Eighth AAAI Conference on Artificial Intelligence},
  2014.

\bibitem{parvin2019tcfaco}
Hashem Parvin, Parham Moradi, and Shahrokh Esmaeili.
\newblock Tcfaco: Trust-aware collaborative filtering method based on ant
  colony optimization.
\newblock {\em Expert Systems with Applications}, 118:152--168, 2019.

\bibitem{lam2008addressing}
Xuan~Nhat Lam, Thuc Vu, Trong~Duc Le, and Anh~Duc Duong.
\newblock Addressing cold-start problem in recommendation systems.
\newblock In {\em Proceedings of the 2nd international conference on Ubiquitous
  information management and communication}, pages 208--211. ACM, 2008.

\bibitem{yu2018joint}
Yonghong Yu, Yang Gao, Hao Wang, and Ruili Wang.
\newblock Joint user knowledge and matrix factorization for recommender
  systems.
\newblock {\em World Wide Web}, 21(4):1141--1163, 2018.

\bibitem{voulodimos2018deep}
Athanasios Voulodimos, Nikolaos Doulamis, Anastasios Doulamis, and Eftychios
  Protopapadakis.
\newblock Deep learning for computer vision: A brief review.
\newblock {\em Computational intelligence and neuroscience}, 2018, 2018.

\bibitem{pouyanfar2018survey}
Samira Pouyanfar, Saad Sadiq, Yilin Yan, Haiman Tian, Yudong Tao, Maria~Presa
  Reyes, Mei-Ling Shyu, Shu-Ching Chen, and SS~Iyengar.
\newblock A survey on deep learning: Algorithms, techniques, and applications.
\newblock {\em ACM Computing Surveys (CSUR)}, 51(5):92, 2018.

\bibitem{sedhain2015autorec}
Suvash Sedhain, Aditya~Krishna Menon, Scott Sanner, and Lexing Xie.
\newblock Autorec: Autoencoders meet collaborative filtering.
\newblock In {\em Proceedings of the 24th International Conference on World
  Wide Web}, pages 111--112. ACM, 2015.

\bibitem{strub2016hybrid}
Florian Strub, Romaric Gaudel, and J{\'e}r{\'e}mie Mary.
\newblock Hybrid recommender system based on autoencoders.
\newblock In {\em Proceedings of the 1st Workshop on Deep Learning for
  Recommender Systems}, pages 11--16. ACM, 2016.

\bibitem{wu2019hierarchical}
Le~Wu, Lei Chen, Richang Hong, Yanjie Fu, Xing Xie, and Meng Wang.
\newblock A hierarchical attention model for social contextual image
  recommendation.
\newblock {\em IEEE Transactions on Knowledge and Data Engineering}, 2019.

\bibitem{song2019session}
Weiping Song, Zhiping Xiao, Yifan Wang, Laurent Charlin, Ming Zhang, and Jian
  Tang.
\newblock Session-based social recommendation via dynamic graph attention
  networks.
\newblock In {\em Proceedings of the Twelfth ACM International Conference on
  Web Search and Data Mining}, pages 555--563. ACM, 2019.

\bibitem{bansal2016ask}
Trapit Bansal, David Belanger, and Andrew McCallum.
\newblock Ask the gru: Multi-task learning for deep text recommendations.
\newblock In {\em Proceedings of the 10th ACM Conference on Recommender
  Systems}, pages 107--114. ACM, 2016.

\bibitem{guo2017deepfm}
Huifeng Guo, Ruiming Tang, Yunming Ye, Zhenguo Li, and Xiuqiang He.
\newblock Deepfm: a factorization-machine based neural network for ctr
  prediction.
\newblock {\em arXiv preprint arXiv:1703.04247}, 2017.

\bibitem{sun2018attentive}
Peijie Sun, Le~Wu, and Meng Wang.
\newblock Attentive recurrent social recommendation.
\newblock In {\em The 41st International ACM SIGIR Conference on Research \&
  Development in Information Retrieval}, pages 185--194. ACM, 2018.

\bibitem{ying2018sequential}
Haochao Ying, Fuzhen Zhuang, Fuzheng Zhang, Yanchi Liu, Guandong Xu, Xing Xie,
  Hui Xiong, and Jian Wu.
\newblock Sequential recommender system based on hierarchical attention
  networks.
\newblock In {\em the 27th International Joint Conference on Artificial
  Intelligence}, 2018.

\bibitem{goodfellow2014generative}
Ian Goodfellow, Jean Pouget-Abadie, Mehdi Mirza, Bing Xu, David Warde-Farley,
  Sherjil Ozair, Aaron Courville, and Yoshua Bengio.
\newblock Generative adversarial nets.
\newblock In {\em Advances in neural information processing systems}, pages
  2672--2680, 2014.

\bibitem{he2018adversarial}
Xiangnan He, Zhankui He, Xiaoyu Du, and Tat-Seng Chua.
\newblock Adversarial personalized ranking for recommendation.
\newblock In {\em The 41st International ACM SIGIR Conference on Research \&
  Development in Information Retrieval}, pages 355--364. ACM, 2018.

\bibitem{wang2017irgan}
Jun Wang, Lantao Yu, Weinan Zhang, Yu~Gong, Yinghui Xu, Benyou Wang, Peng
  Zhang, and Dell Zhang.
\newblock Irgan: A minimax game for unifying generative and discriminative
  information retrieval models.
\newblock In {\em Proceedings of the 40th International ACM SIGIR conference on
  Research and Development in Information Retrieval}, pages 515--524. ACM,
  2017.

\bibitem{wang2018neural}
Qinyong Wang, Hongzhi Yin, Zhiting Hu, Defu Lian, Hao Wang, and Zi~Huang.
\newblock Neural memory streaming recommender networks with adversarial
  training.
\newblock In {\em Proceedings of the 24th ACM SIGKDD International Conference
  on Knowledge Discovery \& Data Mining}, pages 2467--2475. ACM, 2018.

\bibitem{wang2019minimax}
Zongwei Wang, Min Gao, Xinyi Wang, Junliang Yu, Junhao Wen, and Qingyu Xiong.
\newblock A minimax game for generative and discriminative sample models for
  recommendation.
\newblock In {\em Pacific-Asia Conference on Knowledge Discovery and Data
  Mining}, pages 420--431. Springer, 2019.

\bibitem{fan2019graph}
Wenqi Fan, Yao Ma, Qing Li, Yuan He, Eric Zhao, Jiliang Tang, and Dawei Yin.
\newblock Graph neural networks for social recommendation.
\newblock {\em arXiv preprint arXiv:1902.07243}, 2019.

\bibitem{wu2019dual}
Qitian Wu, Hengrui Zhang, Xiaofeng Gao, Peng He, Paul Weng, Han Gao, and Guihai
  Chen.
\newblock Dual graph attention networks for deep latent representation of
  multifaceted social effects in recommender systems.
\newblock {\em arXiv preprint arXiv:1903.10433}, 2019.

\bibitem{jamali2009using}
Mohsen Jamali and Martin Ester.
\newblock Using a trust network to improve top-n recommendation.
\newblock In {\em Proceedings of the third ACM conference on Recommender
  systems}, pages 181--188. ACM, 2009.

\bibitem{jamali2009trustwalker}
Mohsen Jamali and Martin Ester.
\newblock Trustwalker: a random walk model for combining trust-based and
  item-based recommendation.
\newblock In {\em Proceedings of the 15th ACM SIGKDD international conference
  on Knowledge discovery and data mining}, pages 397--406. ACM, 2009.

\bibitem{zhang2017collaborative}
Chuxu Zhang, Lu~Yu, Yan Wang, Chirag Shah, and Xiangliang Zhang.
\newblock Collaborative user network embedding for social recommender systems.
\newblock In {\em 17th SIAM International Conference on Data Mining, SDM 2017},
  pages 381--389. Society for Industrial and Applied Mathematics Publications,
  2017.

\bibitem{wen2018network}
Yufei Wen, Lei Guo, Zhumin Chen, and Jun Ma.
\newblock Network embedding based recommendation method in social networks.
\newblock In {\em Companion of the The Web Conference 2018 on The Web
  Conference 2018}, pages 11--12. International World Wide Web Conferences
  Steering Committee, 2018.

\bibitem{grover2016node2vec}
Aditya Grover and Jure Leskovec.
\newblock node2vec: Scalable feature learning for networks.
\newblock In {\em Proceedings of the 22nd ACM SIGKDD international conference
  on Knowledge discovery and data mining}, pages 855--864. ACM, 2016.

\bibitem{zhao2017social}
Zhou Zhao, Qifan Yang, Hanqing Lu, Tim Weninger, Deng Cai, Xiaofei He, and
  Yueting Zhuang.
\newblock Social-aware movie recommendation via multimodal network learning.
\newblock {\em IEEE Transactions on Multimedia}, 20(2):430--440, 2017.

\bibitem{koren2008factorization}
Yehuda Koren.
\newblock Factorization meets the neighborhood: a multifaceted collaborative
  filtering model.
\newblock In {\em Proceedings of the 14th ACM SIGKDD international conference
  on Knowledge discovery and data mining}, pages 426--434. ACM, 2008.

\bibitem{ahn2018binary}
Kwangjun Ahn, Kangwook Lee, Hyunseung Cha, and Changho Suh.
\newblock Binary rating estimation with graph side information.
\newblock In {\em Advances in Neural Information Processing Systems}, pages
  4272--4283, 2018.

\bibitem{zhang2018social}
Honglei Zhang, Gangdu Liu, and Jun Wu.
\newblock Social collaborative filtering ensemble.
\newblock In {\em Pacific Rim International Conference on Artificial
  Intelligence}, pages 1005--1017. Springer, 2018.

\bibitem{nisha2019social}
CC~Nisha and Anuraj Mohan.
\newblock A social recommender system using deep architecture and network
  embedding.
\newblock {\em Applied Intelligence}, 49(5):1937--1953, 2019.

\bibitem{wang2019trust}
Meiqi Wang, Zhiyuan Wu, Xiaoxin Sun, Guozhong Feng, and Bangzuo Zhang.
\newblock Trust-aware collaborative filtering with a denoising autoencoder.
\newblock {\em Neural Processing Letters}, 49(2):835--849, 2019.

\bibitem{wu2018collaborative}
Le~Wu, Peijie Sun, Richang Hong, Yong Ge, and Meng Wang.
\newblock Collaborative neural social recommendation.
\newblock {\em IEEE Transactions on Systems, Man, and Cybernetics: Systems},
  2018.

\bibitem{deng2016deep}
Shuiguang Deng, Longtao Huang, Guandong Xu, Xindong Wu, and Zhaohui Wu.
\newblock On deep learning for trust-aware recommendations in social networks.
\newblock {\em IEEE transactions on neural networks and learning systems},
  28(5):1164--1177, 2016.

\bibitem{rafailidis2017recommendation}
Dimitrios Rafailidis and Fabio Crestani.
\newblock Recommendation with social relationships via deep learning.
\newblock In {\em Proceedings of the ACM SIGIR International Conference on
  Theory of Information Retrieval}, pages 151--158. ACM, 2017.

\bibitem{wu2019graph}
Liwei Wu, Hsiang-Fu Yu, Nikhil Rao, James Sharpnack, and Cho-Jui Hsieh.
\newblock Graph dna: Deep neighborhood aware graph encoding for collaborative
  filtering.
\newblock {\em arXiv preprint arXiv:1905.12217}, 2019.

\bibitem{ying2018graph}
Rex Ying, Ruining He, Kaifeng Chen, Pong Eksombatchai, William~L Hamilton, and
  Jure Leskovec.
\newblock Graph convolutional neural networks for web-scale recommender
  systems.
\newblock In {\em Proceedings of the 24th ACM SIGKDD International Conference
  on Knowledge Discovery \& Data Mining}, pages 974--983. ACM, 2018.

\bibitem{fan2019deep}
Wenqi Fan, Tyler Derr, Yao Ma, Jianping Wang, Jiliang Tang, and Qing Li.
\newblock Deep adversarial social recommendation.
\newblock {\em arXiv preprint arXiv:1905.13160}, 2019.

\bibitem{xiao2019variational}
Teng Xiao, Hui Tian, and Hong Shen.
\newblock Variational deep collaborative matrix factorization for social
  recommendation.
\newblock In {\em Pacific-Asia Conference on Knowledge Discovery and Data
  Mining}, pages 426--437. Springer, 2019.

\bibitem{bao2018contextual}
Hongfeng Bao, Le~Wu, and Peijie Sun.
\newblock Contextual attention model for social recommendation.
\newblock In {\em Pacific Rim Conference on Multimedia}, pages 630--641.
  Springer, 2018.

\bibitem{liu2018social2}
Chi~Harold Liu, Jie Xu, Jian Tang, and Jon Crowcroft.
\newblock Social-aware sequential modeling of user interests: A deep learning
  approach.
\newblock {\em IEEE Transactions on Knowledge and Data Engineering}, 2018.

\bibitem{gao2017unified}
Junyu Gao, Tianzhu Zhang, and Changsheng Xu.
\newblock A unified personalized video recommendation via dynamic recurrent
  neural networks.
\newblock In {\em Proceedings of the 25th ACM international conference on
  Multimedia}, pages 127--135. ACM, 2017.

\bibitem{geng2015learning}
Xue Geng, Hanwang Zhang, Jingwen Bian, and Tat-Seng Chua.
\newblock Learning image and user features for recommendation in social
  networks.
\newblock In {\em Proceedings of the IEEE International Conference on Computer
  Vision}, pages 4274--4282, 2015.

\bibitem{liu2015learning}
Shaowei Liu, Peng Cui, Wenwu Zhu, and Shiqiang Yang.
\newblock Learning socially embedded visual representation from scratch.
\newblock In {\em Proceedings of the 23rd ACM international conference on
  Multimedia}, pages 109--118. ACM, 2015.

\bibitem{chen2019social}
Chong Chen, Min Zhang, Yiqun Liu, and Shaoping Ma.
\newblock Social attentional memory network: Modeling aspect-and friend-level
  differences in recommendation.
\newblock In {\em Proceedings of the Twelfth ACM International Conference on
  Web Search and Data Mining}, pages 177--185. ACM, 2019.

\bibitem{rafailidis2019neural}
Dimitrios Rafailidis and Gerhard Weiss.
\newblock A neural attention model for adaptive learning of social friends'
  preferences.
\newblock {\em arXiv preprint arXiv:1907.01644}, 2019.

\bibitem{xiao2017neural}
Lin Xiao, Zhang Min, Liu Yiqun, and Ma~Shaoping.
\newblock A neural network model for social-aware recommendation.
\newblock In {\em Asia Information Retrieval Symposium}, pages 125--137.
  Springer, 2017.

\bibitem{fan2018deep}
Wenqi Fan, Qing Li, and Min Cheng.
\newblock Deep modeling of social relations for recommendation.
\newblock In {\em Thirty-Second AAAI Conference on Artificial Intelligence},
  2018.

\bibitem{wang2017item}
Xiang Wang, Xiangnan He, Liqiang Nie, and Tat-Seng Chua.
\newblock Item silk road: Recommending items from information domains to social
  users.
\newblock In {\em Proceedings of the 40th International ACM SIGIR conference on
  Research and Development in Information Retrieval}, pages 185--194. ACM,
  2017.

\bibitem{liu2018social}
Chun-Yi Liu, Chuan Zhou, Jia Wu, Yue Hu, and Li~Guo.
\newblock Social recommendation with an essential preference space.
\newblock In {\em Thirty-Second AAAI Conference on Artificial Intelligence},
  2018.

\bibitem{vincent2010stacked}
Pascal Vincent, Hugo Larochelle, Isabelle Lajoie, Yoshua Bengio, and
  Pierre-Antoine Manzagol.
\newblock Stacked denoising autoencoders: Learning useful representations in a
  deep network with a local denoising criterion.
\newblock {\em Journal of machine learning research}, 11(Dec):3371--3408, 2010.

\bibitem{lai2015recurrent}
Siwei Lai, Liheng Xu, Kang Liu, and Jun Zhao.
\newblock Recurrent convolutional neural networks for text classification.
\newblock In {\em Twenty-ninth AAAI conference on artificial intelligence},
  2015.

\bibitem{pinheiro2014recurrent}
Pedro~HO Pinheiro and Ronan Collobert.
\newblock Recurrent convolutional neural networks for scene labeling.
\newblock In {\em 31st International Conference on Machine Learning (ICML)},
  number CONF, 2014.

\bibitem{wu2016personal}
Sai Wu, Weichao Ren, Chengchao Yu, Gang Chen, Dongxiang Zhang, and Jingbo Zhu.
\newblock Personal recommendation using deep recurrent neural networks in
  netease.
\newblock In {\em 2016 IEEE 32nd International Conference on Data Engineering
  (ICDE)}, pages 1218--1229. IEEE, 2016.

\bibitem{donkers2017sequential}
Tim Donkers, Benedikt Loepp, and J{\"u}rgen Ziegler.
\newblock Sequential user-based recurrent neural network recommendations.
\newblock In {\em Proceedings of the Eleventh ACM Conference on Recommender
  Systems}, pages 152--160. ACM, 2017.

\bibitem{wu2017recurrent}
Chao-Yuan Wu, Amr Ahmed, Alex Beutel, Alexander~J Smola, and How Jing.
\newblock Recurrent recommender networks.
\newblock In {\em Proceedings of the tenth ACM international conference on web
  search and data mining}, pages 495--503. ACM, 2017.

\bibitem{wu2016joint}
Chao-Yuan Wu, Amr Ahmed, Alex Beutel, and Alexander~J Smola.
\newblock Joint training of ratings and reviews with recurrent recommender
  networks.
\newblock 2016.

\bibitem{kingma2013auto}
Diederik~P Kingma and Max Welling.
\newblock Auto-encoding variational bayes.
\newblock {\em arXiv preprint arXiv:1312.6114}, 2013.

\bibitem{karamanolakis2018item}
Giannis Karamanolakis, Kevin~Raji Cherian, Ananth~Ravi Narayan, Jie Yuan,
  Da~Tang, and Tony Jebara.
\newblock Item recommendation with variational autoencoders and heterogeneous
  priors.
\newblock In {\em Proceedings of the 3rd Workshop on Deep Learning for
  Recommender Systems}, pages 10--14. ACM, 2018.

\bibitem{perozzi2014deepwalk}
Bryan Perozzi, Rami Al-Rfou, and Steven Skiena.
\newblock Deepwalk: Online learning of social representations.
\newblock In {\em Proceedings of the 20th ACM SIGKDD international conference
  on Knowledge discovery and data mining}, pages 701--710. ACM, 2014.

\bibitem{luong2015effective}
Minh-Thang Luong, Hieu Pham, and Christopher~D Manning.
\newblock Effective approaches to attention-based neural machine translation.
\newblock {\em arXiv preprint arXiv:1508.04025}, 2015.

\bibitem{xu2015show}
Kelvin Xu, Jimmy Ba, Ryan Kiros, Kyunghyun Cho, Aaron Courville, Ruslan
  Salakhudinov, Rich Zemel, and Yoshua Bengio.
\newblock Show, attend and tell: Neural image caption generation with visual
  attention.
\newblock In {\em International conference on machine learning}, pages
  2048--2057, 2015.

\bibitem{elkahky2015multi}
Ali~Mamdouh Elkahky, Yang Song, and Xiaodong He.
\newblock A multi-view deep learning approach for cross domain user modeling in
  recommendation systems.
\newblock In {\em Proceedings of the 24th International Conference on World
  Wide Web}, pages 278--288. International World Wide Web Conferences Steering
  Committee, 2015.

\bibitem{guo2015librec}
Guibing Guo, Jie Zhang, Zhu Sun, and Neil Yorke-Smith.
\newblock Librec: A java library for recommender systems.
\newblock In {\em UMAP Workshops}, volume~4, 2015.

\bibitem{lecun2015deep}
Yann LeCun, Yoshua Bengio, and Geoffrey Hinton.
\newblock Deep learning.
\newblock {\em nature}, 521(7553):436, 2015.

\bibitem{alonso2019robust}
Santiago Alonso, Jes{\'u}s Bobadilla, Fernando Ortega, and Ricardo Moya.
\newblock Robust model-based reliability approach to tackle shilling attacks in
  collaborative filtering recommender systems.
\newblock {\em IEEE Access}, 7:41782--41798, 2019.

\bibitem{zhang2017robust}
Fuzhi Zhang, Yuanli Lu, Jianmin Chen, Shaoshuai Liu, and Zhoujun Ling.
\newblock Robust collaborative filtering based on non-negative matrix
  factorization and r1-norm.
\newblock {\em Knowledge-based systems}, 118:177--190, 2017.

\bibitem{yu2017novel}
Hongtao Yu, Ruibo Gao, Kun Wang, and Fuzhi Zhang.
\newblock A novel robust recommendation method based on kernel matrix
  factorization.
\newblock {\em Journal of Intelligent \& Fuzzy Systems}, 32(3):2101--2109,
  2017.

\bibitem{chalapathy2019deep}
Raghavendra Chalapathy and Sanjay Chawla.
\newblock Deep learning for anomaly detection: A survey.
\newblock {\em arXiv preprint arXiv:1901.03407}, 2019.

\bibitem{ren2019learning}
Yafeng Ren and Donghong Ji.
\newblock Learning to detect deceptive opinion spam: A survey.
\newblock {\em IEEE Access}, 7:42934--42945, 2019.

\bibitem{zhao2017towards}
Siyuan Zhao, Zhiwei Xu, Limin Liu, and Mengjie Guo.
\newblock Towards accurate deceptive opinion spam detection based on word
  order-preserving cnn.
\newblock {\em arXiv preprint arXiv:1711.09181}, 2017.

\bibitem{zhang2018dri}
Wen Zhang, Yuhang Du, Taketoshi Yoshida, and Qing Wang.
\newblock Dri-rcnn: An approach to deceptive review identification using
  recurrent convolutional neural network.
\newblock {\em Information Processing \& Management}, 54(4):576--592, 2018.

\bibitem{li2015learning}
Luyang Li, Wenjing Ren, Bing Qin, and Ting Liu.
\newblock Learning document representation for deceptive opinion spam
  detection.
\newblock In {\em Chinese Computational Linguistics and Natural Language
  Processing Based on Naturally Annotated Big Data}, pages 393--404. Springer,
  2015.

\bibitem{wang2017detecting}
Xuepeng Wang, Kang Liu, and Jun Zhao.
\newblock Detecting deceptive review spam via attention-based neural networks.
\newblock In {\em National CCF Conference on Natural Language Processing and
  Chinese Computing}, pages 866--876. Springer, 2017.

\bibitem{wang2017handling}
Xuepeng Wang, Kang Liu, and Jun Zhao.
\newblock Handling cold-start problem in review spam detection by jointly
  embedding texts and behaviors.
\newblock In {\em Proceedings of the 55th Annual Meeting of the Association for
  Computational Linguistics (Volume 1: Long Papers)}, pages 366--376, 2017.

\bibitem{chung2014empirical}
Junyoung Chung, Caglar Gulcehre, KyungHyun Cho, and Yoshua Bengio.
\newblock Empirical evaluation of gated recurrent neural networks on sequence
  modeling.
\newblock {\em arXiv preprint arXiv:1412.3555}, 2014.

\bibitem{hochreiter1997long}
Sepp Hochreiter and J{\"u}rgen Schmidhuber.
\newblock Long short-term memory.
\newblock {\em Neural computation}, 9(8):1735--1780, 1997.

\bibitem{ren2016deceptive}
Yafeng Ren and Yue Zhang.
\newblock Deceptive opinion spam detection using neural network.
\newblock In {\em Proceedings of COLING 2016, the 26th International Conference
  on Computational Linguistics: Technical Papers}, pages 140--150, 2016.

\bibitem{wang2018detecting}
Chih-Chien Wang, Min-Yuh Day, Chien-Chang Chen, and Jia-Wei Liou.
\newblock Detecting spamming reviews using long short-term memory recurrent
  neural network framework.
\newblock In {\em Proceedings of the 2nd International Conference on
  E-commerce, E-Business and E-Government}, pages 16--20. ACM, 2018.

\bibitem{wang2016learning}
Xuepeng Wang, Kang Liu, Shizhu He, and Jun Zhao.
\newblock Learning to represent review with tensor decomposition for spam
  detection.
\newblock In {\em Proceedings of the 2016 Conference on Empirical Methods in
  Natural Language Processing}, pages 866--875, 2016.

\bibitem{dong2018opinion}
Manqing Dong, Lina Yao, Xianzhi Wang, Boualem Benatallah, Chaoran Huang, and
  Xiaodong Ning.
\newblock Opinion fraud detection via neural autoencoder decision forest.
\newblock {\em Pattern Recognition Letters}, 2018.

\bibitem{aghakhani2018detecting}
Hojjat Aghakhani, Aravind Machiry, Shirin Nilizadeh, Christopher Kruegel, and
  Giovanni Vigna.
\newblock Detecting deceptive reviews using generative adversarial networks.
\newblock In {\em 2018 IEEE Security and Privacy Workshops (SPW)}, pages
  89--95. IEEE, 2018.

\bibitem{wu2016collaborative}
Yao Wu, Christopher DuBois, Alice~X Zheng, and Martin Ester.
\newblock Collaborative denoising auto-encoders for top-n recommender systems.
\newblock In {\em Proceedings of the Ninth ACM International Conference on Web
  Search and Data Mining}, pages 153--162. ACM, 2016.

\bibitem{wang2016collaborative}
Hao Wang, SHI Xingjian, and Dit-Yan Yeung.
\newblock Collaborative recurrent autoencoder: Recommend while learning to fill
  in the blanks.
\newblock In {\em Advances in Neural Information Processing Systems}, pages
  415--423, 2016.

\bibitem{yuan2019adversarial}
Feng Yuan, Lina Yao, and Boualem Benatallah.
\newblock Adversarial collaborative neural network for robust recommendation.
\newblock 2019.

\bibitem{jhamb2018attentive}
Yogesh Jhamb, Travis Ebesu, and Yi~Fang.
\newblock Attentive contextual denoising autoencoder for recommendation.
\newblock In {\em Proceedings of the 2018 ACM SIGIR International Conference on
  Theory of Information Retrieval}, pages 27--34. ACM, 2018.

\bibitem{zhou2018atrank}
Chang Zhou, Jinze Bai, Junshuai Song, Xiaofei Liu, Zhengchao Zhao, Xiusi Chen,
  and Jun Gao.
\newblock Atrank: An attention-based user behavior modeling framework for
  recommendation.
\newblock In {\em Thirty-Second AAAI Conference on Artificial Intelligence},
  2018.

\bibitem{loyola2017modeling}
Pablo Loyola, Chen Liu, and Yu~Hirate.
\newblock Modeling user session and intent with an attention-based
  encoder-decoder architecture.
\newblock In {\em Proceedings of the Eleventh ACM Conference on Recommender
  Systems}, pages 147--151. ACM, 2017.

\bibitem{liu2018stamp}
Qiao Liu, Yifu Zeng, Refuoe Mokhosi, and Haibin Zhang.
\newblock Stamp: short-term attention/memory priority model for session-based
  recommendation.
\newblock In {\em Proceedings of the 24th ACM SIGKDD International Conference
  on Knowledge Discovery \& Data Mining}, pages 1831--1839. ACM, 2018.

\bibitem{seo2017representation}
Sungyong Seo, Jing Huang, Hao Yang, and Yan Liu.
\newblock Representation learning of users and items for review rating
  prediction using attention-based convolutional neural network.
\newblock In {\em 3rd international workshop on machine learning methods for
  recommender systems (MLRec)(SDM’17)}, 2017.

\bibitem{tay2018latent}
Yi~Tay, Luu Anh~Tuan, and Siu~Cheung Hui.
\newblock Latent relational metric learning via memory-based attention for
  collaborative ranking.
\newblock In {\em Proceedings of the 2018 World Wide Web Conference}, pages
  729--739. International World Wide Web Conferences Steering Committee, 2018.

\bibitem{chen2017attentive}
Jingyuan Chen, Hanwang Zhang, Xiangnan He, Liqiang Nie, Wei Liu, and Tat-Seng
  Chua.
\newblock Attentive collaborative filtering: Multimedia recommendation with
  item-and component-level attention.
\newblock In {\em Proceedings of the 40th International ACM SIGIR conference on
  Research and Development in Information Retrieval}, pages 335--344. ACM,
  2017.

\bibitem{tintarev2011designing}
Nava Tintarev and Judith Masthoff.
\newblock Designing and evaluating explanations for recommender systems.
\newblock In {\em Recommender systems handbook}, pages 479--510. Springer,
  2011.

\bibitem{sharma2013social}
Amit Sharma and Dan Cosley.
\newblock Do social explanations work?: studying and modeling the effects of
  social explanations in recommender systems.
\newblock In {\em Proceedings of the 22nd international conference on World
  Wide Web}, pages 1133--1144. ACM, 2013.

\bibitem{chen2018neural}
Chong Chen, Min Zhang, Yiqun Liu, and Shaoping Ma.
\newblock Neural attentional rating regression with review-level explanations.
\newblock In {\em Proceedings of the 2018 World Wide Web Conference on World
  Wide Web}, pages 1583--1592. International World Wide Web Conferences
  Steering Committee, 2018.

\bibitem{chen2018visually}
Xu~Chen, Yongfeng Zhang, Hongteng Xu, Yixin Cao, Zheng Qin, and Hongyuan Zha.
\newblock Visually explainable recommendation.
\newblock {\em arXiv preprint arXiv:1801.10288}, 2018.

\bibitem{wang2018reinforcement}
Xiting Wang, Yiru Chen, Jie Yang, Le~Wu, Zhengtao Wu, and Xing Xie.
\newblock A reinforcement learning framework for explainable recommendation.
\newblock In {\em 2018 IEEE International Conference on Data Mining (ICDM)},
  pages 587--596. IEEE, 2018.

\bibitem{mcinerney2018explore}
James McInerney, Benjamin Lacker, Samantha Hansen, Karl Higley, Hugues
  Bouchard, Alois Gruson, and Rishabh Mehrotra.
\newblock Explore, exploit, and explain: personalizing explainable
  recommendations with bandits.
\newblock In {\em Proceedings of the 12th ACM Conference on Recommender
  Systems}, pages 31--39. ACM, 2018.

\bibitem{abdollahi2016explainable}
Behnoush Abdollahi and Olfa Nasraoui.
\newblock Explainable restricted boltzmann machines for collaborative
  filtering.
\newblock {\em arXiv preprint arXiv:1606.07129}, 2016.

\bibitem{wu2019context}
Libing Wu, Cong Quan, Chenliang Li, Qian Wang, Bolong Zheng, and Xiangyang Luo.
\newblock A context-aware user-item representation learning for item
  recommendation.
\newblock {\em ACM Transactions on Information Systems (TOIS)}, 37(2):22, 2019.

\bibitem{chen2019dynamic}
Xu~Chen, Yongfeng Zhang, and Zheng Qin.
\newblock Dynamic explainable recommendation based on neural attentive models.
\newblock In {\em Proceedings of the 33rd AAAI Conference on Artificial
  Intelligence}, 2019.

\bibitem{cong2019hierarchical}
Dawei Cong, Yanyan Zhao, Bing Qin, Yu~Han, Murray Zhang, Alden Liu, and Nat
  Chen.
\newblock Hierarchical attention based neural network for explainable
  recommendation.
\newblock In {\em Proceedings of the 2019 on International Conference on
  Multimedia Retrieval}, pages 373--381. ACM, 2019.

\bibitem{costa2018automatic}
Felipe Costa, Sixun Ouyang, Peter Dolog, and Aonghus Lawlor.
\newblock Automatic generation of natural language explanations.
\newblock In {\em Proceedings of the 23rd International Conference on
  Intelligent User Interfaces Companion}, page~57. ACM, 2018.

\bibitem{li2017neural}
Piji Li, Zihao Wang, Zhaochun Ren, Lidong Bing, and Wai Lam.
\newblock Neural rating regression with abstractive tips generation for
  recommendation.
\newblock In {\em Proceedings of the 40th International ACM SIGIR conference on
  Research and Development in Information Retrieval}, pages 345--354. ACM,
  2017.

\bibitem{lu2018like}
Yichao Lu, Ruihai Dong, and Barry Smyth.
\newblock Why i like it: multi-task learning for recommendation and
  explanation.
\newblock In {\em Proceedings of the 12th ACM Conference on Recommender
  Systems}, pages 4--12. ACM, 2018.

\bibitem{ouyang2018improving}
Sixun Ouyang, Aonghus Lawlor, Felipe Costa, and Peter Dolog.
\newblock Improving explainable recommendations with synthetic reviews.
\newblock {\em arXiv preprint arXiv:1807.06978}, 2018.

\bibitem{zhao2018you}
Guoshuai Zhao, Hao Fu, Ruihua Song, Tetsuya Sakai, Xing Xie, and Xueming Qian.
\newblock Why you should listen to this song: Reason generation for explainable
  recommendation.
\newblock In {\em 2018 IEEE International Conference on Data Mining Workshops
  (ICDMW)}, pages 1316--1322. IEEE, 2018.

\bibitem{suzuki2018toward}
Takafumi Suzuki, Satoshi Oyama, and Masahito Kurihara.
\newblock Toward explainable recommendations: Generating review text from
  multicriteria evaluation data.
\newblock In {\em 2018 IEEE International Conference on Big Data (Big Data)},
  pages 3549--3551. IEEE, 2018.

\bibitem{lin2018neural}
Jionghao Lin and Yiren Liu.
\newblock A neural network based explainable recommender system.
\newblock {\em arXiv preprint arXiv:1812.11740}, 2018.

\bibitem{hou2019explainable}
Min Hou, Le~Wu, Enhong Chen, Zhi Li, Vincent~W Zheng, and Qi~Liu.
\newblock Explainable fashion recommendation: A semantic attribute region
  guided approach.
\newblock {\em arXiv preprint arXiv:1905.12862}, 2019.

\bibitem{kang2017visually}
Wang-Cheng Kang, Chen Fang, Zhaowen Wang, and Julian McAuley.
\newblock Visually-aware fashion recommendation and design with generative
  image models.
\newblock In {\em 2017 IEEE International Conference on Data Mining (ICDM)},
  pages 207--216. IEEE, 2017.

\bibitem{kumar2019c}
Sudhir Kumar and Mithun~Das Gupta.
\newblock cgan: Complementary fashion item recommendation.
\newblock {\em arXiv preprint arXiv:1906.05596}, 2019.

\bibitem{bharadhwaj2018layer}
Homanga Bharadhwaj.
\newblock Layer-wise relevance propagation for explainable recommendations.
\newblock {\em arXiv preprint arXiv:1807.06160}, 2018.

\bibitem{tang2018personalized}
Jiaxi Tang and Ke~Wang.
\newblock Personalized top-n sequential recommendation via convolutional
  sequence embedding.
\newblock In {\em Proceedings of the Eleventh ACM International Conference on
  Web Search and Data Mining}, pages 565--573. ACM, 2018.

\bibitem{chen2018sequential}
Xu~Chen, Hongteng Xu, Yongfeng Zhang, Jiaxi Tang, Yixin Cao, Zheng Qin, and
  Hongyuan Zha.
\newblock Sequential recommendation with user memory networks.
\newblock In {\em Proceedings of the eleventh ACM international conference on
  web search and data mining}, pages 108--116. ACM, 2018.

\bibitem{zheng2017joint}
Lei Zheng, Vahid Noroozi, and Philip~S Yu.
\newblock Joint deep modeling of users and items using reviews for
  recommendation.
\newblock In {\em Proceedings of the Tenth ACM International Conference on Web
  Search and Data Mining}, pages 425--434. ACM, 2017.

\bibitem{mikolov2010recurrent}
Tom{\'a}{\v{s}} Mikolov, Martin Karafi{\'a}t, Luk{\'a}{\v{s}} Burget, Jan
  {\v{C}}ernock{\`y}, and Sanjeev Khudanpur.
\newblock Recurrent neural network based language model.
\newblock In {\em Eleventh annual conference of the international speech
  communication association}, 2010.

\bibitem{he2016vbpr}
Ruining He and Julian McAuley.
\newblock Vbpr: visual bayesian personalized ranking from implicit feedback.
\newblock In {\em Thirtieth AAAI Conference on Artificial Intelligence}, 2016.

\bibitem{liu2017deepstyle}
Qiang Liu, Shu Wu, and Liang Wang.
\newblock Deepstyle: Learning user preferences for visual recommendation.
\newblock In {\em Proceedings of the 40th International ACM SIGIR Conference on
  Research and Development in Information Retrieval}, pages 841--844. ACM,
  2017.

\bibitem{wang2017your}
Suhang Wang, Yilin Wang, Jiliang Tang, Kai Shu, Suhas Ranganath, and Huan Liu.
\newblock What your images reveal: Exploiting visual contents for
  point-of-interest recommendation.
\newblock In {\em Proceedings of the 26th International Conference on World
  Wide Web}, pages 391--400. International World Wide Web Conferences Steering
  Committee, 2017.

\bibitem{yu2018aesthetic}
Wenhui Yu, Huidi Zhang, Xiangnan He, Xu~Chen, Li~Xiong, and Zheng Qin.
\newblock Aesthetic-based clothing recommendation.
\newblock In {\em Proceedings of the 2018 World Wide Web Conference on World
  Wide Web}, pages 649--658. International World Wide Web Conferences Steering
  Committee, 2018.

\bibitem{he2016deep}
Kaiming He, Xiangyu Zhang, Shaoqing Ren, and Jian Sun.
\newblock Deep residual learning for image recognition.
\newblock In {\em Proceedings of the IEEE conference on computer vision and
  pattern recognition}, pages 770--778, 2016.

\bibitem{selvaraju2017grad}
Ramprasaath~R Selvaraju, Michael Cogswell, Abhishek Das, Ramakrishna Vedantam,
  Devi Parikh, and Dhruv Batra.
\newblock Grad-cam: Visual explanations from deep networks via gradient-based
  localization.
\newblock In {\em Proceedings of the IEEE International Conference on Computer
  Vision}, pages 618--626, 2017.

\bibitem{hadsell2006dimensionality}
Raia Hadsell, Sumit Chopra, and Yann LeCun.
\newblock Dimensionality reduction by learning an invariant mapping.
\newblock In {\em 2006 IEEE Computer Society Conference on Computer Vision and
  Pattern Recognition (CVPR'06)}, volume~2, pages 1735--1742. IEEE, 2006.

\bibitem{mirza2014conditional}
Mehdi Mirza and Simon Osindero.
\newblock Conditional generative adversarial nets.
\newblock {\em arXiv preprint arXiv:1411.1784}, 2014.

\bibitem{sajjadi2018tempered}
Mehdi~SM Sajjadi, Giambattista Parascandolo, Arash Mehrjou, and Bernhard
  Sch{\"o}lkopf.
\newblock Tempered adversarial networks.
\newblock {\em arXiv preprint arXiv:1802.04374}, 2018.

\bibitem{bach2015pixel}
Sebastian Bach, Alexander Binder, Gr{\'e}goire Montavon, Frederick Klauschen,
  Klaus-Robert M{\"u}ller, and Wojciech Samek.
\newblock On pixel-wise explanations for non-linear classifier decisions by
  layer-wise relevance propagation.
\newblock {\em PloS one}, 10(7):e0130140, 2015.

\bibitem{gatys2015texture}
Leon Gatys, Alexander~S Ecker, and Matthias Bethge.
\newblock Texture synthesis using convolutional neural networks.
\newblock In {\em Advances in neural information processing systems}, pages
  262--270, 2015.

\bibitem{vinagre2015overview}
Jo{\~a}o Vinagre, Al{\'\i}pio~M{\'a}rio Jorge, and Jo{\~a}o Gama.
\newblock An overview on the exploitation of time in collaborative filtering.
\newblock {\em Wiley interdisciplinary reviews: Data mining and knowledge
  discovery}, 5(5):195--215, 2015.

\bibitem{kiperwasser2016simple}
Eliyahu Kiperwasser and Yoav Goldberg.
\newblock Simple and accurate dependency parsing using bidirectional lstm
  feature representations.
\newblock {\em Transactions of the Association for Computational Linguistics},
  4:313--327, 2016.

\bibitem{goldberg2012dynamic}
Yoav Goldberg and Joakim Nivre.
\newblock A dynamic oracle for arc-eager dependency parsing.
\newblock {\em Proceedings of COLING 2012}, pages 959--976, 2012.

\bibitem{li2017social}
Wentao Li, Min Gao, Wenge Rong, Junhao Wen, Qingyu Xiong, Ruixi Jia, and Tong
  Dou.
\newblock Social recommendation using euclidean embedding.
\newblock In {\em 2017 International Joint Conference on Neural Networks
  (IJCNN)}, pages 589--595. IEEE, 2017.

\bibitem{park2017uniwalk}
Haekyu Park, Hyunsik Jeon, Junghwan Kim, Beunguk Ahn, and U~Kang.
\newblock Uniwalk: Explainable and accurate recommendation for rating and
  network data.
\newblock {\em arXiv preprint arXiv:1710.07134}, 2017.

\bibitem{defferrard2016convolutional}
Micha{\"e}l Defferrard, Xavier Bresson, and Pierre Vandergheynst.
\newblock Convolutional neural networks on graphs with fast localized spectral
  filtering.
\newblock In {\em Advances in neural information processing systems}, pages
  3844--3852, 2016.

\bibitem{ouyang2019learning}
Yi~Ouyang, Bin Guo, Xing Tang, Xiuqiang He, Jian Xiong, and Zhiwen Yu.
\newblock Learning cross-domain representation with multi-graph neural network.
\newblock {\em arXiv preprint arXiv:1905.10095}, 2019.

\bibitem{vanschoren2018meta}
Joaquin Vanschoren.
\newblock Meta-learning: A survey.
\newblock {\em arXiv preprint arXiv:1810.03548}, 2018.

\bibitem{yao2019recommendations}
Lina Yao, Xianzhi Wang, Quan~Z Sheng, Schahram Dustdar, and Shuai Zhang.
\newblock Recommendations on the internet of things: Requirements, challenges,
  and directions.
\newblock {\em IEEE Internet Computing}, 23(3):46--54, 2019.

\bibitem{crosby2016blockchain}
Michael Crosby, Pradan Pattanayak, Sanjeev Verma, Vignesh Kalyanaraman, et~al.
\newblock Blockchain technology: Beyond bitcoin.
\newblock {\em Applied Innovation}, 2(6-10):71, 2016.

\bibitem{weng2018deepchain}
Jia-Si Weng, Jian Weng, Ming Li, Yue Zhang, and Weiqi Luo.
\newblock Deepchain: Auditable and privacy-preserving deep learning with
  blockchain-based incentive.
\newblock {\em IACR Cryptology ePrint Archive}, 2018:679, 2018.

\end{thebibliography}


\begin{thebibliography}{1}

\bibitem{kour2014real}
George Kour and Raid Saabne.
\newblock Real-time segmentation of on-line handwritten arabic script.
\newblock In {\em Frontiers in Handwriting Recognition (ICFHR), 2014 14th
  International Conference on}, pages 417--422. IEEE, 2014.

\bibitem{kour2014fast}
George Kour and Raid Saabne.
\newblock Fast classification of handwritten on-line arabic characters.
\newblock In {\em Soft Computing and Pattern Recognition (SoCPaR), 2014 6th
  International Conference of}, pages 312--318. IEEE, 2014.

\bibitem{hadash2018estimate}
Guy Hadash, Einat Kermany, Boaz Carmeli, Ofer Lavi, George Kour, and Alon
  Jacovi.
\newblock Estimate and replace: A novel approach to integrating deep neural
  networks with existing applications.
\newblock {\em arXiv preprint arXiv:1804.09028}, 2018.

\end{thebibliography}

\end{document}